\documentclass[11pt]{scrartcl}
\usepackage[utf8]{inputenc}

\usepackage{comment}

\usepackage{graphicx}
\usepackage{amsmath}
\usepackage{amssymb}
\usepackage{amsthm}

\usepackage{pgfplots}
\usepackage[font=small,labelfont=bf]{caption}

\usepackage[font=footnotesize]{subcaption}
\usepackage{listofitems}

\usepackage{xcolor}
\definecolor{darkgreen}{rgb}{0,.5,0}
\definecolor{darkblue}{rgb}{0,0,.6}
\definecolor{darkred}{rgb}{.6,0,0}
\definecolor{pink}{RGB}{255,99,170}

\colorlet{inputlayercolor}{blue!80!black}
\colorlet{hiddenlayercolor}{pink}
\colorlet{outputlayercolor}{green!50!black}

\DeclareUnicodeCharacter{03F5}{$\varepsilon$}

\usepackage[
pdfauthor={Paul-Hermann Balduf, Kimia Shaban},			 
pdfcreator={LaTeX with hyperref},
colorlinks=true,	 
linkcolor=darkred,	  
citecolor=darkgreen,	 
filecolor=black,	 
urlcolor=darkblue,	 
bookmarks=true,
plainpages=false,
pdfpagelabels=true
]{hyperref} 

\usepackage[nameinlink,noabbrev 
]{cleveref}	
\crefname{figure}{figure}{figures}

\usepackage{placeins} 

\usepackage{csquotes} 

\usepackage{tabularray}

\numberwithin{equation}{section}

\usepackage[
	backend=bibtex,
	style=phys,
	date=year,
	url=true,
	isbn=false
]{biblatex}
\AtEveryBibitem{
	\clearfield{urlyear}
	\clearfield{urlmonth}
}

\usepackage{tikz}
\usetikzlibrary{arrows.meta}
\usetikzlibrary{calc}

\tikzset{>=latex} 

\tikzstyle{input feature}=[thick, inner sep=0.5,outer sep=0.6,minimum width=1cm,minimum height=.6cm,rectangle, draw=inputlayercolor!30!black,fill=inputlayercolor!30,font={\footnotesize}]
\tikzstyle{node in}=[thick,circle,inner sep=0.5,outer sep=0.6,minimum size =.7cm,  draw=inputlayercolor!30!black,fill=inputlayercolor!30,font={\footnotesize}]
\tikzstyle{node hidden}=[thick,circle,inner sep=0.5,outer sep=0.6,blue!20!black,minimum size=.7cm, draw=hiddenlayercolor!30!black, fill=hiddenlayercolor, font={\footnotesize}]
\tikzstyle{node out}=[thick,inner sep=0.5,outer sep=1mm,  minimum width=1cm,minimum height=.6cm, rectangle, draw=outputlayercolor!30!black,fill=outputlayercolor!30 ,font={\small}]

\newcommand{\abs}[1]{\lvert #1 \rvert}
\renewcommand{\d}{\textnormal{d}}

\newcommand{\period}{\mathcal P}
\newcommand{\Aut}{\operatorname{Aut}}

\title{Predicting Feynman periods in $\phi^4$--theory}
\author{Paul-Hermann Balduf \footnote{University of Waterloo, Ontario, Canada.}~\footnote{Associate Postdoc at Perimeter Institute for Theoretical Physics, Waterloo, Ontario, Canada.} , Kimia Shaban\footnotemark[1]}

\begin{document}

\maketitle

\begin{abstract}

	We present efficient data--driven approaches to predict Feynman periods in $\phi^4$--theory from properties of the underlying Feynman graphs. We find that the numbers of cuts and cycles determines the period to approximately 2\% accuracy. Hepp bound and Martin invariant allow to predict the period with accuracy much better than 1\%. In most cases, the period is a multi-linear function of the parameters in question. Besides classical correlation analysis, we also investigate the usefulness of machine-learning algorithms to predict the period. When sufficiently many properties of the graph are used, the period can be predicted with better than $0.05\%$ relative accuracy. 
	
	We use one of the constructed prediction models for   weighted Monte-Carlo sampling of Feynman graphs, and compute the primitive contribution to the beta function of $\phi^4$-theory at $L\in \left \lbrace 13, 14, 15, 16 \right \rbrace $ loops. Our results confirm the previously known numerical estimates of the primitive beta function and improve their accuracy. Compared to uniform random sampling of graphs, our new algorithm reaches 35-fold higher accuracy in fixed runtime, or requires 1000-fold less runtime to reach a given accuracy. 
	
	The data set of all periods computed for this work, combined with a previous data set, is made publicly available. Besides the physical application, it could serve as a benchmark for graph-based machine learning algorithms.
\end{abstract}

\newpage

\tableofcontents

\newpage

\section{Introduction}\label{sec:introduction}

\subsection{Motivation and Background}\label{sec:motivation}

This paper concerns \emph{Feynman periods} \cite{broadhurst_knots_1995,schnetz_quantum_2010} in $\phi^4$-theory in 4-dimensional spacetime. If $I$ is a subdivergence-free vertex-type Feynman integral, then the Feynman period is the coefficient of the linear scale dependence of $I$, or equivalently the residue of the $\frac 1 \epsilon$ pole in dimensional regularization.
The motivation to study periods stems from at least two distinct aspects: Firstly, the sum of all periods constitutes the primitive part of the beta function of $\phi^4$ theory, which is conjectured to be asymptotically dominant at large loop order \cite{brezin_perturbation_1977,mckane_nonperturbative_1984}. Secondly, the period can be considered the simplest version of a Feynman integral and serves as a canonical example to study distributions of Feynman integrals, number theory, novel integration algorithms, or divergence and resummation of perturbation theory.  

The period can be expressed in many equivalent forms, for example as a parametric integral based on the first Symanzik polynomial $\psi_G$ of a subdivergence-free graph $G$,
\begin{align}\label{def:period}
	\period (G)&=  \left( \prod_{e\in E_G} \int \limits_0^\infty \d a_e  \right)   \; \delta \left( 1-\sum_{e\in E_G}  a_e \right) \frac{1} {\psi_G^{ 2}}.
\end{align} 
The period is a real number, independent of the kinematics of the graph. We use Euclidean spacetime, but $\period(G)$ has the same value in Minkowski spacetime. For more details and references see \cite{schnetz_quantum_2010,kompaniets_minimally_2017,schnetz_numbers_2018,balduf_statistics_2023}. 

Recently, a new  algorithm developed by Borinsky and collaborators \cite{borinsky_tropical_2023a,borinsky_tropical_2023}, based on the recursive structure of the Hepp bound \cite{hepp_proof_1966,panzer_hepp_2022}, was used by the first author to numerically compute millions of periods in $\phi^4$-theory and examine their statistical properties \cite{balduf_statistics_2023}. 
The present paper concerns two of the central observations of \cite{balduf_statistics_2023}, namely
\begin{enumerate}
	\item Although the period  is given by a complicated integral, its value is  correlated with various properties of the graph which can be computed much more easily.	
	\item The distribution of periods at a fixed loop order has  large standard deviation and extreme outliers. This implies that in order to estimate physically relevant sums and averages by uniform random samples, one needs impractically large samples. On the other hand, one can not resort to computing \emph{all} graphs of a given loop order since the number of Feynman graphs  grows factorially with the loop order, reaching $\gg 1$ million period graphs at $L=14$ loops.  
\end{enumerate}
The goal of the present paper is to exploit the first point to construct an   approximation function $\bar \period(G)\approx \period(G)$ which can be evaluated as fast as possible while predicting the period as accurately as possible. Such a function allows to draw a non-uniform sample of graphs, weighted by their approximated period, and to compute accurate averages from this sample even for relatively small sample sizes. This will help to tackle the second point above because, as we demonstrate in \cref{sec:implementation}, the total time needed for approximation, non-uniform sampling, and integrating the selected graphs is much smaller than integrating all graphs of an (appropriately larger) uniform sample. We conclude that, although the number of graphs grows quickly with loop order, the numerical effort for approximating their sum does not necessarily grow at the same rate.

The idea to approximate scattering amplitudes without explicitly solving all Feynman integrals is not new.
There is significant literature on the approximation or prediction of various types of scattering amplitudes in quantum theory with modern statistical and machine-learning methods, see e.g. \cite{badger_using_2020,maitre_factorisationaware_2021,badger_loop_2023,ilten_modeling_2023,chahrour_comparing_2022,mizera_scattering_2023,maitre_oneloop_2023}.
Although the scattering amplitude in quantum field theory is expressible as a sum of all Feynman integrals, these works on a technical level are relatively different from ours:    The relevant input for a prediction of a scattering amplitude would be the kinematics of the scattering process, such as   momenta of the involved particles, and the output the amplitude, which is a (at least piece-wise continuous) function of these continuous input parameters. Conversely, in our case, the input to the predictions is a discrete quantity, a single Feynman graph, and the output is a single number, the period,  without notion of continuity.

Also physically, the focus of our own contribution is somewhat orthogonal to most of the literature: By approximating the period, we aim to find the anomalous scale-dependence of a theory, which is independent of any other kinematics except from the scale, whereas existing work has mostly focused on the very complicated dependence of amplitudes on kinematic parameters such as angles, spins, or relative energies, but at a fixed scale.

\subsection{Content and Results}\label{sec:results}
The present \cref{sec:results} gives a summary of the obtained results, and
\cref{sec:discussion} is a discussion of qualitative takeaways and possible next steps.

All periods computed in the present work, combined with those from \cite{balduf_statistics_2023},   will be made publicly available, this data set is described in \cref{sec:data_set}.  

\medskip 
\noindent
In \cref{sec:estimation}, we introduce several ways to approximate the period. 
\begin{itemize}
	\item In \cref{sec:resistance}, we show that if the graph is interpreted as an electrical network, then the average resistance of this network allows to predict the period with an accuracy of about 5\% relative error. (Definitions of error measures and fit procedures are explained in \cref{sec:measures}).
	\item The number of triangles is weakly correlated with the period, but a multi-linear function of the cycles of higher length gives a prediction with approximately 1.5\% accuracy (\cref{sec:cycles}). 
	\item One can introduce a \enquote{cycle polynomial}, a generating function of cycles (\cref{def:cycle_polynomial}). Its evaluation at 0.364 is almost, but not exactly, invariant under period symmetries, and gives an approximation with 5\% accuracy that works (with the same parameters) across all loop orders $8 \leq L \leq 18$ examined in this work. 
	\item Using the logarithm of the number of 6-, 8-, and 10-edge cuts, we obtain an an accuracy of around 2.5\% (\cref{sec:cuts}). The accuracy can be improved to 1.3\% if one uses a multi-linear function of cycles and logarithms of cuts  simultaneously. 
	\item For all approximations based on cuts and cycles,  allowing for non-linear terms results in almost no improvements in the accuracy.
	\item In \cref{sec:hepp}, we give new fit coefficients for the approximation of the period by the Hepp bound. The accuracy of this prediction is around 1\%, as known from the literature, and it can not be improved by adding higher order monomial terms. However, the accuracy increases by a factor of roughly 5 if we include the number of 6-edge cuts into the model. With 6,-, 8-, and 10-edge-cuts included, the Hepp approximation performs consistently better than 0.2\%. 
	\item The first Martin invariant allows to approximate the period to 4\% accuracy. Higher Martin invariants  improve the accuracy, to 0.4\% for a multi-linear function including $M^{[4]}$ (\cref{sec:martin}). 
	\item Unlike the cut and cycle models, the Martin approximation benefits from introducing non-linear monomials of low degree, giving an accuracy of around  0.1\% for a cubic function including $M^{[4]}$ (\cref{fig:martin_quadratic_relative_standard_deviation}). 
	\item A  linear function of Martin invariants does not benefit from using the number of cuts. Conversely, for a non-linear function of Martin invariants, including the cuts  improves the accuracy to 0.05\% (\cref{fig:martin_cut}). 
\end{itemize}

\medskip
\noindent
In \cref{sec:sampling} we examine the speed of the approximation models, and   use them in a weighted sampling algorithm for Feynman periods. 
\begin{itemize}
	\item \Cref{sec:accuracy_performance} is a short review of importance sampling in Monte Carlo calculations.  Both the relative standard deviation $\delta$, and the time  $t_a$ required for predicting one period, are limiting factors for the reachable accuracy. To first order, the accuracy scales linearly in $\delta$, and with the square root $\sqrt{t_a}$ (\cref{relative_uncertainty_approximation}).
	\item For all approximations of \cref{sec:estimation}, the time $t_a$ per graph grows exponentially with the loop order (\cref{fig:ta}). 
	\item The 5-cycle prediction and the resistance prediction are much faster than generating the primitive graphs, where prediction times $t_a$ range in the microseconds. Cut and cycle predictions  take a few milliseconds per graph, which is similar to the time needed to test a graph for primitiveness. For  loop orders $L >10$, Hepp and Martin approximations are much slower than the other approximations, and scale worse with increasing loop order (\cref{fig:ta,fig:ta_generated}).
	\item Hepp and Martin approximation are the only models which require a large cache of subgraphs. This makes it hard to parallelize these models, further decreasing their usability in practice (\cref{sec:performance}).
	\item The combined cut and cycle model is overall the best choice for Monte Carlo sampling of periods. We implemented this model in a proof-of-concept program (\cref{sec:implementation}).
	\item We spent roughly one week at 100 cores to estimate the primitive beta function for each of $L\in \left \lbrace 13,14,15,16 \right \rbrace $ loops (\cref{tab:beta_results}). Results are compatible with, and have better accuracy than, the ones obtained in   \cite{balduf_statistics_2023}.
	\item  A  comparison of the runtimes and accuracies indicates that the new weighted sampling algorithm produces approximately 35-times better accuracy in a given runtime, or it requires approximately 1000-times less runtime to reach a given accuracy (\cref{sec:implementation}), compared to naive random sampling of periods.
\end{itemize}

\medskip 
\noindent
Finally, in \cref{sec:machine_learning}, we present several different approaches of how general-purpose machine learning methods can be used to obtain  approximations for the period. 

\begin{itemize}
	\item In \cref{sec:ML_introduction}, we review the challenges when using graphs as input to machine-learning models, and common techniques to overcome them. 
	\item Apart from the graph itself, we consider overall  194 input features (described in \cref{sec:features}) constructed from the graph, including all quantities used in \cref{sec:estimation}.
	\item A multi-linear regression of these 194 features approximates the period with an accuracy $\delta \approx 0.1\%$ (\cref{sec:linear_regression}).
	\item A quadratic regression of a large subset of these features reaches $\delta \approx 0.03\%$ (\cref{sec:quadratic_regression}), better than any hand-crafted model of \cref{sec:estimation}, but at the expense of having more than $10^3$ fit parameters.
	\item With the same input features, a multi-layer feed forward neural network reached accuracy around $5\% \leq \delta \leq 10\%$ (\cref{sec:basicmodel}).
	\item In \cref{sec:graph_structure}, we examine three other common types of machine learning algorithm which to some extent take the graph itself, not just its features, as an input. None of these models shows a clear advantage over the standard feed-forward network, and all neural networks are inferior to the multi-linear regression.
\end{itemize}

\subsection{Discussion}\label{sec:discussion}

The central outcome of this paper is that the period is closely correlated with numerous properties of the underlying Feynman graph. These correlations often allow to approximate the period of a given graph with an accuracy around 2\% if the correlation parameters are know. Fine-tuned models achieve better than 0.1\% average relative uncertainty. 

Remarkably, the relations in many cases are such that the logarithm of the period is an  almost  \emph{linear} function of the parameters in question.  Only for some of the models,   logarithms of input parameters or a low-order polynomial function can increase the accuracy.  This is confirmed by the observation that a multi-linear regression of \emph{all} input parameters (\cref{sec:linear_regression})  predicts the period to 0.1\% relative accuracy, while the intrinsically non-linear neural networks (\cref{sec:NN}) fail to deliver higher accuracy than the low-order fits in \cref{sec:estimation}.

Besides being almost linear,  the correlations are also \emph{regular} in the sense that the fit parameters for the various models are often simple functions of loop orders, allowing for the construction of \emph{global} models that take the loop order as an input, such as \cref{cuts_global_model,cycle5_global,cycle10_fj}. This suggests that the correlations are not merely coincidence caused by large data sets with many parameters, but they probably arise from an underlying statistical feature of the Feynman integral for periods. Understanding those effects more systematically in future work can both inform future numerical calculations, and improve our understanding of the asymptotics of large Feynman integrals.

The authors of \cite{panzer_feynman_2023}  conjectured that the sequence of all Martin invariants $M^{[k]}$ is a perfect period invariant, that is, that it coincides for two graphs if and only if the periods coincide. The numerical results of \cref{sec:martin} support this conjecture, but the actual situation is even better: Not only do the Martin invariants coincide for all examined graphs with the same period, but also, a low-order polynomial of the first few $M^{[k]}$ already gives an excellent numerical approximation for the period. 

From a purely numerical perspective, the proof-of-concept results of \cref{sec:implementation} show the potential of weighted sampling of Feynman integrals at high loop order. It has long been known that the perturbation series in quantum field theory can not reasonably be computed at very high order due to the factorial growth of the number of Feynman graphs. With the  weighted sampling technique introduced in the present work, together with the fact that Feynman integrals at high loop order approach a fairly regular distribution \cite{balduf_statistics_2023}, the actual computational effort to obtain (numerical values of) high-order amplitudes is far lower than naively estimated, even if it is still significant.

Regarding the neural network models, it is remarkable that the sophisticated  models (\cref{sec:graph_structure})  did not significantly exceed the accuracy of the most naive feed-forward network, and both approaches are clearly inferior to standard machine learning techniques such as linear or quadratic regression (\cref{sec:linear_regression}). Some of the input features used for the regression models, e.g. the Martin invariants or Hepp bound, arise from physical considerations, and others are generic such as counts of cycles. 
That manually selecting relevant input parameters improves a model is not per se surprising, and it is in line with other work (e.g. in \cite{maitre_oneloop_2023}, the accuracy of predicting amplitudes improves considerably by inputting the known factorization  explicitly, instead of letting the neural network determine this structure on its own). What is surprising is just how good, in absolute numbers, a linear regression of these 194  parameters is at predicting periods. Conversely, the more general machine learning models appear to be \enquote{too nonlinear} and too much susceptible to outliers in the distribution of Feynman graphs, to reliably reproduce the accuracy of a linear model. Moreover, the GCN and SAGE models in \cref{sec:graph_structure}, despite working explicitly with the graph structure, are designed for \emph{large} graphs, and for processing information \emph{on} the graph, such as data associated to vertices, which is not how we used them in the present work. Nevertheless, a more refined version of \enquote{pure} machine learning models -- which does not need the other input features -- would be very desirable because in particular the Hepp bound and Martin invariants take too long to compute in order to effectively use them for graph sampling. Neural networks, once trained, typically produce their output almost instantaneously.

In a completely different direction,  it might also be interesting to use our periods data set as a benchmark for  graph-related machine learning models. On the one hand, the period \emph{is} entirely determined by the graph alone, hence, a sufficiently advanced model could in principle reach perfect accuracy. On the other hand, through the results of \cref{sec:estimation}, we know several concrete graph-theoretic properties that influence the period, which could be used to specifically test the models' ability to recover such abstract properties from a given graph. As described in \cref{sec:data_set}, the data set for such experiments is freely available.

\subsection{Numerical periods dataset}\label{sec:data_set}

\begin{table}[htb]
	\centering
	\begin{tblr}{
			vlines,
			hline{1}={solid},
			hline{2,Z}={solid},
			rowsep=1pt,
			columns={halign=r},
			column{1}={halign=r,mode=math },
			row{1}={halign=c,rowsep=2pt},	
		}
		L & {known irr.}   & known periods & $\frac{1}{\abs{\Aut}}$ proportion &  decompletions &  rel. error  \\
		 3 &      1 &        1 &        1 &         1 &   2.1 \\
		 4 &      1 &        1 &        1 &         1 &   2.6 \\
		 5 &      1 &        2 &        1 &         3 &   5.0 \\
		 6 &      4 &        5 &        1 &        10 &   4.5 \\
		 7 &     11 &       14 &        1 &        44 &   6.0 \\
		 8 &     41 &       49 &        1 &       248 &   7.8 \\
		 9 &    190 &      227 &        1 &     1688  &  21.5 \\
		10 &   1182 &     1354 &        1 &     13094 &  27.1 \\
		11 &   8687 &     9722 &        1 &    114016 &  35.9 \\
		12 &  74204 &    81305 &        1 &   1081529 &  55.7 \\
		13 & 700242 &   755643 &        1 &  11048898 & 215 \\
		14 & 674843 &   858163 &   10.5\% &  13443457 & 341 \\
		15 & 376831 &  4724135 &    3.4\% &  75602166 & 201 \\
		16 &  23179 &  7674289 &   0.54\% & 131646012 & 233 \\
		17 &   8038 & 15371563 &   0.11\% & 285548076 & 187 \\
		18 &   1314 & 20822615 &  0.013\% & 410972405 & 181 
	\end{tblr}
	\caption{Counts of graphs where the period is available in the data set.  Number of 3-vertex irreducible completions; number of not necessarily irreducible completions; proportion of symmetry factors of known graphs relative to total symmetry factor of that loop order; number of decompletions; average relative uncertainty in ppm. All counts refer to graphs, not to distinct periods. Counts for distinct periods were given in \cite{balduf_statistics_2023}.}
	\label{tab:graphs_count}
\end{table}

All periods computed in the present work and in \cite{balduf_statistics_2023} will be made publicly available in the near future. This document will be updated with a hyperlink once the data is uploaded.

The number of periods computed in the present work, for   loop orders $L\in \left \lbrace 8, \ldots, 17 \right \rbrace $, are shown in  \cref{tab:beta_results}.   
We merged these results with the ones of \cite{balduf_statistics_2023}, and used overlaps and symmetries of periods to increase the individual accuracy.

Secondly, we wrote a new tool which produces all possible 3-vertex-product periods from a given input list. We used this tool on the above combined data set and obtained a large number of product periods, in particular all products up to and  including $L=15$ loops. The product symmetry had not been used in the previous merge step, consequently the so-obtained numerical values are an additional, independent, numerical value of the period in question and can be used to improve accuracy. Compared to \cite{balduf_statistics_2023}, this increases the average accuracy only marginally,   since the proportion of product graphs is low and the estimate we obtain from combining lower loop order estimates is relatively inaccurate. However, for $L\geq 15$, the number of so-constructed product periods exceeds the number of numerically computed periods at that loop order.  We refrained from constructing iterated product graphs where the factors themselves were only inferred from product symmetry. Moreover, for $L \geq 16$, we stopped construction of products after a few days to avoid excessively large files. The counts of available periods are summarized in \cref{tab:graphs_count}.

\begin{figure}[htb]
	\centering
	\includegraphics[width=.6\linewidth]{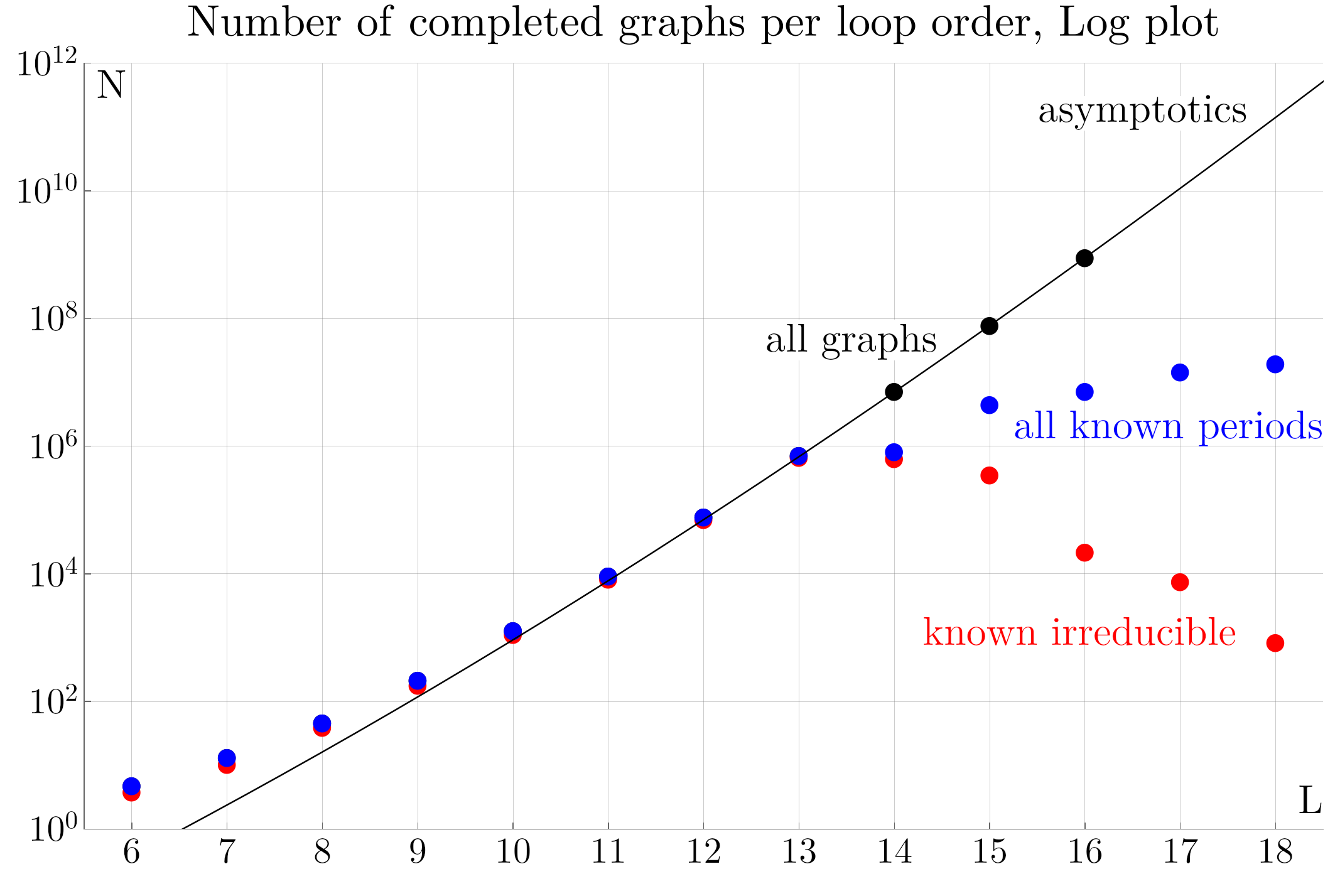}
	\caption{Counts of periods as a function of the loop order $L$. The black dots are the number of completions, the black line is the asymptotics \cite{borinsky_renormalized_2017} including $\frac 1 L$ correction. Red points indicate the number of computed 3-vertex irreducible periods. Blue dots are the number of numerically known periods after using all symmetries, see \cref{tab:graphs_count}.  }
	\label{fig:number_of_graphs}
\end{figure}

For example, by exploiting the product symmetry at 16 loops we know the periods for over 7.9 million completions, or 131 million decompletions, but they represent less than 1\% of all subdivergence-free 16-loop  graphs. This once again illustrates the enormous growth of the number of Feynman graphs with loop order, see \cref{fig:number_of_graphs}.

We want to stress again that for $L \geq 14$, the data set does not include \emph{all} graphs, and the available periods do not follow any particular distribution. Therefore, one can not obtain accurate sums or averages, such as the primitive beta function, by summing all graphs in the data set. Also other statistics, such as the number of decompletions per completion, are skewed and not representative of the entire set of $L$-loop primitives for $L\geq 14$. 

\FloatBarrier

\subsection{Accuracy measures and algorithmic details} \label{sec:measures}

Throughout the paper, the term \enquote{graph $G$}, unless otherwise stated,  refers to a  vacuum graph of $\phi^4$ theory without subidvergences, also called a \emph{completion}. Such $G$ can be turned into a subdivergence-free vertex-type graph (a \emph{decompletion}) upon removing any one vertex. The non-isomorphic decompletions of a fixed completion give rise to different Symanzik polynomials and hence different period integrals (\cref{def:period}), but the value of all these integrals arising from the same completion is  the same. Consequently, by $\period(G)$, we denote the period of any, and hence all, decompletions of $G$. 

The goal of the present work is to discover functions $\bar \period(G)$ which   approximate  the period $\period(G)$.
We will be concerned with two different measures for the quality of an approximation function, where by $\left \langle f \right \rangle $ we denote the arithmetic mean of $f(G)$  over all graphs in question. Firstly,  the average    relative error,
\begin{align}\label{def:relative_difference}
	\Delta :=\left \langle \frac{\abs{  \period - \bar \period}}{\bar \period} \right \rangle  = \left \langle   \sqrt{ \left(\frac{  \period}{\bar \period}-1 \right)^2 }\right \rangle  ,
\end{align}
is a measure for  the reliability for approximating \emph{individual} periods.
Rewritten for the logarithm of the period, assuming $\period \approx \bar \period$, the average relative error is
\begin{align}\label{relative_difference_log}
	\Delta &= \left \langle  \abs{ e^{\ln \period - \ln \bar \period} -1} \right \rangle  \approx \left \langle \abs{\ln \period - \ln \bar \period} \right \rangle . 
\end{align}

Secondly, for importance sampling (to be discussed in \cref{sec:sampling}), we are interested in approximating the  relative  magnitude of periods. In this case, we aim to minimize the  relative standard deviation $\sigma$ of the ratio $ \frac{\period}{\bar\period}$,  given by 
\begin{align}\label{def:relative_standard_deviation}
	\delta:= \frac{\sigma \left( \frac{\period}{\bar \period} \right) }{\left \langle \frac{\period}{\bar \period} \right \rangle  }   =  \sqrt{ \frac{\left \langle  \left(  \frac{\period}{\bar \period } -\left \langle  \frac{\period}{\bar \period } \right \rangle   \right) ^2 \right \rangle}{\left \langle \frac{\period}{\bar \period } \right \rangle^2  }} = \sigma \left(   \frac{\frac{\period}{\bar \period}-\left \langle \frac{\period}{\bar \period} \right \rangle }{\left \langle \frac{\period}{\bar \period} \right \rangle  }  \right)   .
\end{align}
Even in the special case where the approximation is on average exact, $\left \langle  \period \cdot \bar\period^{-1}\right \rangle  = 1$, the average  error  (\ref{def:relative_difference}) and the standard deviation (\ref{def:relative_standard_deviation}) are still generally   different, the former being the arithmetic mean and the latter the  geometric  mean of the individual errors. 
For logarithms, the relative standard deviation is
\begin{align}\label{relative_standard_deviation_log}
	\delta = \frac{\sigma \left( e^{\ln \period - \ln \bar \period} \right)  }{\left \langle e^{\ln \period - \ln \bar \period} \right \rangle  } \approx \sigma \left( \ln \period - \ln \bar \period \right) ,
\end{align}
where  the second step is valid only if $\bar \period \approx c \period$ with an arbitrary constant $c$.
Note that for $\delta$, but not for $\Delta$, it is irrelevant whether the approximation $\bar \period$ differs from $\period$ by   an overall constant factor. 

To determine the numerical constants in an approximation function, we aim to minimize the above errors, using a data set of approximately 2 million periods computed in \cite{balduf_statistics_2023}. For both error measures, \cref{def:relative_difference,def:relative_standard_deviation}, we are comparing the prediction $\bar \period$ with the numerical integration result $\period$ graph-by-graph. For $L \geq 14$ loops, our data set does not contain all graphs of the given loop order. We will assume throughout that the samples are large enough to be representative for the respective loop order.

For a fixed loop order, the values of $\period$ span several orders of magnitude, which makes a naive least-squares optimization unstable. Therefore, we construct most of the approximation functions for the logarithm $\ln \period$, and use the logarithmic versions   \cref{relative_difference_log,relative_standard_deviation_log} of the error terms.
Moreover, the leading asymptotic growth with the loop order $L$ is known analytically \cite{mckane_nonperturbative_1984}. In many cases, the numerical stability of our prediction can be improved if we scale the period, and analogously the Hepp bound (see \cite{panzer_hepp_2022} and  \cref{sec:hepp}),   according to
\begin{align}\label{period_scaling}
	\period(G) := \frac{ \period'(G)}{\left( \frac 3 2  \right) ^L L^{\frac 5 2}}, \qquad  \mathcal{H(G)}:= \frac{\mathcal{H}'(G)}{4^L}.
\end{align}
To not clutter notation, we will usually omit the prime $'$ and explain in the text if our functions refer to the scaled or unscaled quantities.

Unless otherwise mentioned, all optimization was done with Wolfram  Mathematica\textsuperscript{\textregistered{}} 13.3. 
The average relative error $\Delta$ can be minimized by a conventional least-squares fit,  while $\delta$ requires a non-linear optimization. We  compared both approaches  and found that the resulting best-fit parameters are very similar for both cases. However, minimizing $\delta$ is much more unstable and requires fine-tuning of start parameters, while rarely resulting in a significant improvement of the achieved minimum $\delta$, compared to the achieved value of $\delta$ when $\Delta$ was minimized. Consequently, unless otherwise mentioned, our results arise from a least-squares minimization of $\Delta$ with respect to logarithms of periods.

\subsection{Acknowledgments and attributions}

We thank Karen Yeats, Michael Borinsky, Willie Lei, Andrea Favorito and Erik Panzer for discussions and comments on the draft.
Many of the electrical properties considered in \cref{sec:resistance,sec:features} were inspired  by a talk by David Wagner. Examining edge cuts in \cref{sec:cuts} was motivated by a talk by Erik Panzer, announcing that $c_6$ is invariant (joint work with Jacob Mercer).

Computations for the Monte Carlo integration in \cref{sec:implementation} and for many of the models of \cref{sec:machine_learning} ran on the servers of University of Waterloo in autumn 2023. Research at Perimeter Institute is supported in part by the Government of Canada through the Department of
Innovation, Science and Economic Development and by the Province of Ontario through the Ministry of Colleges and Universities.

\Cref{sec:estimation,sec:sampling} were mostly contributed by PHB, \cref{sec:machine_learning} was mostly contributed by KS.

\newpage

\section{Period approximation by graph properties}\label{sec:estimation}

\subsection{Average resistance}\label{sec:resistance}
In \cite[Sec. 6]{balduf_statistics_2023}, it was found that the average of the distances between pairs of vertices in a Feynman graph $G$ correlates with the period, consistently across loop orders. Although this implies a neat intuitive picture --- the period roughly being a measure for the spatial extension of the graph --- the correlation is too weak to be useful as an approximation function. This changes if instead of the average distance, we consider the average resistance, also known as Kirchhoff index.

If the graph $G$ were an electrical network where every edge has a resistance of 1 Ohm, the electrical resistance $r_{v_1,v_2}$  between any two vertices $v_1$ and $v_2$ is given by Kirchhoff's laws \cite{kirchhoff_ueber_1847}. An algebraic algorithm to compute $r_{v_1,v_2}$ is based on  the  Laplace matrix $\mathbb L$ of the graph. Its Moore-Penrose pseudoinverse  
$\mathbb L^+$ can be computed analytically for certain classes of graphs \cite{azimi_explicit_2023}, and a numerical calculation or $\mathbb L^+$, given $\mathbb L$, is readily available in linear algebra software packages. 
The resistance between two vertices is then  given by 
\begin{align*}
	r_{v_i,v_j}= \mathbb L^+_{v_i,v_i}+\mathbb L^+_{v_j,v_j}-\mathbb L^+_{v_i,v_j}-\mathbb L^+_{v_j,v_i}.
\end{align*}

There is also an equivalent perspective on the resistance in graphs, which is more in the spirit of the intuition that a Feynman graph represents a \enquote{discretization} of a process in spacetime: The Laplace matrix $\mathbb L$ of a graph can be  viewed as a discretization of a continuous Laplace operator. Then, $\mathbb L^+$ is a corresponding (discretized) Green function, and $r_{v_i,v_j}$ its (discretized) derivative  \cite{kassel_transfer_2015}. Using resistance as a measure of distance is not new to graph theory, see e.g. \cite{klein_resistance_1993,kirkland_distances_1997}. 
The \emph{Kirchhoff index} of a graph $G$ is the average of the resistances between all pairs of vertices $v\in V_G$,
\begin{align}\label{def:kirchhoff_index}
	R(G) := \frac{2}{\abs{V_G} \left( \abs{V_G}-1 \right)  }\sum_{v_1 <v_2 \in V_G} r_{v_1,v_2}.
\end{align}

\begin{figure}[htb]
	\begin{subfigure}{ .49 \linewidth}
		\centering
		\includegraphics[width=\linewidth]{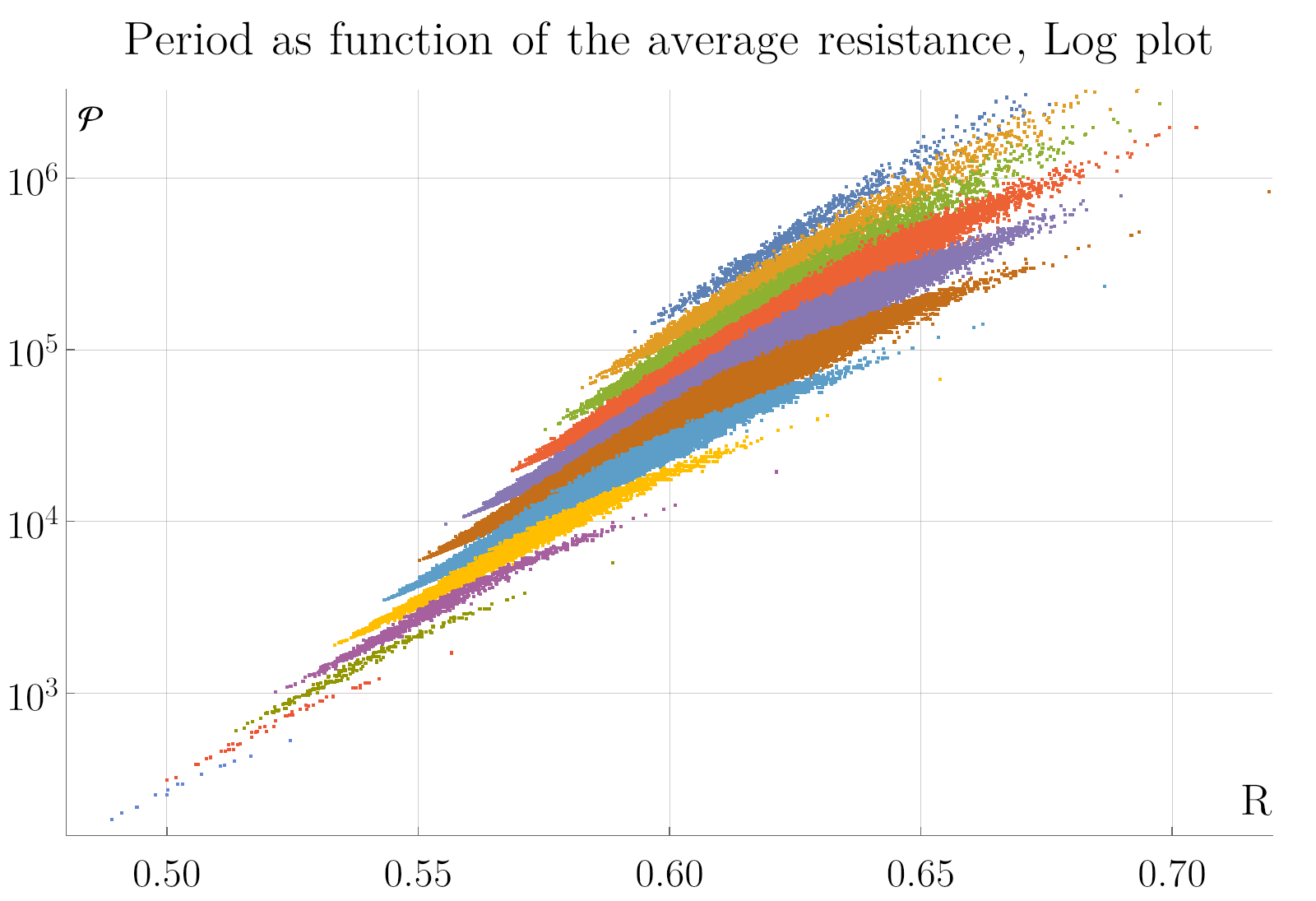}
		\subcaption{}
		\label{fig:resistance_plot}
	\end{subfigure}
	\begin{subfigure}{ .49 \linewidth}
		\centering
		\includegraphics[width=\linewidth]{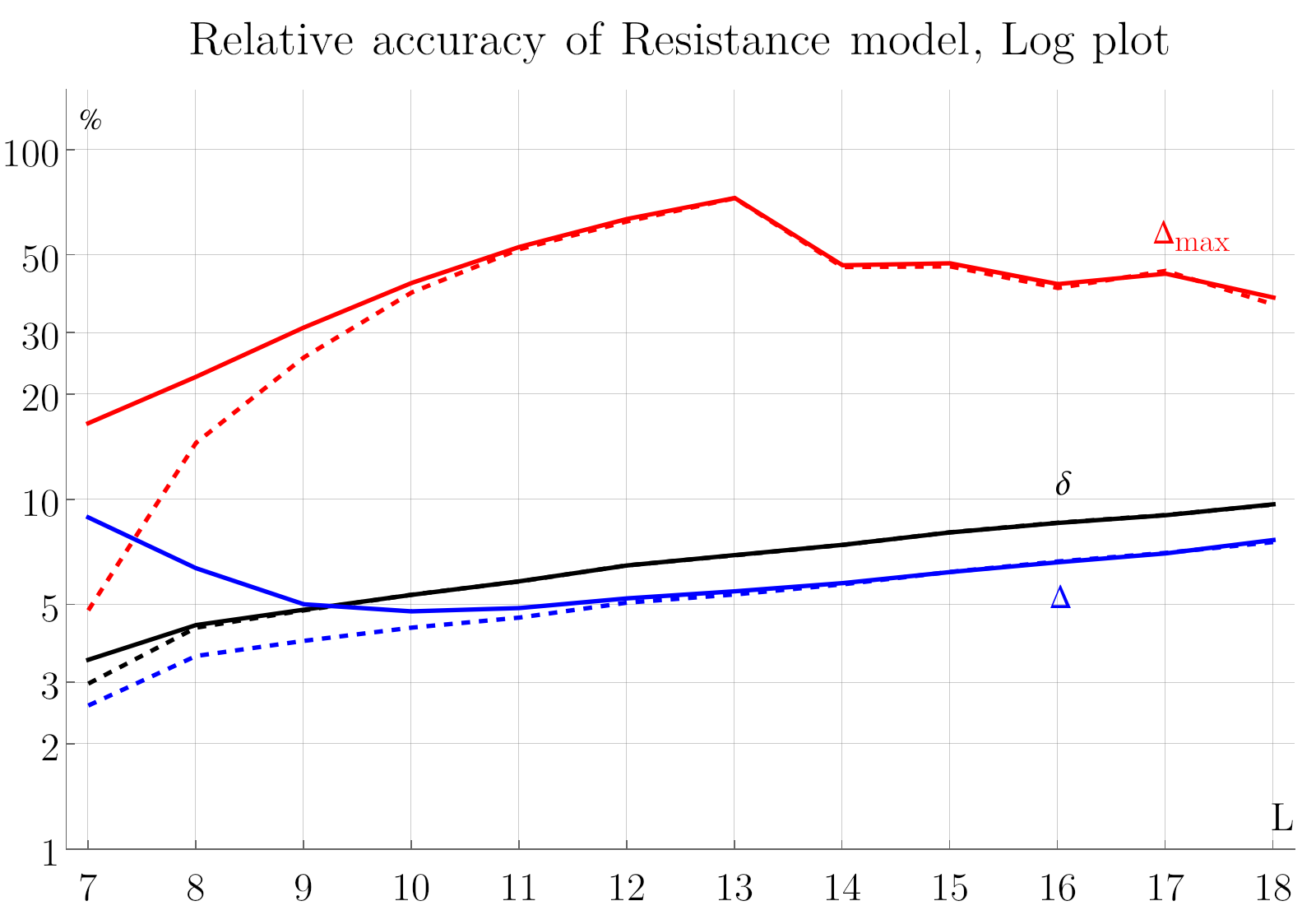}
		\subcaption{}
		\label{fig:resistance_accuracy}
	\end{subfigure}

	\caption{\textbf{(a)} Period as function of the average resistance, log plot.  $\ln \period$ can be approximated by a linear function of $R(G)$. Colors indicate loop order $7 \leq L \leq 18$. Thin lines are the best fit functions \cref{def:resistance_approximation}. \textbf{(b)} Relative accuracy of the resistance approximation. For $L \geq 10$ loops, the global model \cref{resistance_global} (thick lines) shows similar performance as the best fit (dashed lines). The maximum error $\Delta_\text{max}$ (\cref{def:relative_difference}) decays for $L \geq 14$ because the samples are non-complete and do not including the most problematic graphs.  }
\end{figure}

As seen in \cref{fig:resistance_plot}, the logarithm of the period is approximated by a linear function of the Kirchhoff index,
\begin{align}\label{def:resistance_approximation}
	\ln \bar \period(G) := a_L \cdot R(G) + b_L. 
\end{align}
The parameters $a_L$ and $b_L$ in \cref{def:resistance_approximation} change with the loop order. We find empirically that for $L \geq 10$ they are approximately given by
\begin{align}\label{resistance_global}
	a_L &=  -0.020(L-30.7)^2 + 43.2, \qquad b_L =   0.014(L-17.8)^2 -11.95 .
\end{align}
\Cref{resistance_global} constitutes a \enquote{global model} for the resistance approximation of the period, that is, a model  which accepts the loop order $L$ as an input parameter and can therefore be used to predict periods of arbitrary (not too large) loop order. 

\Cref{def:resistance_approximation,resistance_global} approximate the period of individual graphs with mediocre accuracy, typical errors are $5\% \leq \delta \leq 10\%$, see \cref{fig:resistance_accuracy}. The pivotal advantage of the resistance approximation, compared to most of the following models, is that it is extremely fast to compute  since it only requires simple linear algebra operations.  

Note in \cref{fig:resistance_accuracy}  that   the average individual error $\Delta$ (\cref{def:relative_difference}) behaves almost identically to the average standard deviation $\delta$. This   phenomenon is also observed for the other approximations to be constructed later in this work. To not clutter plots and tables, in the following we will often leave out $\Delta$ and only report $\delta$.

Besides the Kirchhoff index, we further examined various other \enquote{electrical} properties of the graph. 
The \emph{transfer current matrix} of an electrical network expresses how the current through an edge  changes if a unit current is imposed parallel to another edge. The \emph{Ursell functions} \cite{ursell_evaluation_1927} are correlation functions between sets of these edge currents. We found these functions to be correlated with the period, but not as strongly as the Kirchhoff index. All these quantities are being used as input for the machine learning models considered in \cref{sec:machine_learning}.

Another benefit of \enquote{electrical} properties of a graph are that they come with various interesting alternative interpretations, for example   in terms of random walks, see e.g. \cite{doyle_random_2006}, or in terms of random spanning tree models, see e.g. \cite{kassel_transfer_2015}. Curiously,  average distance and  Kirchhoff index have been employed to predict physical properties associated with graphs already long ago, but in a completely different setting: If the graph is a hydrocarbon molecule, then these quantities predict the boiling point  \cite{wiener_structural_1947}.

\subsection{Cycles}\label{sec:cycles}

A \emph{circuit} in a graph $G$ is a sequence of edges that form a closed path, where edges and vertices are allowed to be visited more than once, and not all edges or vertices of $G$ have to be used.
A \emph{cycle} is a circuit that visits no vertex more than once. This implies that a cycle is connected and does not contain any edge more than once. For us, two cycles will be considered identical if they only differ by the order of edges they contain. A cycle is the same as what is called \emph{loop} in the physics literature. Recall that the physics   \emph{loop number} $L$ amounts to the dimension of cycle space, that is, the number of \emph{linearly independent} cycles of a graph, not the total number of cycles. 

Let $n_j(G)$ be the total number of cycles of length $j\in \mathbb N$ contained in a graph $G$.   It is not required that these cycles are mutually disjoint.  There is no simple relation between $n_j$ and the loop number $L$, in particular,     $n_j$ can be much larger than  $L$. However, a $L$-loop completed graph has $L+2$ vertices, therefore a cycle can have at most $L+2$ edges, and $n_j=0$ for all $j>L+2$. Furthermore, the graphs considered in the present paper are free of subdivergences, therefore they have no multiedge-subgraphs and consequently no cycles of length 2, $n_2=0$. The first non-trivial cycle count is $n_3$, the number of triangles.

\begin{figure}[htb]
	\centering 
	\begin{subfigure}[b]{.32 \textwidth}
		\includegraphics[width=\linewidth]{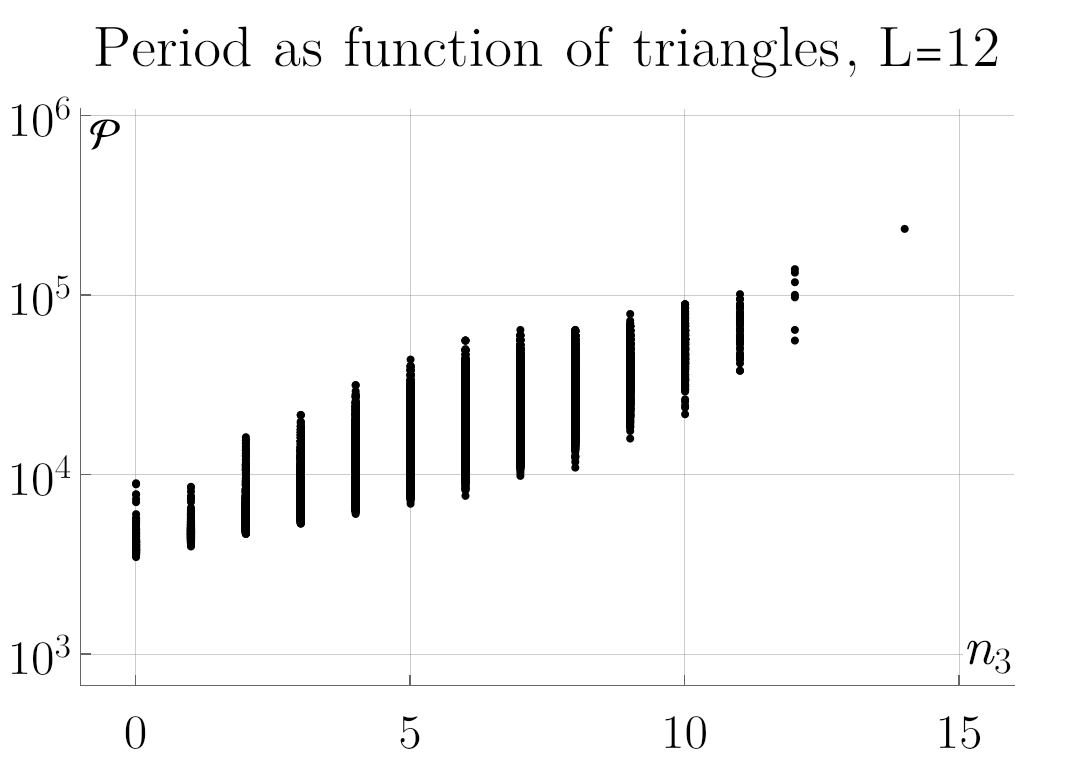}
		\subcaption{}
		\label{fig:period_triangles}
	\end{subfigure}
	\begin{subfigure}[b]{.32 \textwidth}
		\includegraphics[width=\linewidth]{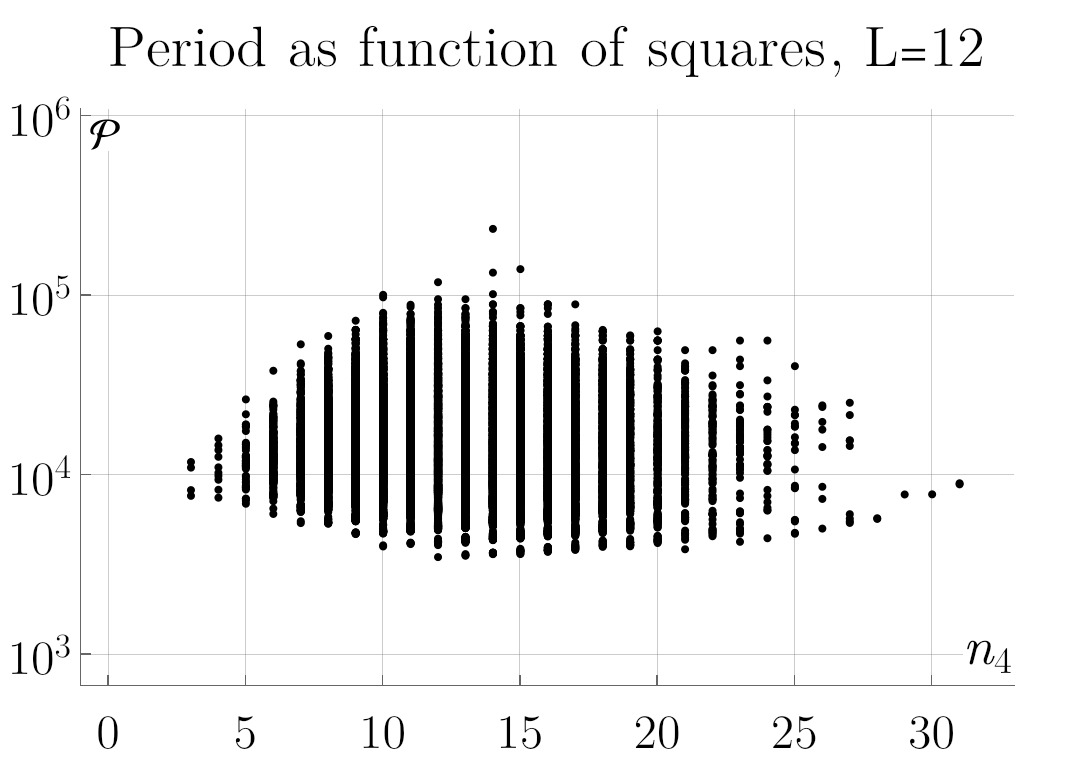}
		\subcaption{}
		\label{fig:period_triangles_squares}
	\end{subfigure}
	\begin{subfigure}[b]{.32 \textwidth}
		\includegraphics[width=\linewidth]{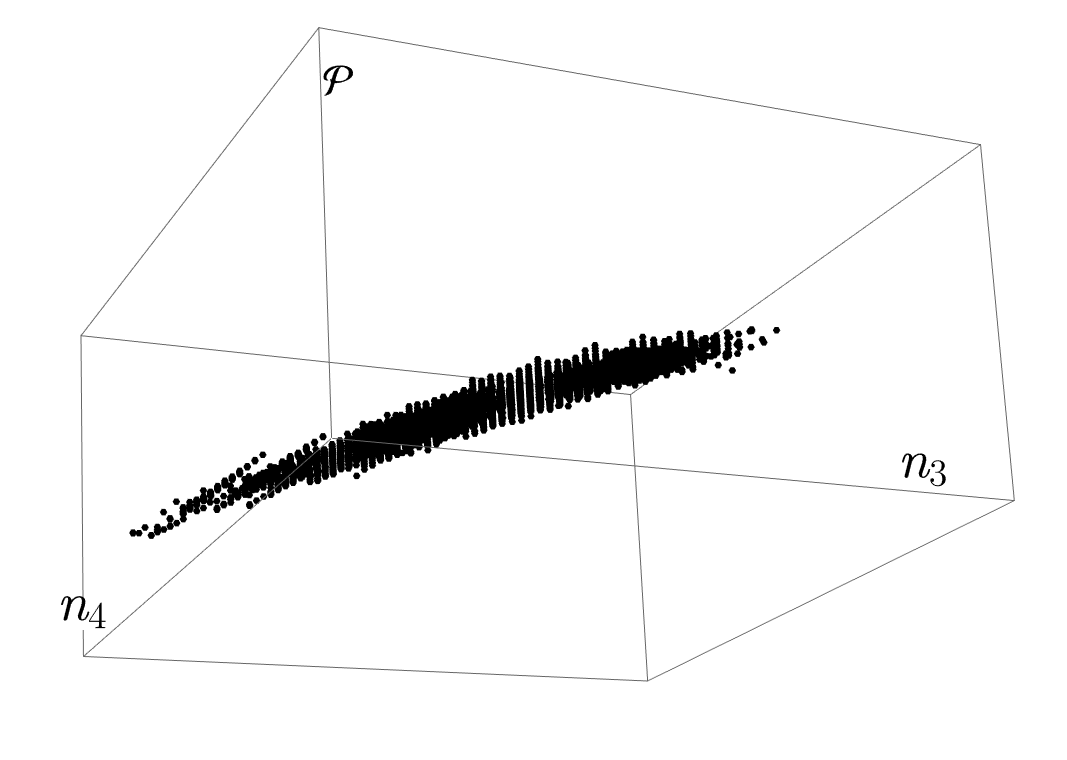}
		\subcaption{}
		\label{fig:period_cycles_3d}
	\end{subfigure}
	
	\caption{ \textbf{(a)} The (logarithm of the) period is correlated with the number of triangles $n_3$, but with significant fluctuations. \textbf{(b)} There is almost no correlation between the period and the number of squares $n_4$.  \textbf{(c)} The period is strongly correlated with $n_3$ and $n_4$ \emph{simultaneously}. }
	 	\label{fig:cycles_34}
\end{figure}

It had been observed in \cite{balduf_statistics_2023}  that the period $\period$ is correlated with the number of triangles $n_3$, but almost uncorrelated with the number of squares $n_4$, see \cref{fig:period_triangles,fig:period_triangles_squares} for the $L=11$ case. The picture changes drastically if one considers triangles and squares simultaneously: The values of $\ln \period$ almost lie in a tilted plane in the $(n_3,n_4,\ln \period)$-space, see \cref{fig:period_cycles_3d}.
This pattern continues  in higher dimension: To a good approximation, $\ln \period$ is a multi-linear function of the numbers of cycles $\left \lbrace n_3,n_4,\ldots, n_{j_\text{max}} \right \rbrace   $.

\begin{figure}[htb]
	\centering 
	\begin{subfigure}[b]{.48 \textwidth}
		\includegraphics[width=\linewidth]{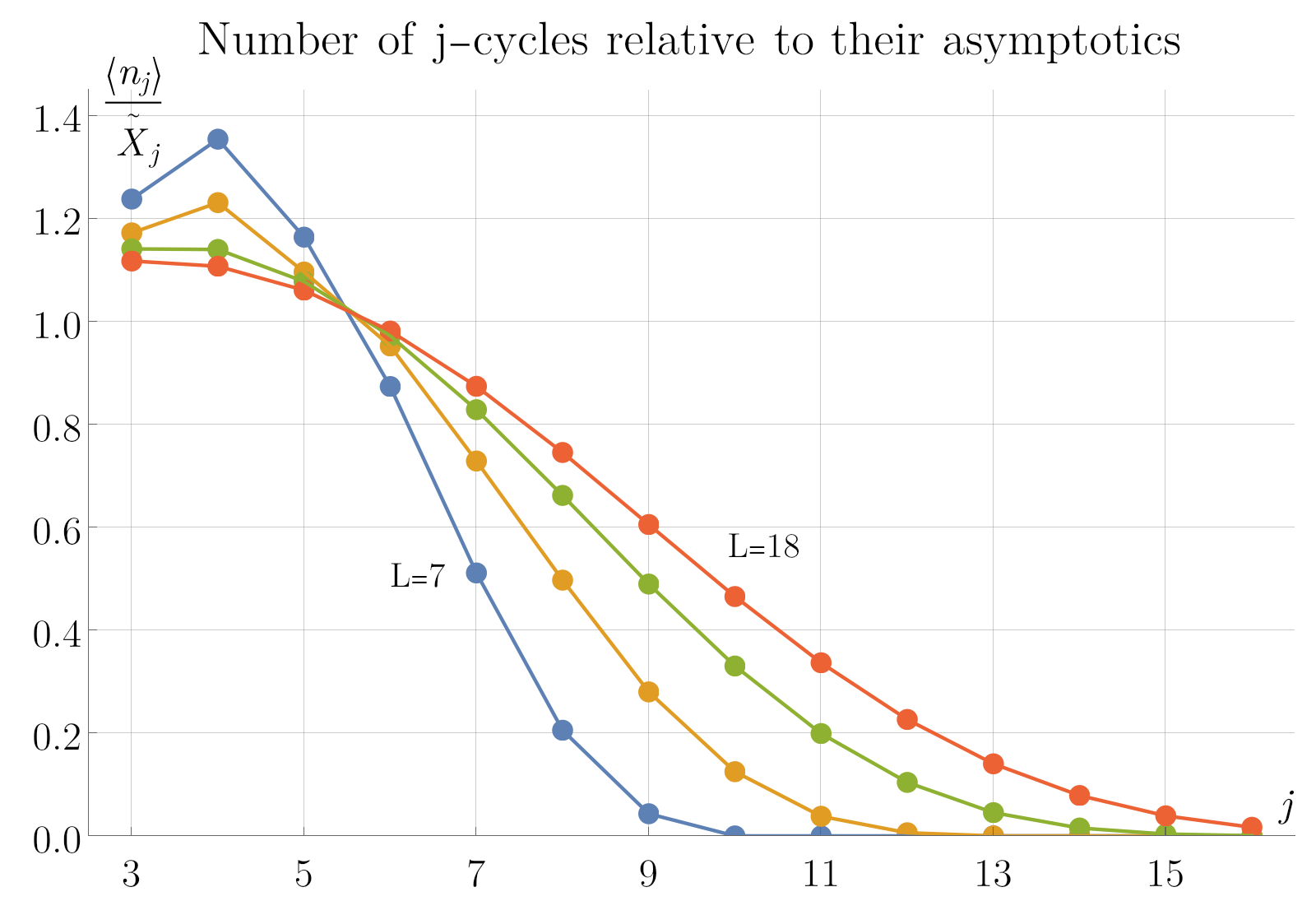}
		\subcaption{}
		\label{fig:cycle_length}
	\end{subfigure}
	\begin{subfigure}[b]{.48 \textwidth}
		\includegraphics[width=\linewidth]{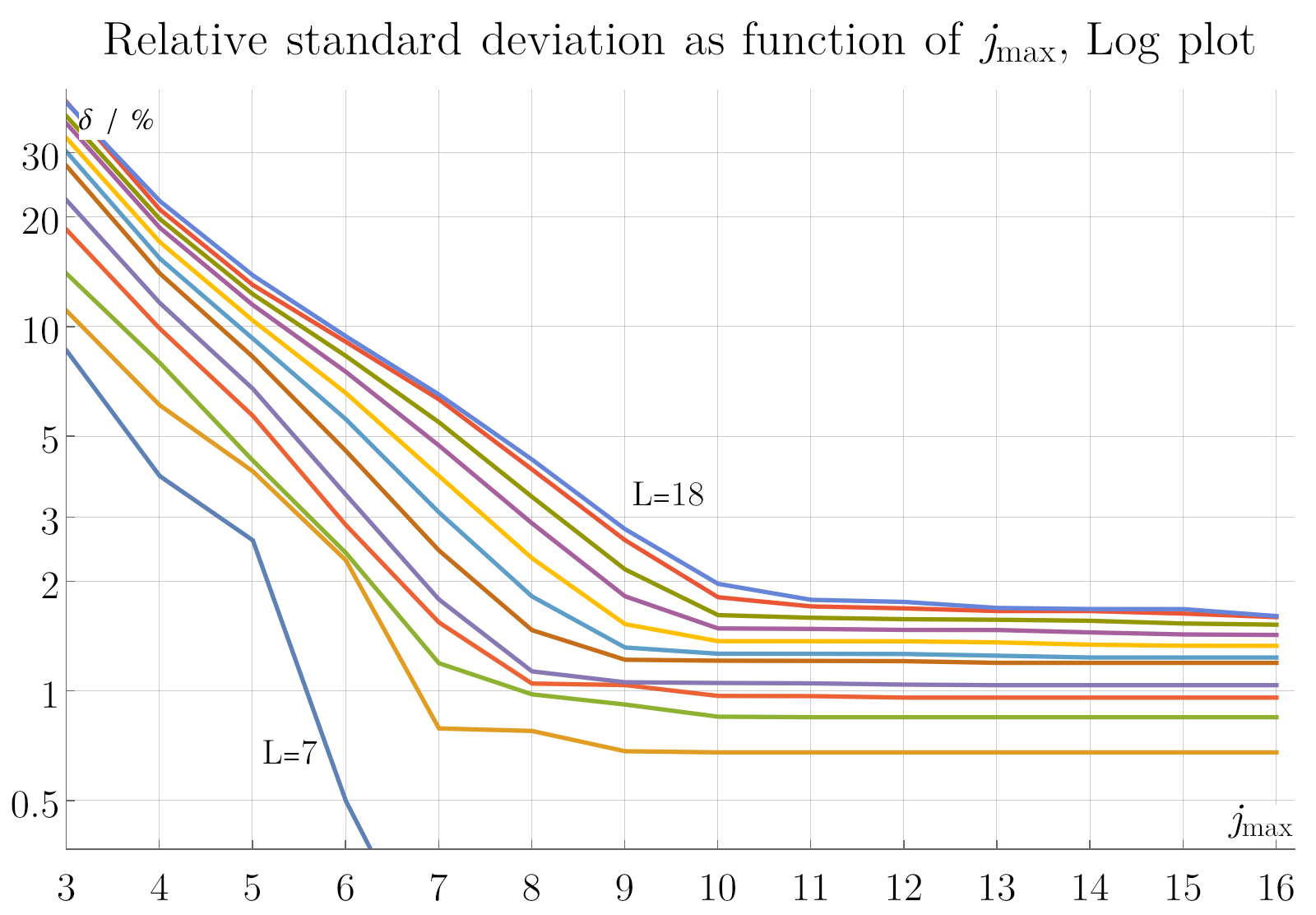}
		\subcaption{}
		\label{fig:cycles_delta}
	\end{subfigure}
	
	\caption{\textbf{(a)} Average number $\left \langle n_j \right \rangle $ of cycles of length $j$, scaled relative to the asymptotics \cref{Xj_prediction}, for loop orders 7,10,14, and 18. This number is close to unity as long as the cycle is much smaller than the size of the graph. We do not know any explicit reason for why the lines appear to cross near $j=6$. \textbf{(b)} Relative standard deviation (\cref{def:relative_standard_deviation}) for the cycle approximation (\cref{def:cycle_approximation}) for irreducible graphs as a function of the maximum used cycle length $j_\text{max}$, logarithmic plot.   }

\end{figure}

For reasons of numerical stability, we scale the parameters $n_j$ such that they take roughly similar values for different $j$. 
As discussed in \cite{balduf_statistics_2023}, the space of random 4-regular $L$-loop graphs in the limit $L \rightarrow\infty$ is a useful mathematical model that is similar to, but not identical with, the set of primitive graphs. Let $X_j$ be the number of cycles of length $j$ in such a random graph. For distinct $j$, the $X_j$ are  independent Poisson distributed variables with mean  \cite{bollobas_probabilistic_1980,mckay_short_2004}
\begin{align}\label{Xj_prediction}
	\tilde   X_j   &= \frac{3^j}{2j}, \qquad L \rightarrow \infty. 
\end{align}
Note that, perhaps counter intuitively, these numbers are finite in the limit of  infinitely large graphs.  As seen from \cref{fig:cycle_length}, the expected averages from \cref{Xj_prediction} are a reasonably  good approximation for our data as long as the cycle length $j$ is significantly smaller than $L$.

Scaling the number of cycles according to  \cref{Xj_prediction}, we define the  multi-linear approximation model for the logarithm of the period  as
\begin{align}\label{def:cycle_approximation}
	\ln \bar \period  &:=f_0 + \sum_{j=3}^{j_{\text{max}}} f_j \frac{2 j \cdot n_j}{3^j}.
\end{align}
Here, $j_\text{max}\in \left \lbrace 3, \ldots, L+2 \right \rbrace   $ is the maximum cycle length to be considered, and the parameters $f_j$ depend on $L$ and $j_\text{max}$, and will be determined numerically.

As expected from the cases $j_\text{max}=3$ and $j_\text{max}=4$ (\cref{fig:cycles_34}), the approximation \cref{def:cycle_approximation} gets better with increasing $j_\text{max}$. The relative standard deviation $\delta$ (\cref{def:relative_standard_deviation})  is shown in \cref{fig:cycles_delta}.  We find that  higher $j_\text{max}$ increase the accuracy for high loop orders, but the improvements are very small beyond $j_\text{max}\approx 10$. On the other hand, the effort for counting cycles increases with growing $j_\text{max}$. From an algorithmic perspective, $j_\text{max}=10$ is therefore the most relevant choice. Secondly, the case $j_\text{max}=5$ is interesting  because $\left \lbrace n_3,n_4,n_5 \right \rbrace $ can be obtained very efficiently from traces of the adjacency matrix of a graph, without constructing the full cycle space. The two models of  \cref{def:cycle_approximation} for $j_\text{max}=5$ and $j_\text{max}=10$, respectively,  will be examined in the following.

\begin{figure}[htb]
	\begin{subfigure}{ .49 \linewidth}
		\centering
		\includegraphics[width=\linewidth]{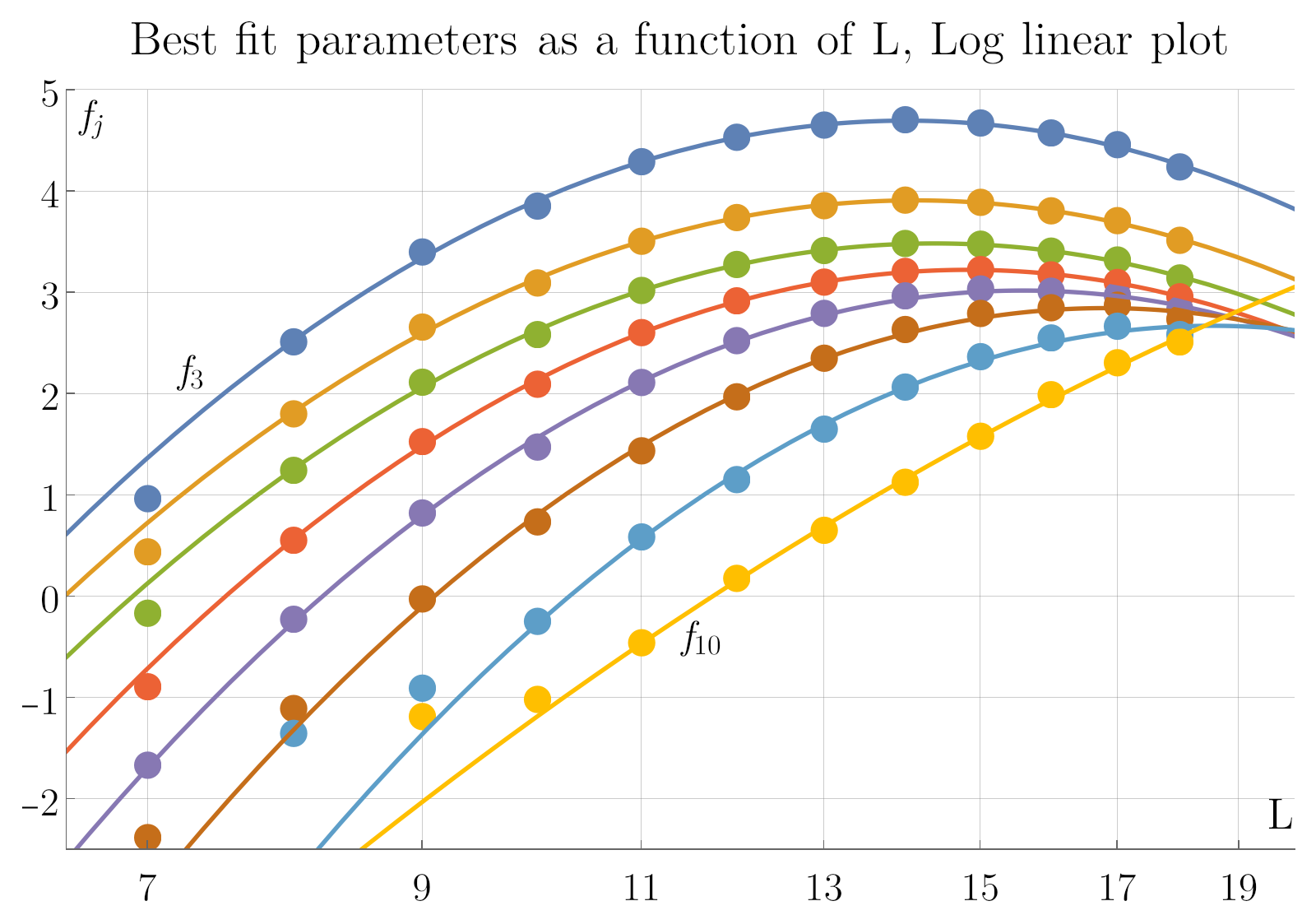}
		\subcaption{}
		\label{fig:cycle10_parameters}
	\end{subfigure}
	\begin{subfigure}{ .49 \linewidth}
		\centering
		\includegraphics[width=\linewidth]{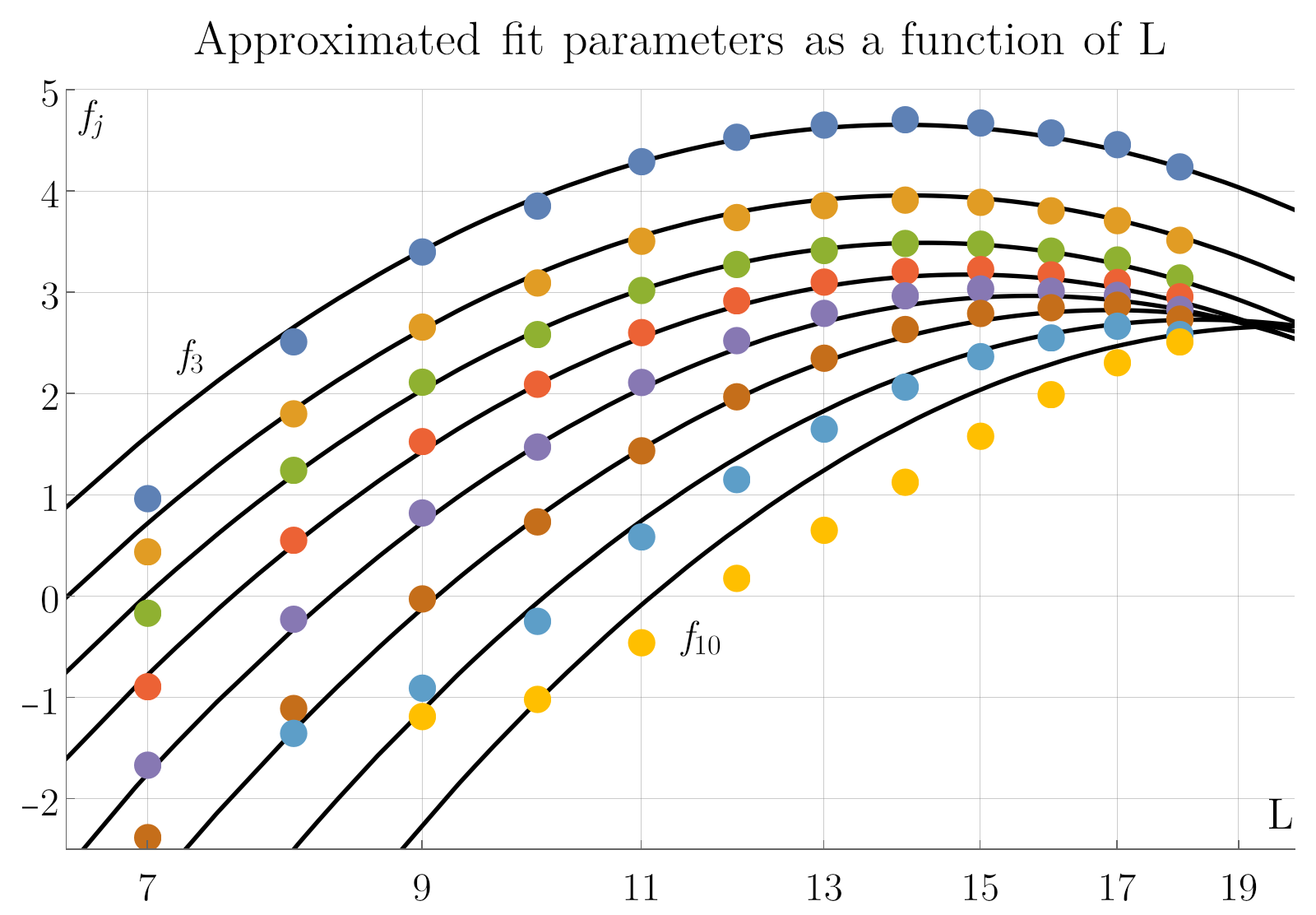}
		\subcaption{}
		\label{fig:cycle10_parameters_approximation}
	\end{subfigure}
	
	\caption{\textbf{(a)} Best fit parameter for the cycle model \cref{def:cycle_approximation} for $j_\text{max}=10$. The parameters approximately lie on quadratic functions of $\ln (L)$ (\cref{cycle10_fj_ansatz}, solid lines). \textbf{(b)} Black lines: Empirical function \cref{cycle10_fj}. Apart from $f_{10}$, this model is has acceptable accuracy for all data points for $L \geq 8$. }
	
\end{figure}

For  $j_\text{max}=10$,  \cref{def:cycle_approximation} contains 9 free parameters $\left \lbrace f_0, f_3, \ldots, f_{10} \right \rbrace $, each of which depends on the loop order. We determined their best-fit value as described in \cref{sec:measures}. 
As seen in \cref{fig:cycle10_parameters}, the parameters  $f_j(L)$ are very regular in $L$ and can be approximated  by a quadratic function of $\ln(L)$, 
\begin{align}\label{cycle10_fj_ansatz}
	f_j(L) &= - k_j \left( \ln \frac{L}{l_j}\right) ^2 + m_j.
\end{align}

If the best fit parameters $\left \lbrace k_j,l_j,m_j \right \rbrace $ are known for each $j\in \left \lbrace 3, \ldots, 10 \right \rbrace $, one can construct the cycle approximation \cref{def:cycle_approximation} for each loop order $L\in \left \lbrace7, \ldots, 18 \right \rbrace $. These are 27 parameters which need to be hard-coded into a corresponding program. Alternatively, we found empirically that  $\left \lbrace k_j,l_j,m_j \right \rbrace $ in \cref{cycle10_fj_ansatz} can be approximated by simple regular functions of $j$, such that \cref{cycle10_fj_ansatz} turns into the global model
\begin{align}\label{cycle10_fj}
	f_j(L) &= -\left( 0.19 j + 5.9 \right)  \left(  \ln \frac{L}{j \left(  e^{ -0.54 j+ 2.62} +1.93\right)  } \right) ^2 + 2.54 + e^{-0.40 j + 1.95}.
\end{align}

The parameter $f_0$ in \cref{def:cycle_approximation} represents a multiplicative constant in the approximated period $\bar \period$, it does not influence the standard deviation $\delta$ (\cref{def:relative_standard_deviation}).  To predict the absolute value of the period, the value of $f_0$ is of crucial importance and \cref{cycle10_fj_ansatz} is not accurate enough. A better description is
\begin{align}\label{cycle10_c0}
	f_0 &=   39.6 \left(  \ln \frac{L}{15.8} \right) ^2 -24.4+ \ln \left(1.40 +  \frac{\left( L-13.6 \right) ^3}{340} -\frac{L}{29} \right) .
\end{align}

For the case $j_\text{max}=5$, an ansatz similar \cref{cycle10_fj_ansatz} works to describe the individual $f_j$. A global model for $j_\text{max}=5$ is given by
\begin{align}\label{cycle5_global}
	f_3 &= -0.96 \left( \ln \frac{L}{12} \right) ^2 +1.85 \qquad \qquad f_4  = -1.06 \left( \ln \frac{L}{12.75} \right) ^2+1.30\nonumber  \\
	f_5 &=-1.37 \left( \ln \frac{L}{14.4} \right) ^2 +0.96\\
	f_0 &= 3.39\left( \ln \frac{L}{11.35} \right) ^2 -6.286 + \ln \Big( 1.013 - (0.035 L-0.49)^2 \Big) . \nonumber 
\end{align}
\Cref{fig:cycle_models_relative_standard_deviation} shows the accuracy of the two cycle models. $j_\text{max}=5$   is obviously less accurate than the choice $j_\text{max}=10$, but this is compensated for by an increased speed, see \cref{sec:performance}.

Conceptually, one can even add another layer of generalization by modeling the parameters in \cref{cycle10_fj} as functions of the upper limit $j_\text{max}$. The present data suggests that this dependence, too, is smooth. It would be interesting to understand the asymptotic behavior of such a generalized cycle approximation as $j_\text{max}\rightarrow \infty$ and $L\rightarrow\infty$, but this is out of reach with the present data.

Beyond the multi-linear model \cref{def:cycle_approximation}, we have also examined models that use non-linear polynomials, or logarithms of $n_j$. Despite introducing several additional parameters, the improvements over the multi-linear model were so small that we do not deem it useful to report any further details.

\begin{figure}[htb]
	\begin{subfigure}{ .49 \linewidth}
		\centering
		\includegraphics[width=\linewidth]{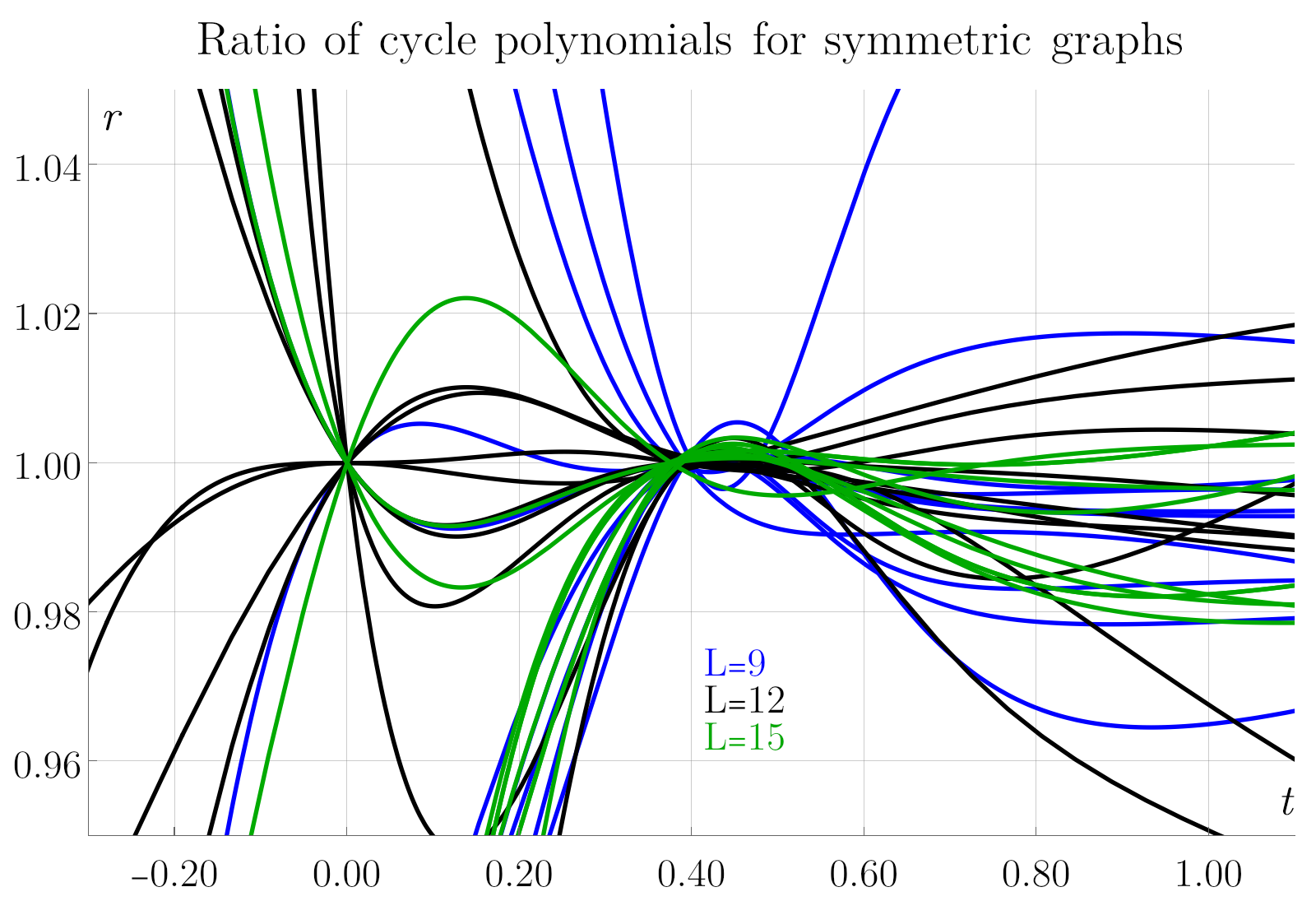}
		\subcaption{}
		\label{fig:cycle_polynomial_ratio}
	\end{subfigure}
	\begin{subfigure}{ .49 \linewidth}
		\centering
		\includegraphics[width=\linewidth]{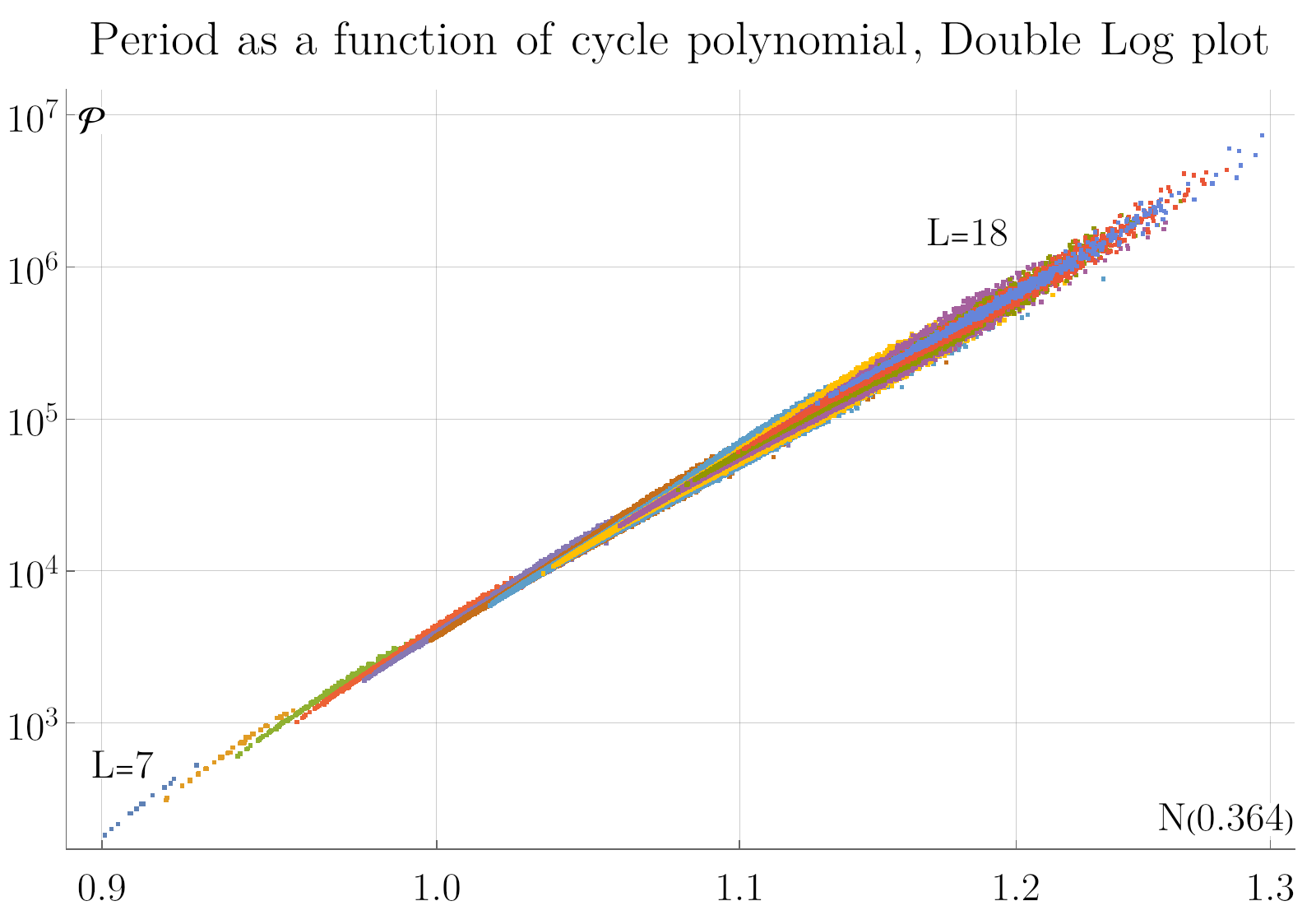}
		\subcaption{}
		\label{fig:cycle_polynomial_approximation}
	\end{subfigure}
	
	\caption{\textbf{(a)} Ratio $r(t)$ of cycle polynomials for pairs of graphs which have the same period. This ratio is generally not unity, indicating that the cycle polynomial (\cref{def:cycle_polynomial}) does not respect period symmetries. At $t\approx 0.37$, the ratio is very close to unity. Note that not all of the lines go through the point $(0,1)$. The plot shows 20 randomly selected pairs of periods per loop order. \textbf{(b)} Relation between $ \period$ and $\ln N_G(0.364)$. The overall trend is clearly linear. This plot shows the periods of different loop orders without the scaling of \cref{period_scaling}, i.e. the same approximate relation holds across all loop orders. Due to overplotting, it is invisible that many data points lie on the central line, and relatively few on the margins.  }
	
\end{figure}

\begin{figure}[htb]
	\begin{subfigure}{ .49 \linewidth}
		\centering
		\includegraphics[width=\linewidth]{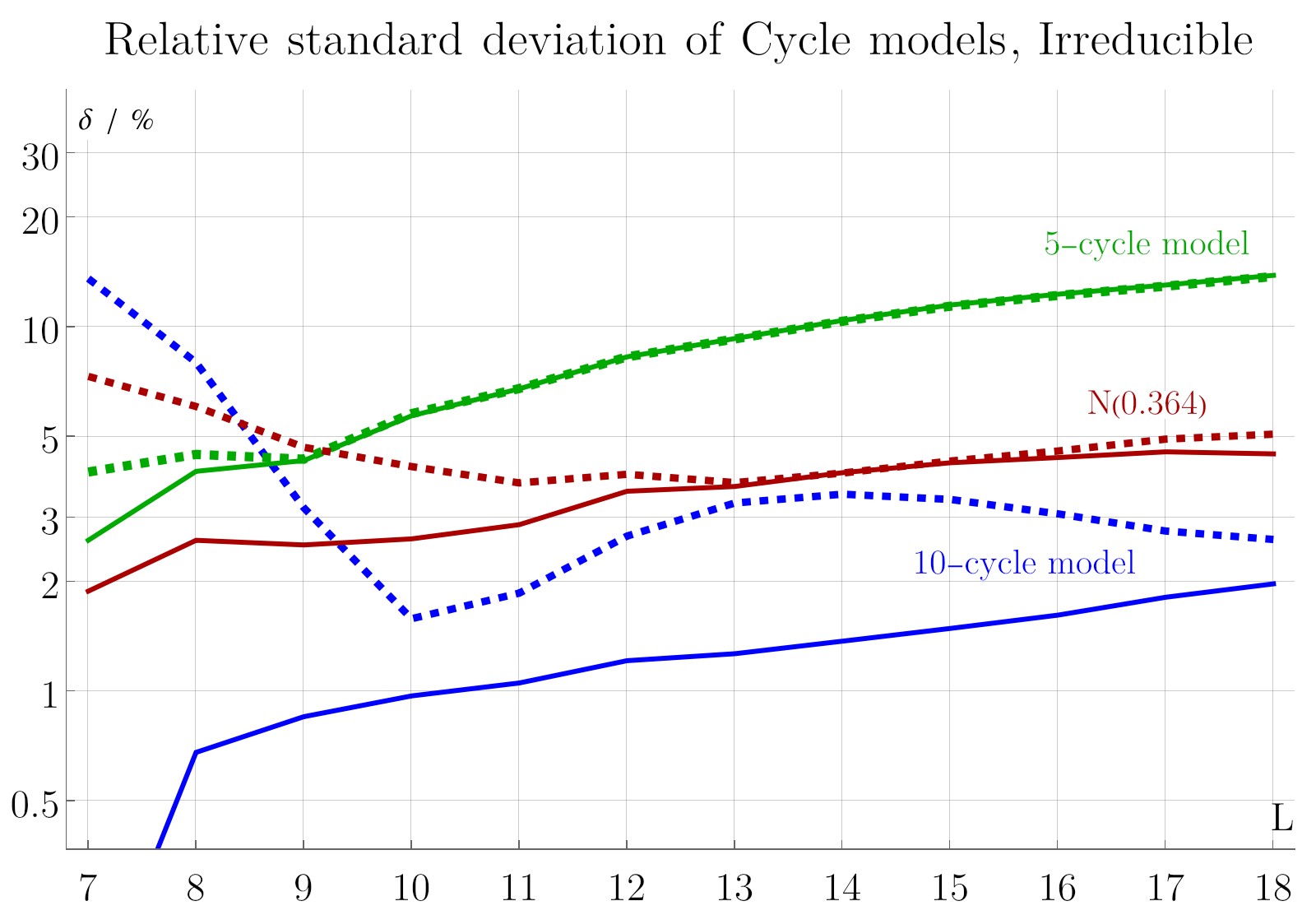}
		\subcaption{}
		\label{fig:cycle_models_relative_standard_deviation}
	\end{subfigure}
	\begin{subfigure}{ .49 \linewidth}
		\centering
		\includegraphics[width=\linewidth]{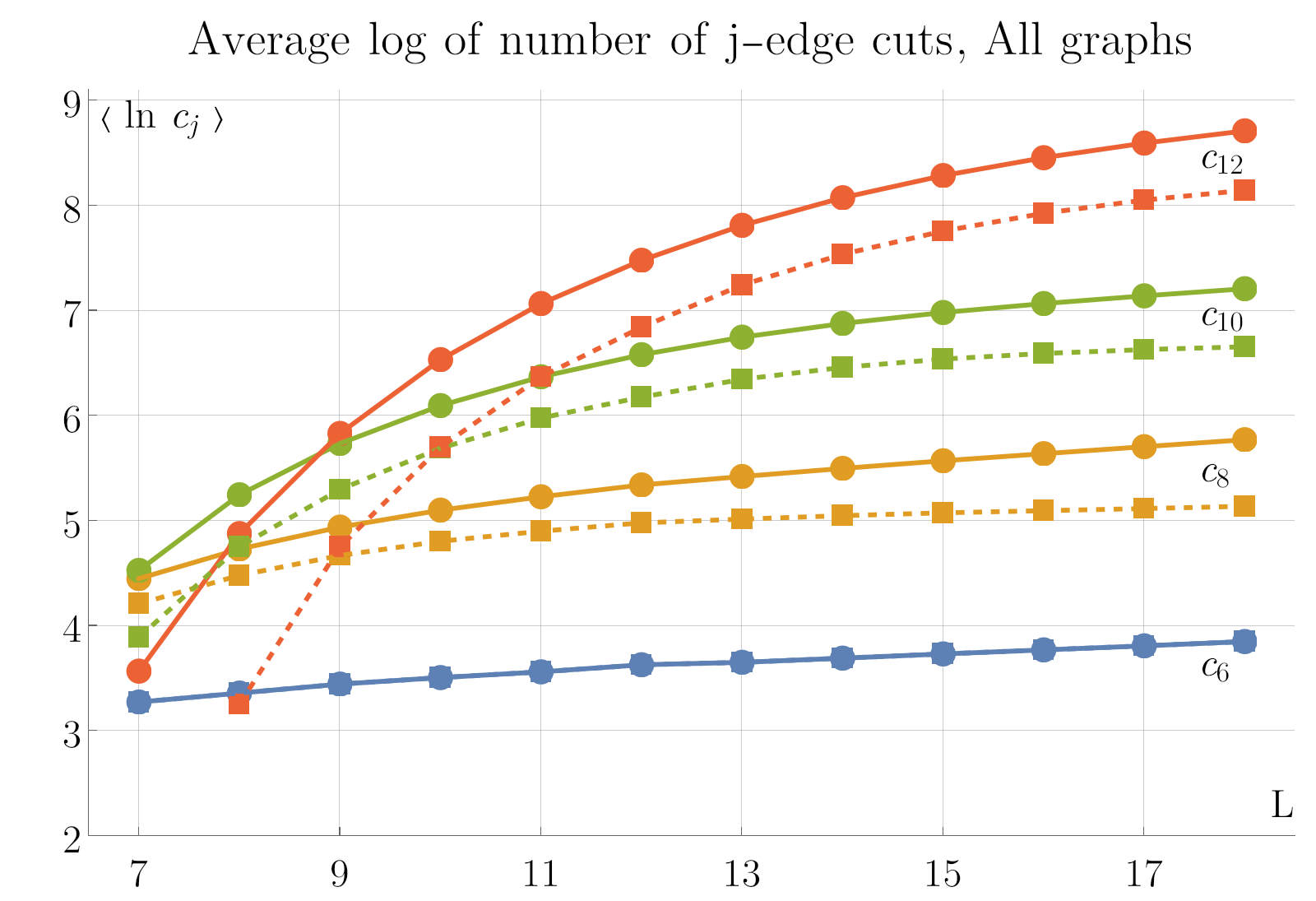}
		\subcaption{}
		\label{fig:cut_averages}
	\end{subfigure}

	\caption{\textbf{(a)} Relative standard deviation of the various cycle models, evaluated for irreducible graphs. Dashed lines are the global models, solid lines are best fit models for the individual loop orders. 
		\textbf{(b)}  Average of the number of $k$-edge cuts, depending on the loop order, log plot. Solid lines are cuts that produce arbitrarily many components, dashed lines are cuts into exactly two components. We see that the average of $  \ln   c_k    $ changes slowly with the loop order.
	}
	
\end{figure}

Conversely, the insight that the period can be approximated by a function which is \emph{linear} in the numbers of cycles  suggests to consider a \emph{cycle polynomial}, defined as
\begin{align}\label{def:cycle_polynomial}
	N_G(t) &:= \sum_{j=3}^{\abs{V_G}} t^j n_j (G) .
\end{align}
Like the individual counts of cycles $n_j$, the polynomial $N_G (t)$ is not invariant under period symmetries. However, one finds empirically that it is closed to being invariant when evaluated near $t= 0.364$. More precisely, if $G_1,G_2$ are two graphs related by a period symmetry, then $r(t):=\frac{N_{G_1}(t)}{N_{G_2}(t)}$ in most cases intersects the line $r(t)=1$ at $t\approx 0.364$, see \cref{fig:cycle_polynomial_ratio}. This pattern is broadly consistent across the loop orders we consider. But even within a single loop order, we have not been able to find a $t\in \mathbb C$ which  respects the symmetries exactly.

For the choice $t=0.364$,  the logarithm of the cycle polynomial is a remarkably good approximation of the logarithm of the period, see \cref{fig:cycle_polynomial_approximation}. In particular, this quantity approximates all periods without scaling for different loop orders according to \cref{period_scaling}, similar to the average distance examined in \cite[Sec. 6.2]{balduf_statistics_2023}. A fit including all loop orders results in 
\begin{align}\label{def:cycle_polynomial_approximation}
	\ln \bar \period (G) &= 28.21 \ln N_G ( 0.364) + 8.258.
\end{align}
This model has an accuracy (\cref{def:relative_standard_deviation}) of $\delta \approx 5\%$, see \cref{fig:cycle_models_relative_standard_deviation}, which can be further improved if one introduces a fine-tuned evaluation point $t$, and fit parameters, for each loop order. Nonetheless,  this model is not as useful as the 10-cycle model in practice because computing the cycle polynomial always takes longer than just the first 10 cycles, which already give rise to the better approximation \cref{cycle10_fj}.

\subsection{Cuts}\label{sec:cuts}

A \emph{cut} is  a subset $C \subset E_\Gamma$ of the edges such that the graph induced by $E_\Gamma \setminus C$ is not connected.   As with the cycles in \cref{sec:cycles}, there can be different conventions regarding what exactly is being counted. We generally consider vertex-induced cuts, that is, we select a subset of vertices and determine the number $n$ of edges that connects this subset to the remainder, making this an $n$-edge cut. We consider cuts as different when they arise from selecting different vertices, regardless of whether the obtained set of connected components is isomorphic to some other such set.

\begin{figure}[htb]

	\begin{subfigure}{ .49 \linewidth}
		\centering
		\includegraphics[width=\linewidth]{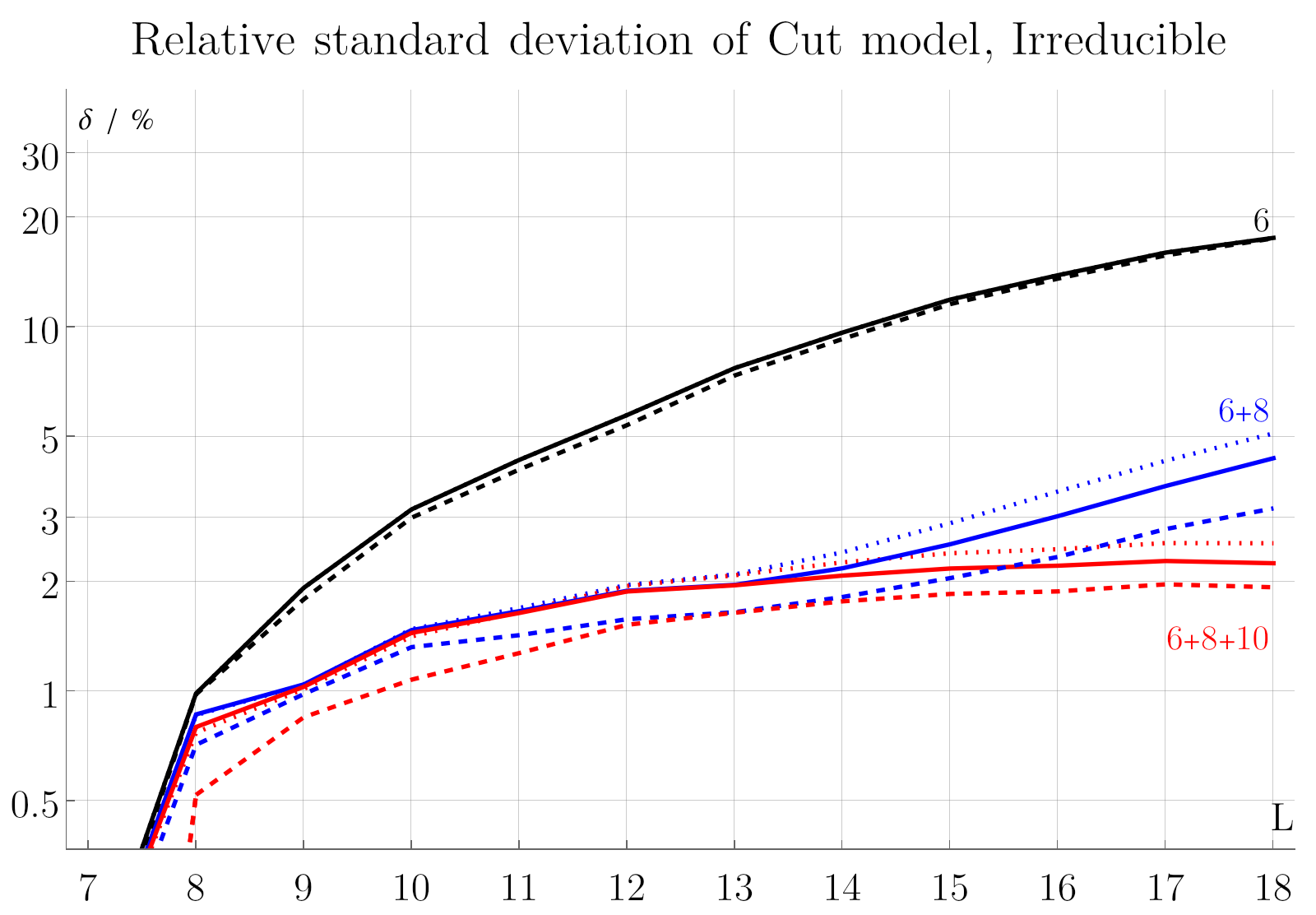}
		\subcaption{}
		\label{fig:cut_relative_standard_deviation}
	\end{subfigure}
		\begin{subfigure}{ .49 \linewidth}
		\centering
		\includegraphics[width=\linewidth]{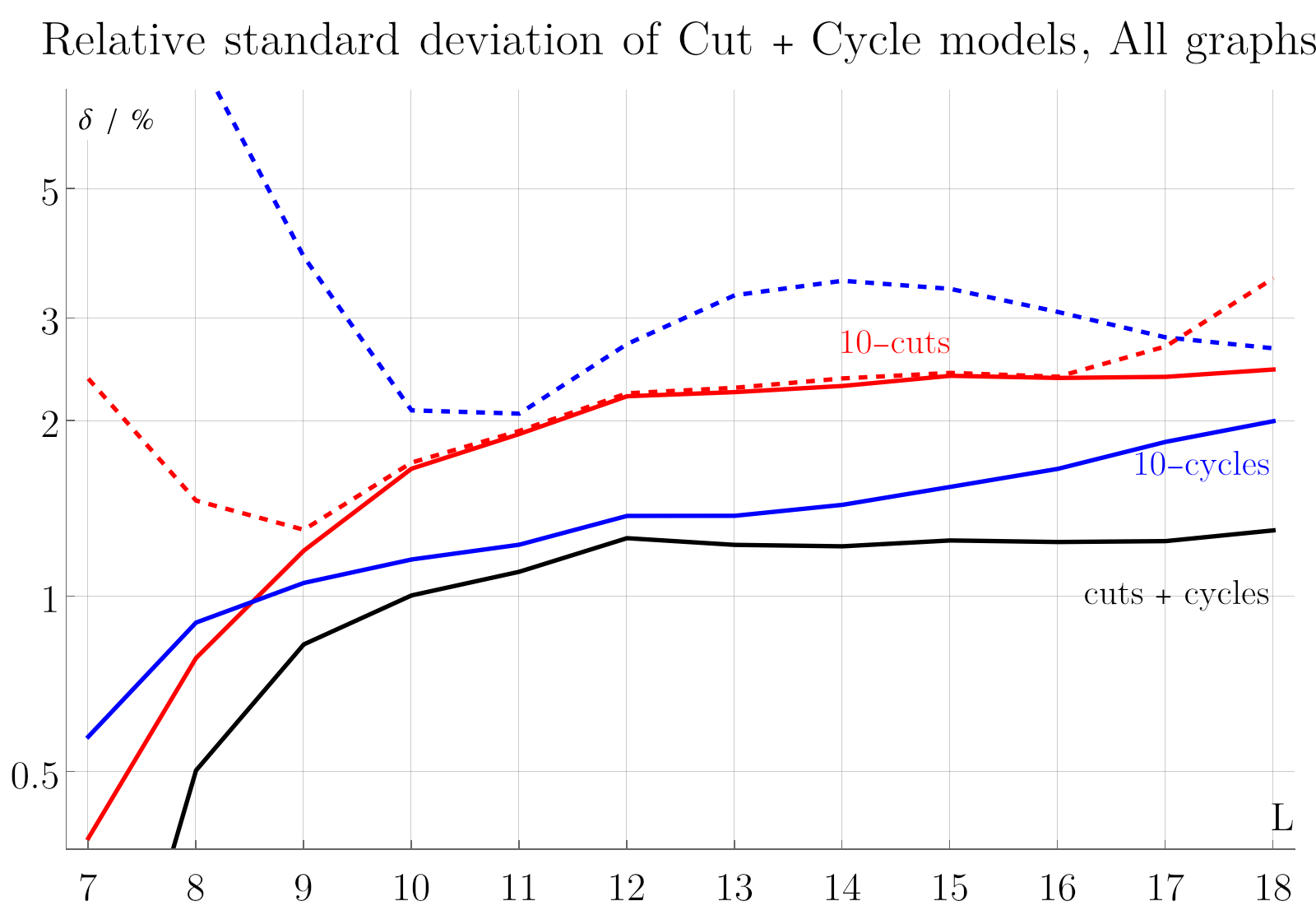}
		\subcaption{}
		\label{fig:cut_cycle_relative_standard_deviation_reducible}
	\end{subfigure}
	
	\caption{
		\textbf{(a)} Relative standard deviation $\delta$ (\cref{def:relative_standard_deviation}) of the cut model for irreducible periods. Thick: Linear model \cref{def:cut_approximation_linear}, based on connected cuts.  The 6-edge cuts (black) are a relatively poor prediction, including the 8-edge cuts (blue) gives significant improvements. The 10-edge cuts (red)  are useful only for $L \geq 14$ loops.  If we allow cuts into more than 2 components (dotted), the model gets slightly worse. Allowing quadratic functions of $\ln(c_j)$ (dashed) slightly improves the quality at the expense of introducing many new fit parameters. 	\textbf{(b)} Comparison of the performance of linear cut- and cycle models (\cref{def:cycle_approximation,def:cut_approximation_linear,def:cut_cycle_approximation}) for all (not only irreducible) graphs. Bold lines show best fits (using all graphs). Dashed lines are the global models, evaluated for all graphs.}
	\label{fig:cuts}
\end{figure}

The Feynman graphs considered in this work are 4-edge connected, and they not contain subgraphs that are connected to the remainder graph  with an odd number of edges.  
Hence, a 4-edge cut in these graphs can only arise from  cutting all 4 edges adjacent to a single vertex; the number of 4-edge cuts is simply the number of vertices $c_4=\abs{V_G}=(L+2)$ and thus uninteresting for predicting periods. The first nontrivial cuts are 6-edge cuts. The number $c_6$ of 6-edge cuts is known\footnote{Erik Panzer and Jacob Mercer, in preparation. Announced in a talk by Erik Panzer at HU Berlin in December 2021. All our numerical data confirms invariance of $c_6$.} to respect the symmetries of the Feynman period. Starting from  8-edge cuts, there is the possibility that a cut produces more than two connected components. We call a $j$-edge cut \emph{connected} if it gives rise to exactly two connected components, and otherwise \emph{unrestricted}. The number of connected cuts (dashed lines in \cref{fig:cut_averages}) is lower than the number of unrestricted cuts. 
The average number of $j$-edge cuts grows with the loop order, but  the logarithm $\ln  c_j $ grows slowly for the loop orders under consideration, see \cref{fig:cut_averages}, so that we do not introduce any further normalization.

We find empirically that the logarithm of the period is an almost linear function of not $c_j$ itself, but its logarithm $\ln  c_j $, and let
\begin{align}\label{def:cut_approximation_linear}
	\ln \bar \period &:= g_0 + g_6 \ln c_6  + g_8 \ln  c_8  + g_{10} \ln  c_{10} .
\end{align}

Although the $(j\geq 8)$-edge cuts do not respect the symmetries of the period, they prove useful in predicting its numerical value. We find that the connected cuts deliver slightly better accuracy than the unrestricted ones. For $L \leq 12$ loops,  including higher than $8$-edge cuts has almost no influence, for larger loop orders, the 10-edge cuts slightly improve accuracy, as shown in \cref{fig:cut_relative_standard_deviation}.  
We determined the parameters $g_j$ as a best fit to the irreducible periods, and found them to be smooth functions of $L$. A possible global model of \cref{def:cut_approximation_linear} is
\begin{align}\label{cuts_global_model}
	g_0 &= 19.1566 -17.8347 e^{ 0.06049 L}, \quad &g_6 &= -3.457 + 2.434 \ln L + 0.119 (\ln L)^2, \nonumber\\
	g_8 &= -1.249 + 0.1734 L , \qquad &g_{10} &= 1.348 + 0.2752 L + 0.01347 L^2.
\end{align}
The best-fit 10-cycle model \cref{def:cycle_approximation} is significantly more accurate than the best-fit 10-cut model \cref{def:cut_approximation_linear}. Nevertheless, the \emph{global} 10-cut model (\cref{cuts_global_model}) is superior to the \emph{global} 10-cycle model (\cref{cycle10_fj}), see \cref{fig:cut_cycle_relative_standard_deviation_reducible}. This is because the cut model contains overall fewer parameters, and is thus easier to describe as a global model.

Besides \cref{def:cut_approximation_linear}, we  examined quadratic functions of the $\ln  c_j $. Moreover, we constructed a multi-linear and a quadratic model based on the number of cuts into exactly three connected components. Like for the generalizations of cycle models in \cref{sec:cycles}, the accuracy gained in comparison to \cref{def:cut_approximation_linear} was so small that we do not discuss these models further. 

The approximation can be further improved by combining the cut model (\cref{def:cut_approximation_linear}) and the cycle model (\cref{def:cycle_approximation}) according to 
\begin{align}\label{def:cut_cycle_approximation}
	\ln \bar \period &:= g_0 + g_6 \ln c_6 + g_8 \ln c_8 + g_{10} \ln c_{10} +\sum_{j=3}^{10} f_j \frac{2 j \cdot n_j}{3^j}
\end{align}
where $c_j$ is the number of connected $j$-edge cuts, and $n_j$ is the number of $j$-edge cycles. In this case, the performance is almost identical if one uses $c_j$ instead of $\ln(c_j)$.   \Cref{def:cut_cycle_approximation}   contains the 8 parameters $\left \lbrace f_3, \ldots, f_{10} \right \rbrace   $ for the cycle part, and 4 parameters $\left \lbrace g_0, g_6, g_8, g_{10} \right \rbrace   $ for the cuts part. All 12 parameters depend on the loop order in a fairly regular fashion, but we refrain from introducing empirical functions, and instead quote the parameters explicitly in \cref{tab:cuts_cycles_parameters} in the appendix.  The accuracy obtained by \cref{def:cut_cycle_approximation} is $\delta \approx 1.2\%$ for all graphs.

\subsection{Hepp bound}\label{sec:hepp}

The Hepp bound \cite{hepp_proof_1966,bloch_motives_2006,panzer_hepp_2022} is a simplified version of the period integral (\cref{def:period}), where the Symanzik polynomial $\psi$ is replaced by   the dominant monomial for each particular value of the Schwinger parameters $\left \lbrace a_e \right \rbrace $. Unlike most of the above parameters, the Hepp bound is a period invariant and respects the 3-vertex cut symmetry. This implies that if a graph $G=G_1\cup G_2$ has a 3-vertex cut, we know that $\mathcal H(G)=\mathcal H(G_1)\mathcal H(G_2)$, see \cite{panzer_hepp_2022} for details. 

It has been observed in \cite{panzer_hepp_2022}, and examined numerically in  \cite{kompaniets_minimally_2017,balduf_statistics_2023}, that the Hepp bound alone predicts the period with an accuracy of $\Delta \approx 1\%$, given the correct fit parameters (which depend on the loop order $L$). 
It was found that the period approximately follows a power-law of the Hepp bound, $\period \sim \mathcal H^b(c_1 + \mathcal H c_2)$, where $b \approx 1.33$. Higher polynomials improve the fit accuracy only marginally. 

Our present work extends the analysis of \cite{kompaniets_minimally_2017,balduf_statistics_2023} by including the information about 6-edge cuts.
Like for all other models in \cref{sec:estimation}, we  compute the least-squares approximation for the Hepp model with respect to logarithms $\left \lbrace \ln \period, \ln \mathcal H \right \rbrace   $. To be consistent with \cref{def:cycle_approximation,def:cut_cycle_approximation}, we use an ansatz
\begin{align}\label{def:hepp_approximation}
	\ln \bar \period &= h_0 + h_1 \ln \mathcal{H} + h_2 \left( \ln \mathcal H \right) ^2 + \ldots. 
\end{align}
It turns out that including the term $h_2$ improves the accuracy by approximately a factor of 2 for large loop order, while higher order monomials provide almost no additional accuracy.  Note that the particular ansatz in \cref{def:hepp_approximation} is neither identical with the one in \cite{balduf_statistics_2023} nor the one in \cite{kompaniets_minimally_2017}, even if each of them contains 3 free fit parameters. As a global model for the period scaled to its leading asymptotics (\cref{period_scaling}), we find
\begin{align}\label{Hepp_global_fit}
	h_0 (L) &= -2.04 L^{1.24} + \ln \Big( 0.277 (L-10.7)- 0.008 (L-15.6)^3 \Big),  \nonumber \\
	h_1(L) &= 1.867 \ln \left( L-4.23 \right)  -0.231 \ln \left( L-4.26 \right) ^2, \\
	h_2(L) &=  (0.184\ln( L) -0.457)^2 - 0.036.\nonumber 
\end{align}

The residues of the approximation \cref{def:hepp_approximation} form \enquote{bands}, see \cite[Fig. 28]{balduf_statistics_2023}. The points in each band share the same number of 6-edge cuts $c_6$.  Unfortunately, the slope and position of these bands is not constant, such that they would require non-linear fit functions. The situation changes dramatically if we include a cubic term $h_3$ into the Hepp bound approximation \cref{def:hepp_approximation}: in that case, this term warps the residues in such a way that the bands become parallel. An approximation function which is cubic in $\ln \mathcal H$ and linear in $c_6$ is significantly more accurate than \cref{def:hepp_approximation} alone. For completeness, we include the higher cuts $c_8,c_{10}$ as well, and define
\begin{align}\label{def:hepp_cut_approximation}
	\ln \bar \period &= h_0 + h_1 \ln \mathcal{H} + h_2 \left( \ln \mathcal H \right) ^2 + h_3 \left( \ln \mathcal H \right) ^3 +  g_6 \ln  c_6  + g_8 \ln  c_8 + g_{10} \ln  c_{10}  .
\end{align}
Unlike for the cuts+cycles model (\cref{def:cut_cycle_approximation}), using the logarithm $\ln(c_j)$ in \cref{def:hepp_cut_approximation} results in better accuracy than using $c_j$ itself. The best fit parameters for \cref{def:hepp_cut_approximation} are quoted in \cref{tab:hepp_parameters} in the appendix.

\begin{figure}[htb]
	\centering
	\includegraphics[width=.6\linewidth]{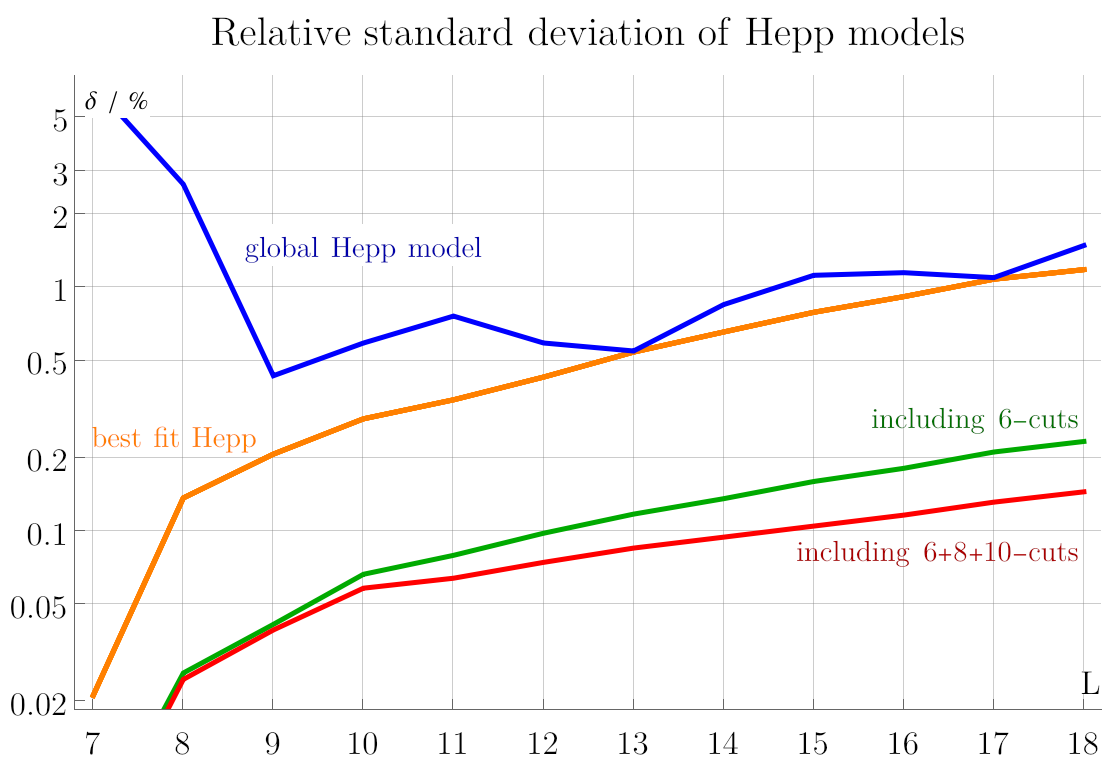}
	\caption{  Relative standard deviation of the various Hepp models. The orange line shows the best fit according to \cref{def:hepp_approximation}, the blue line is the global model for the fit parameters according to \cref{Hepp_global_fit}. This model reaches $\delta \approx 1\%$ accuracy for most loop orders. Including the 6-edge cuts and the third order of the Hepp bound (green) improves the accuracy by roughly a factor of five. The red line shows the model of \cref{def:hepp_cut_approximation}, which also includes 8- and 10-edge cuts.
	}
	\label{fig:hepp_standard_deviation}
	
\end{figure}

The performance of the different Hepp approximations are shown in \cref{fig:hepp_standard_deviation}. Overall, the Hepp bound alone delivers similar performance than the combined cuts+cycles model shown in \cref{fig:cut_cycle_relative_standard_deviation_reducible}. If the Hepp bound is combined with the number of cuts, the accuracy can be increased by a factor of roughly five, where the largest improvement comes from including $c_6$, and we reach $\delta < 0.2\%$ for all loop orders under consideration.

\subsection{Martin invariant}\label{sec:martin}

The Martin invariant \cite{panzer_feynman_2023} is a recently discovered period invariant which is closely related to  $O(N)$-symmetry of $\phi^4$-theory. Concretely, the Martin polynomial of a graph $G$ is $M(G,x)=3^{\abs{V_G}}(x-2)^{-1}  T(G,x-2) $, where  $T(G,N)$ is the $O(N)$-symmetry factor of $G$, see \cite{balduf_statistics_2023}. The Martin invariant $M^{[1]}(G)$ is the linear term   $[x^1]M(G,x)$ of the Martin polynomial. The higher Martin invariants $M^{[k]}$ are computed like $M^{[1]}$, but for a graph $G^{[k]}$ where every edge of $G$ is replaced by $k$ parallel edges.

\begin{figure}[htb]
		\begin{subfigure}{ .49 \linewidth}
		\centering
		\includegraphics[width=\linewidth]{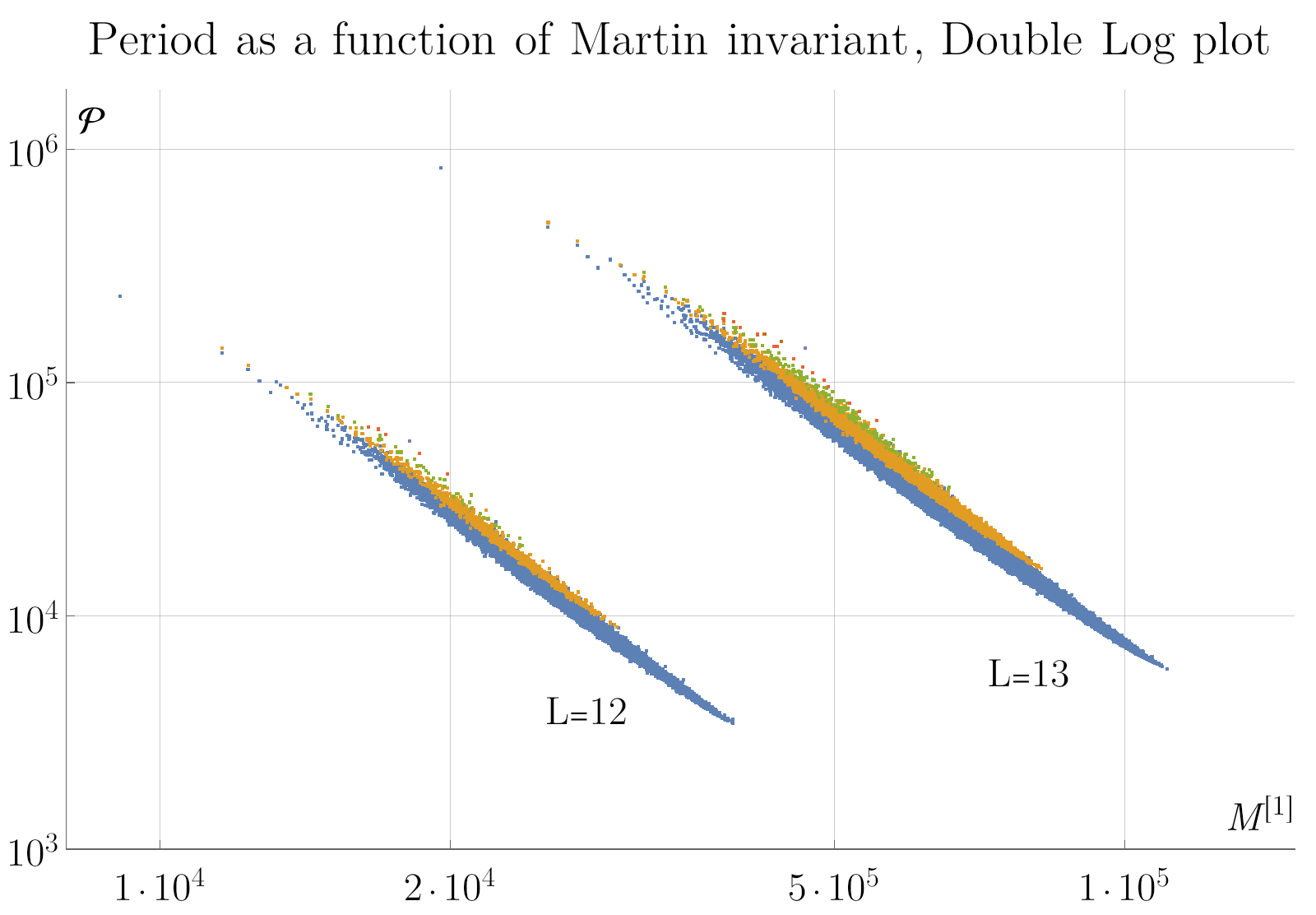}
		\subcaption{}
		\label{fig:martin_period}
	\end{subfigure}
	\begin{subfigure}{ .49 \linewidth}
		\centering
		\includegraphics[width=\linewidth]{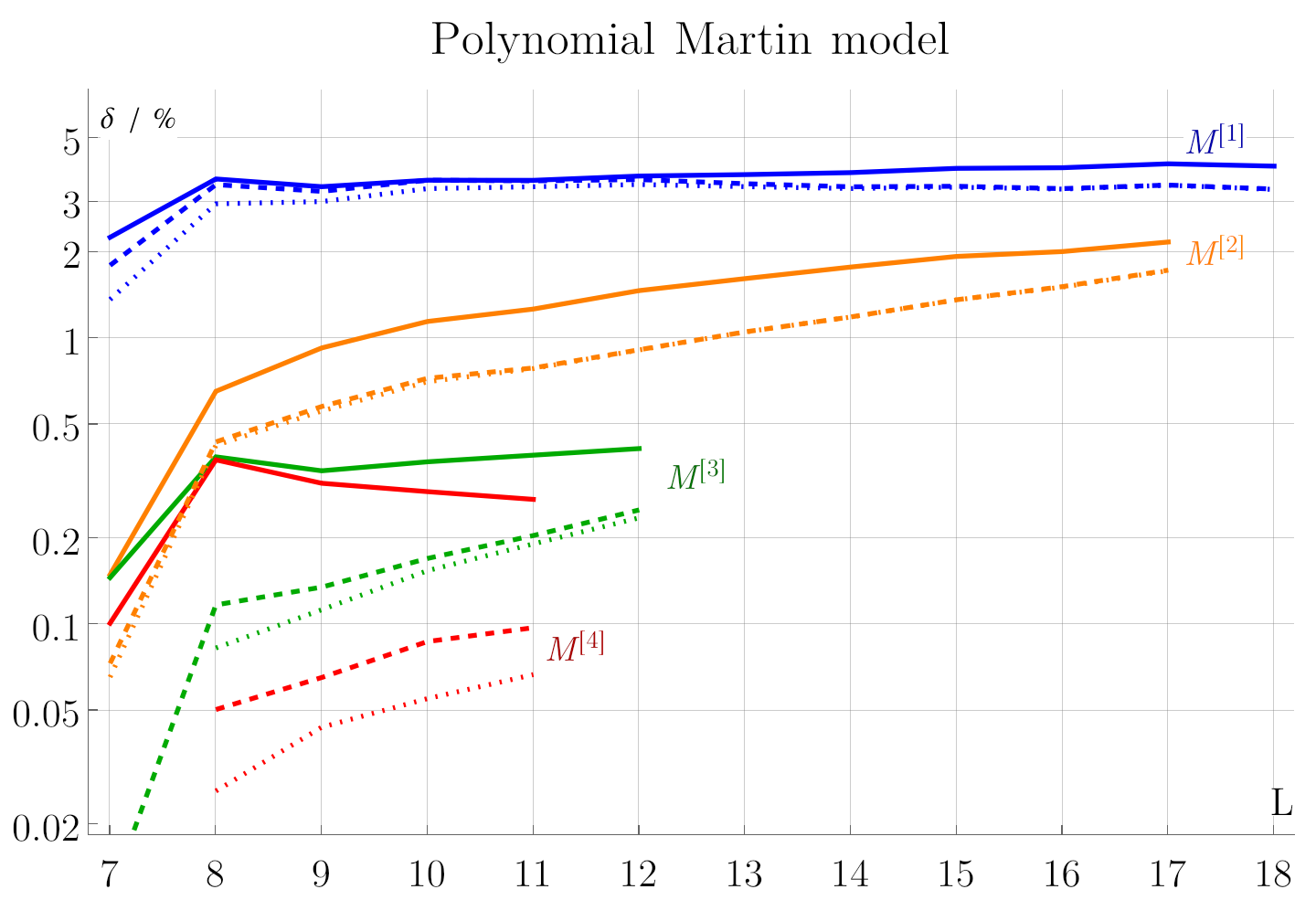}
		\subcaption{}
		\label{fig:martin_quadratic_relative_standard_deviation}
	\end{subfigure}

	\caption{ 	\textbf{(a)} Correlation between the first Martin invariant and the period, double log plot. The relation is almost linear, indicating a power-law dependence of $\period$ on $M^{[1]}$. The relation is visibly different for the 3-vertex irreducible graphs (blue) than for the ones with one 3-vertex cuts (orange), two 3-vertex cuts (green) etc. 
		\textbf{(b)} Relative standard deviation $\delta$ (\cref{def:relative_standard_deviation}) of the non-linear Martin model (\cref{def:martin_approximation}). All lines refer to irreducible graphs only. Solid lines indicate a linear model. The dashed line is a quadratic model, it is superior as soon as at least $M^{[2]}$ is being included. The cubic model (dotted line) provides further improvements as soon as $M^{[3]}$ included.  
	}
	
\end{figure}

Numerically, one finds that $M^{[k+1]} \gg M^{[k]}$. It has been remarked already in \cite{panzer_feynman_2023,balduf_statistics_2023} that the Martin invariant approximates the period by a power law, similar to the Hepp bound, but with opposite slope and lower accuracy.   
A closer examination, see \cref{fig:martin_period},  shows that for the Martin invariant, the best fit parameters depend on the number of 3-vertex cuts. This effect does exist for other approximations as well, but is   much smaller there. Since the Martin invariant respects 3-vertex products of periods, it is sufficient to know an approximation for the irreducible graphs, and we therefore restrict ourselves to approximating 3-vertex irreducible graphs.

As seen in  \cref{fig:martin_period}, we should expect an almost linear relation between $\ln \period$ and $\ln M^{[1]}$. In order to cover potential non-linearities arising from the higher Martin invariants $\ln M^{[k]}$, we use the polynomial ansatz
\begin{align}\label{def:martin_approximation}
	\ln \bar \period &= m_0 + \sum_{k=1}^{k_\text{max}} \sum_{p=1}^{p_\text{max}} m_{k,p} \left(\ln  M^{[k]} \right)^p  .
\end{align}
The relative standard deviation $\delta$ of this model is shown in \cref{fig:martin_quadratic_relative_standard_deviation}. 
Recall that for $L=7$ loops, there are only 9   independent irreducible graphs. The quadratic model including $M^{[4]}$ has 9 free parameters, it therefore reproduces all irreducible periods exactly. 
We observe that a non-linear model, $p_\text{max}>1$ in \cref{def:martin_approximation}, only provides very small improvements over a linear one as long as only $M^{[1]}$ is being used. The more of the higher $M^{[k]}$ are available, the larger the advantage of non-linear models is. Including $M^{[2]}$ makes the quadratic model better than the linear one, but the cubic model is hardly any better than the quadratic one. However, the cubic model is clearly superior to the quadratic one as soon as one includes at least $M^{[3]}$. At 11 loops, a linear model including $M^{[3]}$ reaches $\delta<0.5\%$, while a cubic one is well below $\delta =0.1\%$. We conclude that the Martin invariant gives the best approximation of all models considered, but only if the higher Martin invariants are being included. Unfortunately, it is computationally demanding to determine those higher invariants for large graphs.

\begin{figure}[htb]
	
	\begin{subfigure}{ .49 \linewidth}
		\centering
		\includegraphics[width=\linewidth]{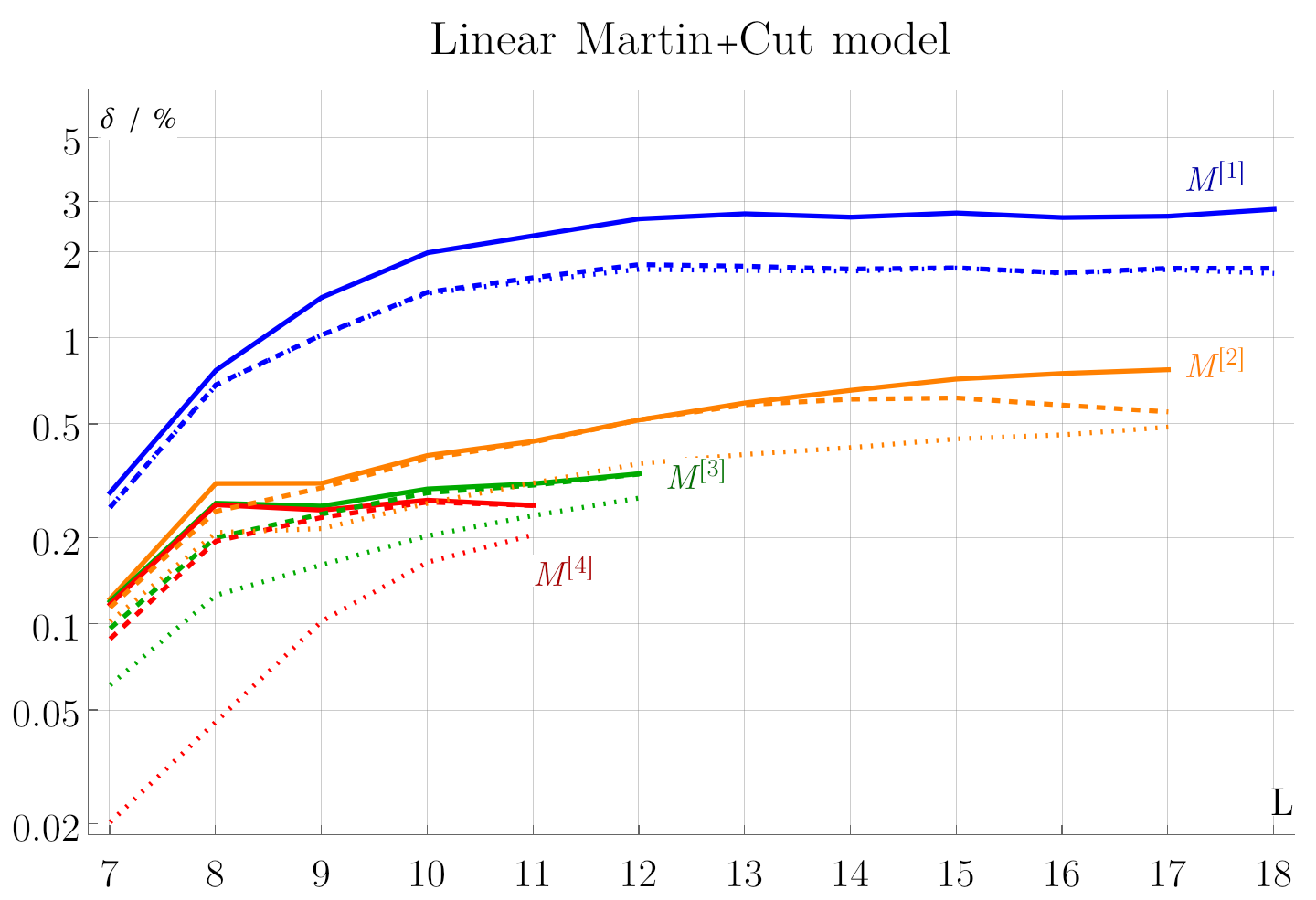}
		\subcaption{}
		\label{fig:martin_cut_linear}
	\end{subfigure}
	\begin{subfigure}{ .49 \linewidth}
		\centering
		\includegraphics[width=\linewidth]{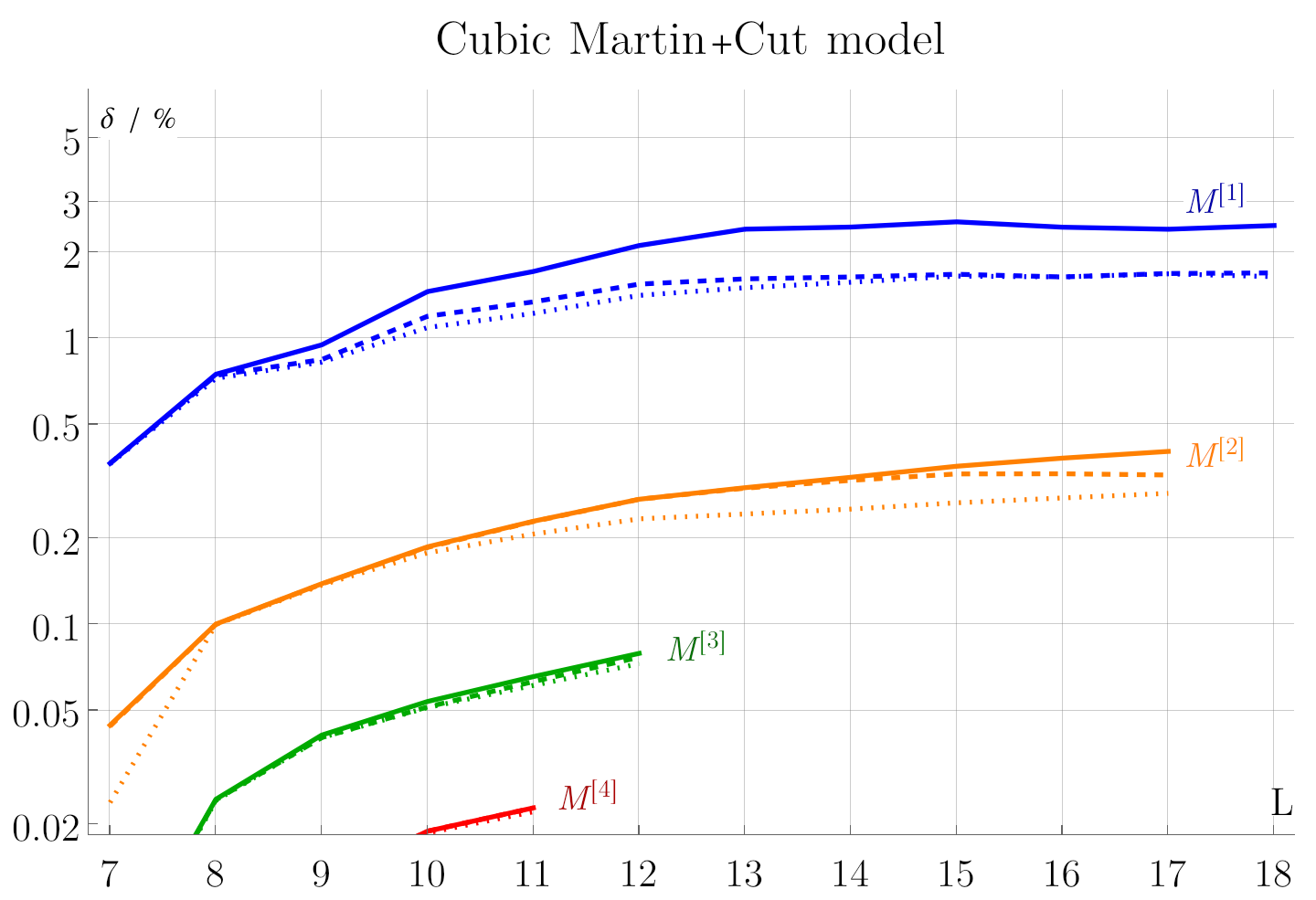}
		\subcaption{}
		\label{fig:martin_cut_cubic}
	\end{subfigure}
	\caption{ 
	 	\textbf{(a)} Linear Martin Model including edge cuts (\cref{def:martin_cut_approximation} with $p_\text{max}=1$).  Using 6-edge cuts (thick) significantly improves accuracy compared to (a). With 6+8-edge cuts (dashed) or 6+8+10-edge cuts (dotted), the accuracy is even better.
		\textbf{(b)} Cubic Martin Model including edge cuts (\cref{def:martin_cut_approximation} with $p_\text{max}=3$). This model is by far superior to both the linear model including cuts (\cref{fig:martin_cut_linear}), and to the cubic model without cuts (\cref{fig:martin_quadratic_relative_standard_deviation}). }
	\label{fig:martin_cut}
\end{figure}

As a remark,  one can get a slightly better approximation by including  combinations of higher order logarithms such as $\ln \left( \ln   M^{[2]} \right)   $. We did not pursue this further. 
However, a considerable improvement is possible by including edge cuts.  Again, we consider a model which is linear in the logarithm of the number of cuts, 
\begin{align}\label{def:martin_cut_approximation}
	\ln \bar \period &= m_0 +\sum_{k=1}^{k_\text{max}} \sum_{p=1}^{p_\text{max}} m_{k,p} \left( \ln M^{[k]} \right)^p   + \sum_{j=6}^{c_\text{max}} g_j \ln  c_j .
\end{align}
First, restricting to the linear model $p_\text{max}=1$, the accuracy for irreducible graphs  is shown in \cref{fig:martin_cut_linear}. Combining $M^{[1]}$ with the  10-edge cuts, we reach consistently below $\delta =2\%$, which is considerably better than $M^{[1]}$ alone (\cref{fig:martin_quadratic_relative_standard_deviation}). However, the 10-edge cuts alone already give around 2.2\% (see \cref{fig:cuts}), therefore, the advantage of including the Martin invariant is relatively small. This changes as soon as we include higher $M^{[k]}$, already with $M^{[2]}$, the error is below 1\%, which is impossible to reach with cuts alone.

The prediction can be improved further by allowing non-linear terms  $p_\text{max}>11$, while the model is still linear in $\ln c_j $. The cubic model, $p_\text{max}=3$, is shown in \cref{fig:martin_cut_cubic}. Again, the advantage of non-linear models grows with higher $M^{[k]}$. Using $M^{[3]}$,   we obtain $\delta<0.1\%$ at $L=12$. This is comparable to the best Hepp bound + cuts model (\cref{fig:hepp_standard_deviation}), while by including $M^{[4]}$ the accuracy of non-linear Martin models is considerably better than any other model.

\FloatBarrier

\newpage

\section{Importance graph sampling}\label{sec:sampling}

\subsection{Accuracy and performance of Monte Carlo sampling}\label{sec:accuracy_performance}
For physical predictions such as the beta function, we don't need individual Feynman integrals, but their sum  weighted with their symmetry factor. 
Let $N_L^{(C)}$ be the number of $L$-loop completions. The beta function can be written as a sum of these completions, see \cite{panzer_hepp_2022}, and, knowing $N_L^{(C)}$, we can equivalently consider the average
\begin{align}\label{PAut_average}
	\left \langle\frac{ \period}{\abs{\Aut}} \right \rangle &:= \frac{1}{N_L^{(C)}} \sum_G\frac{ \period(G)}{\abs{\Aut(G)}}.
\end{align}
For $L \geq 14$ loops, $N_L^{(C)}$ is so large (see counts in \cite{balduf_statistics_2023} and asymptotics in \cite{borinsky_renormalized_2017}) that we can not compute the periods of \emph{all} graphs in question. We are therefore interested in \emph{random samples} of $N_s\ll N_L^{(C)}$ graphs. 

Let $\sigma(\period)$ be the standard deviation of the distribution of periods. The standard deviation of the sample average will be denoted $\Delta_{\text{samp}}$, this quantity measures the uncertainty which is introduced by a non-complete sample, regardless of numerical uncertainties of the individual summands. 
For our application, the sample size $N_s$ is larger than, say, 100, but at the same time much smaller than the population size $N^{ (C )}_L$. In this case, the sampling uncertainty is given by (see the appendix of \cite{balduf_statistics_2023} for details)
\begin{align}\label{sampling_uncertainty}
	\Delta_{\text{samp}} \left \langle \period \right \rangle  =   \frac{1}{\sqrt {N_s} }  \sigma(\period) = \frac{1}{\sqrt {N_s} }  \delta (\period) \left \langle \period \right \rangle  .
\end{align}
Here, $\delta$ is the  relative standard deviation  from \cref{def:relative_standard_deviation}, but for the periods themselves, not for an approximation. For the periods, $\delta$ is almost unity, consequently  
\begin{align}\label{naive_sample_average}
	\frac{\Delta_{\text{samp}}\left \langle \period \right \rangle  }{\left \langle \period \right \rangle  } = \frac{1}{\sqrt{N_s}} \delta(\period)  \approx \frac{1}{\sqrt{N_s}}.
\end{align}
For an accuracy of three significant digits, that is, $10^{-3}$ relative uncertainty, we would need to sample $N_s \approx 10^6$ periods, which is very challenging even with fast integration algorithms. To overcome the problem, we use importance sampling, based on an approximation function $\bar \period(G)$.

A typical random graph generation algorithm produces graphs weighted by their symmetry factor, denoted by $G \sim \frac{1}{\abs{\Aut}}$. We will denote averages of this distribution by $\left \langle \right \rangle  _\text{Aut} $.  These averages are related by \cite{balduf_statistics_2023} 
\begin{align*}
	\left \langle \frac{\period}{\abs{\Aut}} \right \rangle &= \left \langle \frac{1}{\abs{\Aut}} \right \rangle  \left \langle \period \right \rangle _\text{Aut}, \qquad \qquad 	N^{(C)}_L \left \langle \frac{\period}{\abs{\Aut}} \right \rangle  =  N^{(\text{Aut})}_L  \left \langle \period \right \rangle _\text{Aut},
\end{align*}
where $N^{(\text{Aut})}_L$ is the sum of all symmetry factors at $L$ loops. With this, we can rewrite the average from \cref{PAut_average} as
\begin{align}\label{PAut_average_product}
N^{(C)}_L	\left \langle  \frac{\period}{\abs{\Aut}} \right \rangle &= N^{(\Aut)}_L \sum_{G \sim \frac{1}{\abs{\Aut}}}  \bar \period(G) \frac{\period(G)}{\bar \period(G)} \nonumber\\
	&= N^{(\Aut)}_L \left( \sum_{G' \sim \frac{1}{\abs{\Aut}}}  \bar \period(G') \right) \sum_{G \sim \frac{1}{\abs{\Aut}}}  \frac{ \bar \period(G) }{\left( \sum_{G' \sim \frac{1}{\abs{\Aut}}}  \bar \period(G') \right)} \frac{\period(G)}{\bar \period(G)} \nonumber \\
	&= N^{(\Aut)}_L   \cdot \left \langle \bar \period \right \rangle _{\Aut} \cdot  \left \langle \frac{\period}{\bar \period} \right \rangle _{\Aut, \bar \period}.
\end{align}
The first factor $N^{(\text{Aut})}_L$ can be determined exactly \cite{cvitanovic_number_1978,borinsky_renormalized_2017}.
The last two factors are independent from each other and can be computed with different sample sizes. We  use $N_s$ samples for the third factor, but $F_s\cdot N_s  \gg N_s$ samples for the second one. 
Using \cref{sampling_uncertainty} and Gaussian error propagation, the relative sampling uncertainty of the product is
\begin{align}\label{relative_uncertainty_general}
	U:=\frac{\Delta_\text{samp} N^{(C)}_L \left \langle \frac{\period}{\abs{\Aut}} \right \rangle } {N^{(C)}_L \left \langle \frac{\period}{\abs{\Aut}} \right \rangle}  
	&= \frac{1}{\sqrt{N_s}} \left( \delta \left( \frac{\period}{\bar \period}  \right)  + \frac{1}{\sqrt{F_s}}  \delta \left( \bar \period \right)     \right) .
\end{align}
Here, $\delta$ denotes the relative standard deviation from \cref{def:relative_standard_deviation}, but computed for the non-uniform sample, not for the total population (which, in practice, is close to the population value $\delta(\bar \period)$). 
The overall scaling with the sample size $N_s$ is the same as for a naive sample average \cref{naive_sample_average}. However, in \cref{relative_uncertainty_general} a more complicated expression appears in place of the relative standard deviation of the distribution, $\delta(\period)$. Assuming that $\bar \period$ is a good approximation, the standard deviation $\delta(\bar \period)$ will be approximately equal to $\delta(\period)$. But we are free to chose the sampling factor $F_s$, consequently the second summand in parentheses can be made small. A the same time, for a good approximation function $\bar \period$, the ratio $\frac{\period}{\bar \period}$ is almost constant, and therefore $\delta \left( \frac{\period}{\bar \period} \right) $ is much smaller than $\delta(\period)$. Consequently, \cref{relative_uncertainty_general} can yield a smaller total uncertainty than a naive sample of $N_s$ elements.

We want to chose $F_s$ such that the accuracy of \cref{relative_uncertainty_general} is best possible in a fixed runtime. Let $t_i$ be the time it takes to compute one numerical integral, and $t_a$ the time to compute the approximation $\bar \period$, then the total runtime (ignoring time spent on graph generation) is
\begin{align*}
	T = N_s ( t_i + F_s t_a), \qquad \Rightarrow \quad N_s = \frac{T }{t_i + F_s t_a}.
\end{align*}
Inserting this expression for $N_s$ into \cref{relative_uncertainty_general}, we obtain a function $U(T,  F_s)$, which we want to minimize for a fixed $T $. Thus, we demand $\partial_{F_s}U(T,F_s)=0$, and obtain 
\begin{align*}
F_s &= \left(  \frac{\delta(\bar \period) \cdot t_i}{\delta \left( \frac{\period}{\bar \period} \right) \cdot t_a} \right) ^{\frac 23}, \qquad N_s = \frac{T}{t_i} \frac{1}{1+ \left( \frac{\delta(\bar \period)^2}{\delta \left( \frac{\period}{\bar \period} \right) ^2}  \frac{t_a}{t_i}\right) ^{\frac 1 3}}.
\end{align*}
For good approximations, we expect the two ratios $w:= \frac{\delta \left( \frac{\period}{\bar \period} \right)  }{\delta ( \bar \period)}$ and $v:= \frac{t_a}{t_i}$ to be small. Hence, we insert   $F_s =  w^{-\frac 23} v^{-\frac 23}$ and $N_s = \frac{T}{t_i(1+ v^{\frac 13} w^{-\frac 23})}$ into \cref{relative_uncertainty_general} and expand to first order. We find that the relative uncertainty  of the weighted sum  \cref{PAut_average_product} of periods is
\begin{align}\label{relative_uncertainty_approximation}
U &= \delta(\bar \period) \sqrt{\frac{t_i}{T}} \left( w^{\frac 23} + v^{\frac 13} \right) ^{\frac 32} \approx\delta(\bar \period) \sqrt{\frac{t_i}{T}} \left(  w + \sqrt v   \right) .
\end{align} 
For a fixed loop order, $\delta(\bar \period)\approx \delta(\period)$ as well as $t_i$ and $T$ are constants. We see that the quality of an approximation depends linearly on its accuracy $\delta(\frac{\period}{\bar \period})$, and on the square root $\sqrt{t_a}$ of the time it takes to obtain it.

\subsection{Comparison of the  approximation algorithms}\label{sec:performance}

The accuracy $\delta$ (\cref{def:relative_standard_deviation}) of the various approximations has been discussed exhaustively in \cref{sec:estimation}. We wrote a C++ program to compute these quantities, using the \texttt{GNU Eigen} library version 3.4.0 for linear algebra operations. The computation time $t_a$ depends on  details of the particular implementation, we view our program as a proof-of-concept which still has room for algorithmic improvements.

\begin{figure}[htb]
	
	\begin{subfigure}{ .49 \linewidth}
		\centering
		\includegraphics[width=\linewidth]{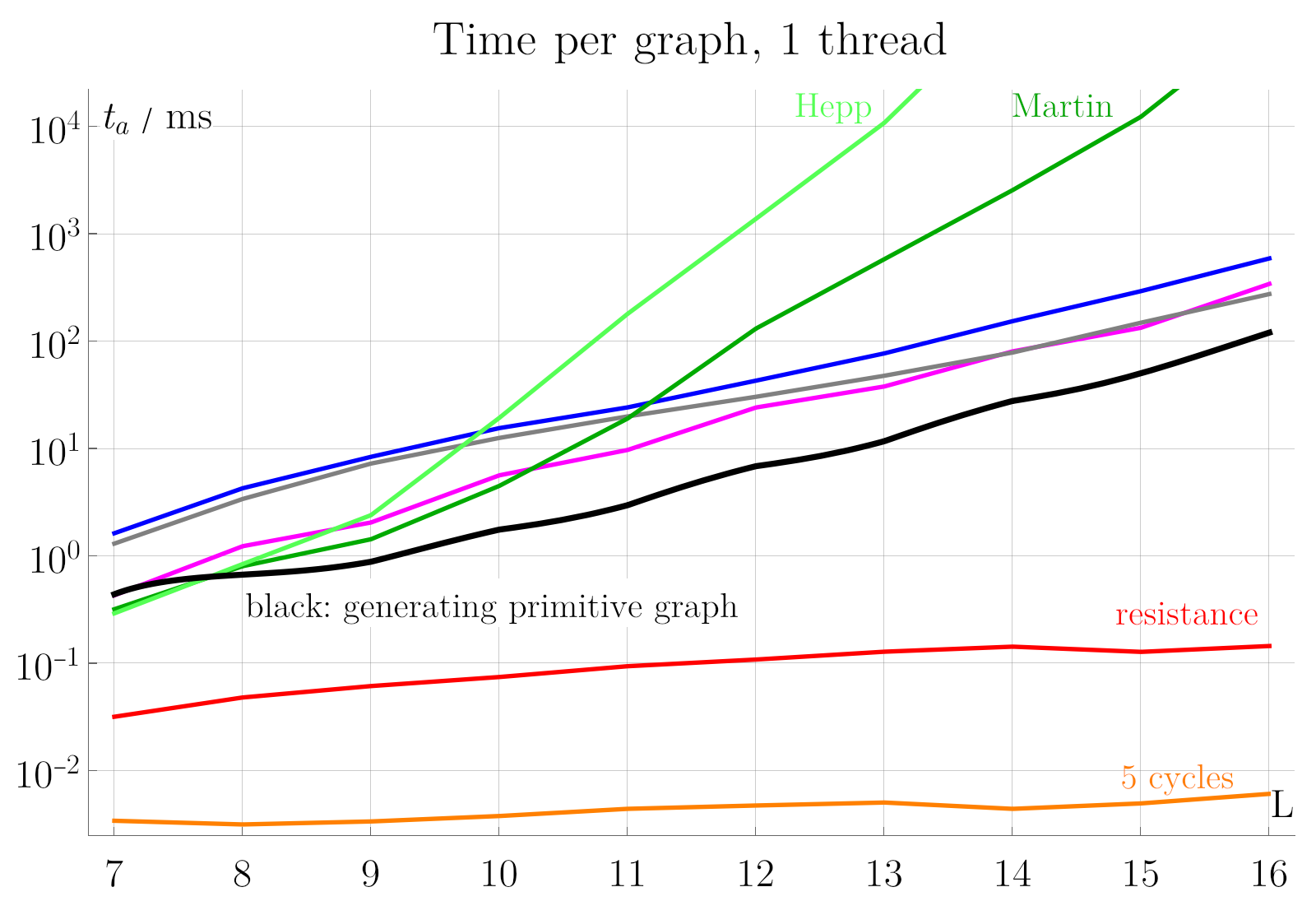}
		\subcaption{}
		\label{fig:ta}
	\end{subfigure}
	\begin{subfigure}{ .49 \linewidth}
		\centering
		\includegraphics[width=\linewidth]{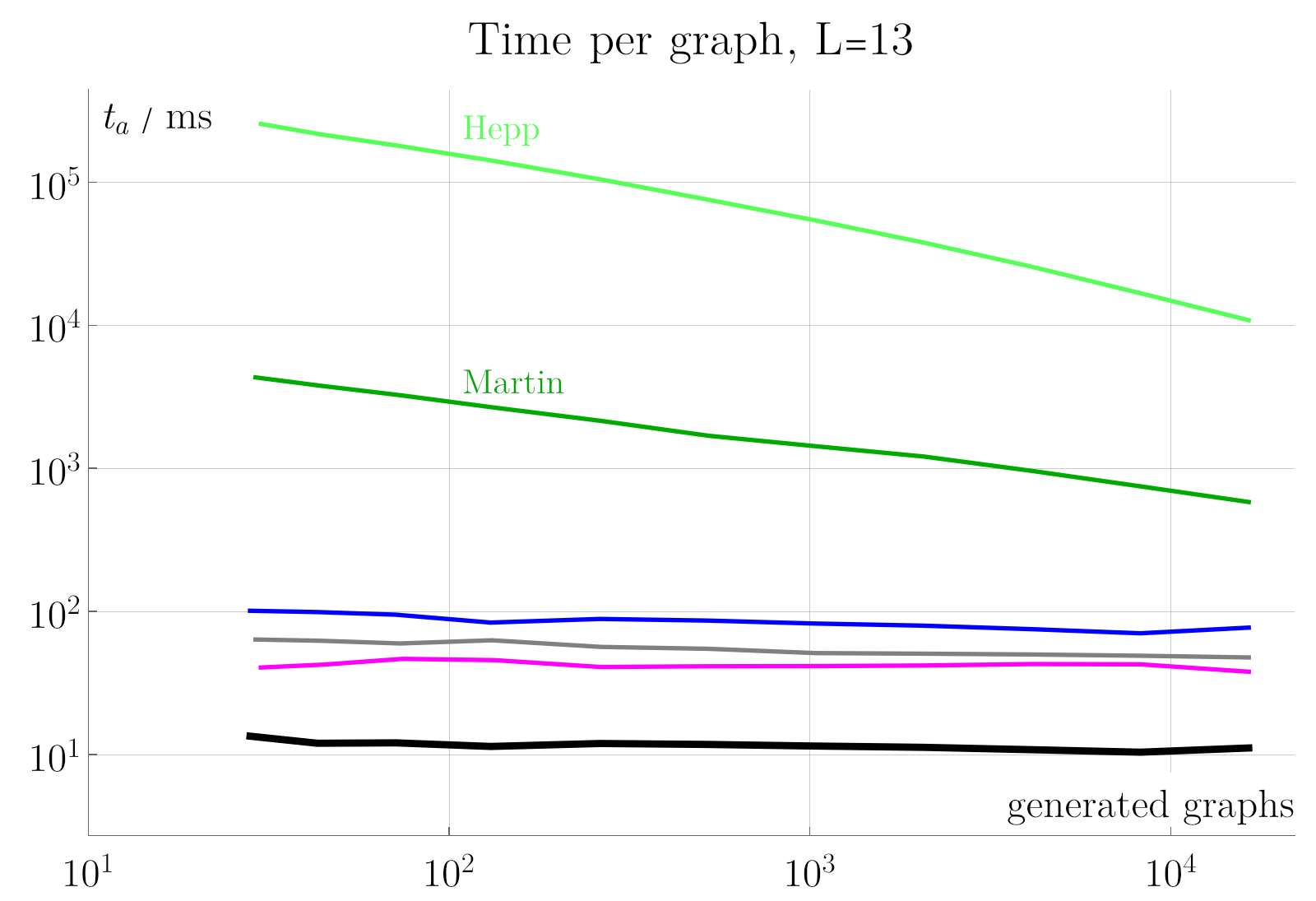}
		\subcaption{}
		\label{fig:ta_generated}
	\end{subfigure}
	\caption{ 
		\textbf{(a)} Time $t_a$ to compute the approximation for one graph. The 5-cycle approximation and the resistance approximation are much faster than any other approximation. The 10-edge-cut (magenta), 10-cycle (gray) and cut+cycle (blue) approximations are similar in speed, and comparable to checking a graph for primitivity (black). The Hepp and Martin approximations are fast for low loop orders, but quickly become slower as $L>10$.   
		\textbf{(b)} The Hepp- and the Martin approximations use caching, therefore the average time required per graph decreases as more graphs are in the cache. All other approximations were implemented without caching.}
	
\end{figure}

The plot \cref{fig:ta} shows the time needed per graph when the approximation is computed for a total of 16000 graphs, where, at low loop order, the graphs will repeat. All computations ran on a single thread of a Intel Xeon Gold 6320 CPU at 2.10 GHz. Several observations are noteworthy:
\begin{itemize}
	\item In all cases, the time $t_a$  scales roughly exponentially with the loop order.
	\item The generation of a random primitive graph takes a few milliseconds (black line in \cref{fig:ta}), this is almost entirely due to testing 6-edge connectedness.
	\item There are largely three classes of runtimes:
	\begin{enumerate}
		\item The 5-cycle approximation and the resistance approximation are much faster than generating the graphs. This is because they merely require linear algebra operations with the graph matrices.
		\item The approximations based on cycles and cuts are similar in speed as checking for 6-edge connectedness. These approximations require some sort of iteration over elements of the graph.
		\item Computing the Martin invariant or the Hepp bound is much slower than generating the graphs. These algorithms require recursive examination of subgraphs and  caching.
	\end{enumerate}
\end{itemize}

The caching implies that the prediction based on Martin and Hepp bound become faster  the more graphs have been computed, as shown in \cref{fig:ta_generated}. Unfortunately, the cache needs to include all subgraphs, and  for $L \geq 14$ loops the cache becomes so large ($\gg 10\text{GB}$) that lookups are increasingly slow and the size starts to run into hardware limits. Note that it would be possible  for all other approximations, too, to store the results in cache. Therefore, caching will never result in a situation where the Hepp or Martin approximations are genuinely faster than the other models when cached. Moreover, cache lookups require to transform graphs into their canonical isomorphic form, which already takes longer than computing the 5-cycle approximation. Also, the caching poses additional challenges for parallelization, while the non-cached models can be parallelized with essentially zero performance loss.

\begin{figure}[htb]
	
	\begin{subfigure}{ .49 \linewidth}
		\centering
		\includegraphics[width=\linewidth]{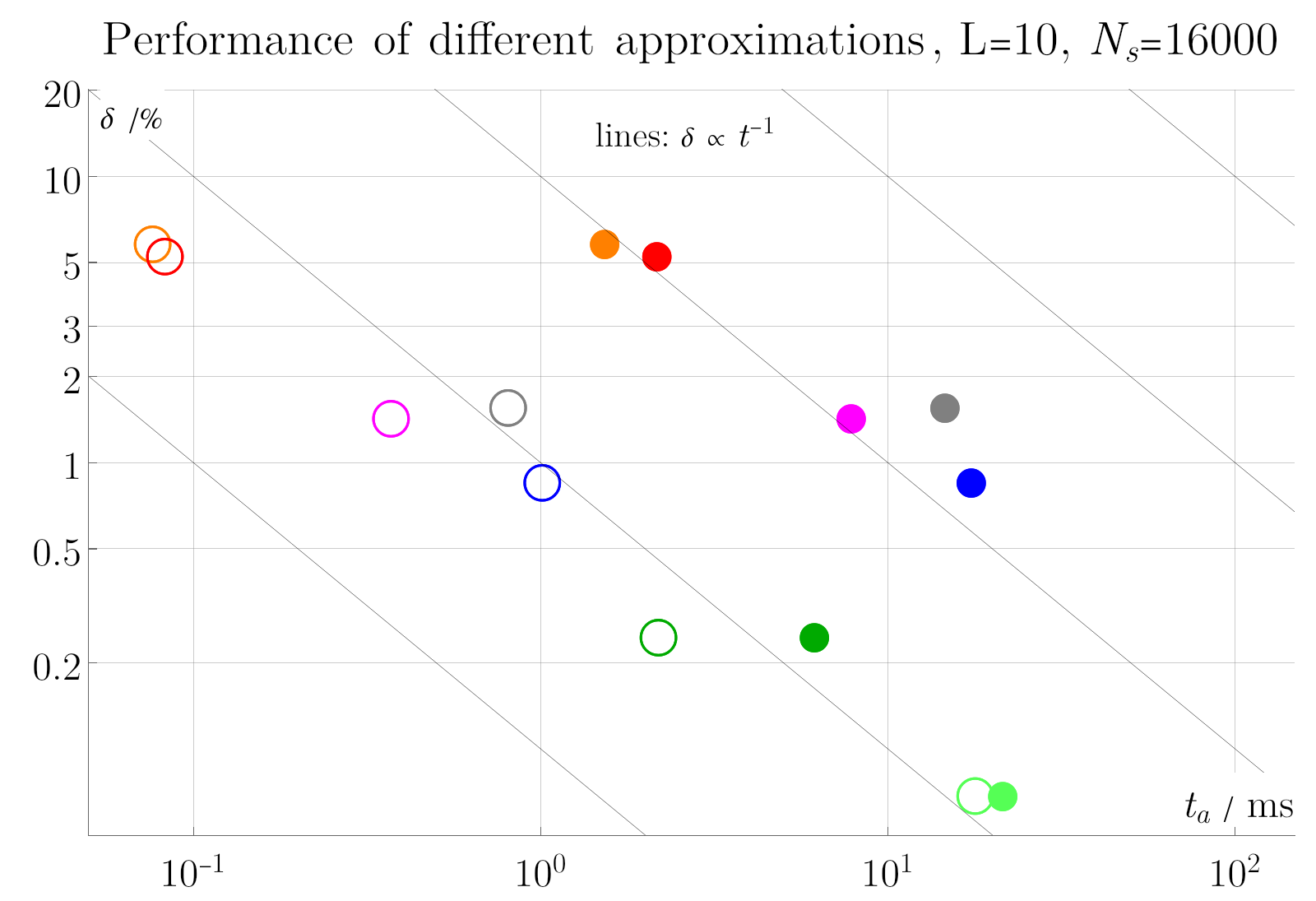}
		\subcaption{}
		\label{fig:performance_L10}
	\end{subfigure}
	\begin{subfigure}{ .49 \linewidth}
		\centering
		\includegraphics[width=\linewidth]{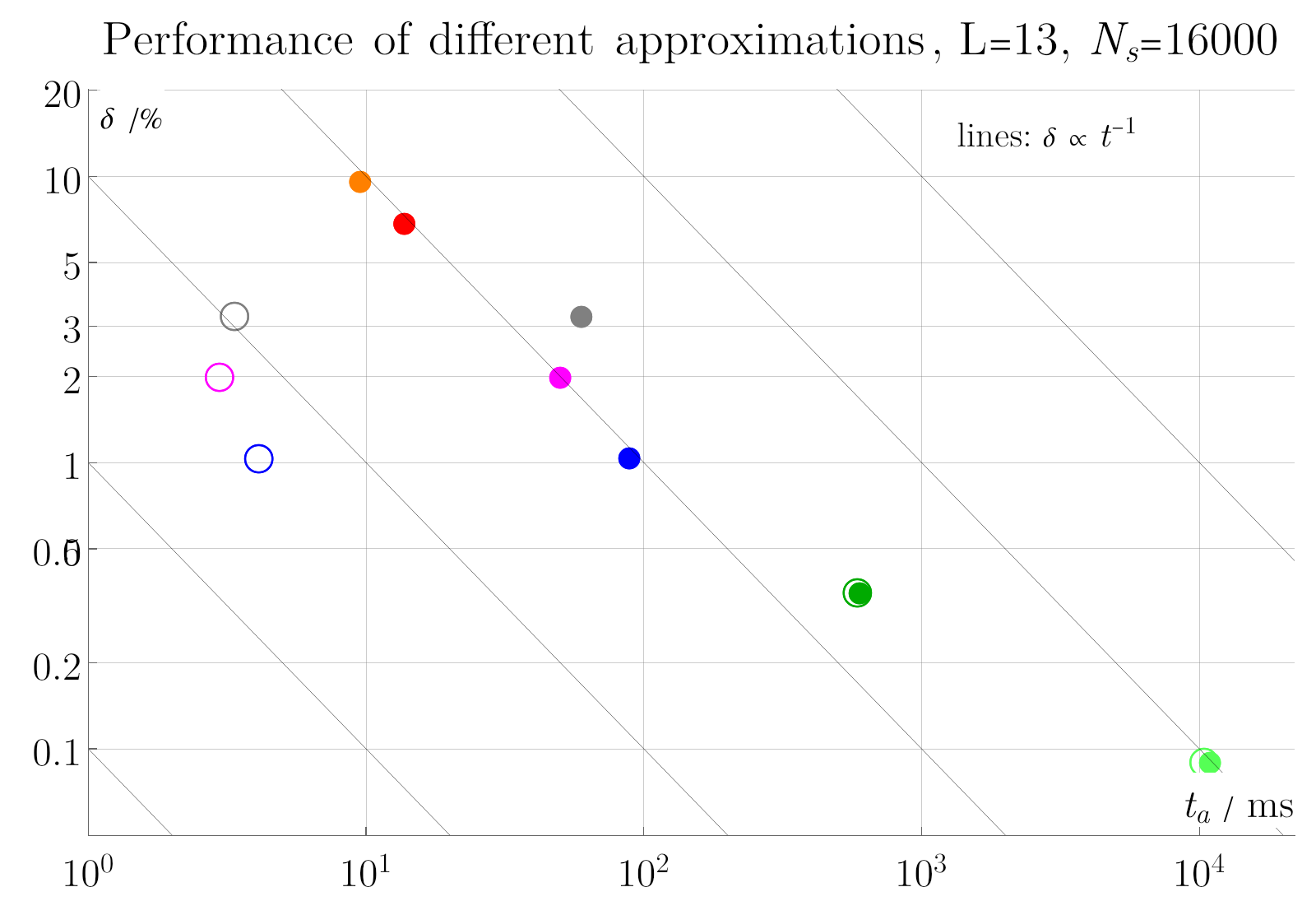}
		\subcaption{}
		\label{fig:performance_L13}
	\end{subfigure}
	\caption{ 
		\textbf{(a)} Comparison of $(\delta, t_a)$ for the different approximations at $L=10$ loops. Filled symbols: 1 thread, open symbols: 20 threads. For the Hepp (light green) and Martin (dark green) approximations, the performance gain by of multithreading is small. Note that there is an overall trend $\delta \propto t_a^{-1}$, indicated by lines. In view of \cref{relative_uncertainty_approximation}, this suggests that the predictions with smaller $\delta$ are overall superior.  
		\textbf{(b)} At $L=13$ loops, all approximations take longer than at $L=10$, but the penalty is disproportionally larger for the Hepp and Martin approximations (green), which thereby become impractical.  }
	\label{fig:performance}
\end{figure}

Following \cref{sec:accuracy_performance},   the quality of an approximation model is linear in $\delta$ and in $\sqrt{t_a}$, where smaller values are better. To visualize the performance of our approximation functions, we plot their parameters in a double-logarithmic coordinate system with axes $\delta$ and $ t_a$, see \cref{fig:performance}.
We find that the points $(\delta, t_a)$ of the various approximations (although these approximations rely on entirely different algorithms) roughly follow the trend $\delta \propto \frac{1}{t_a}$.  In view of \cref{relative_uncertainty_approximation}, this suggests that the higher time demand is more than offset by the higher accuracy, i.e. it is advisable to use the approximation with the smallest $\delta$. However, in \cref{fig:performance_L13} we see that already for $L=13$, the Hepp and Martin approximations deviate from the $\delta \propto \frac{1}{t_a}$ trend, and this effect becomes significantly worse for higher loop orders. We therefore conclude that from all predictions considered in \cref{sec:estimation}, the combined cut and cycle model is the most useful one for $L \geq 14$ loops because it delivers best accuracy among all models that do not suffer from the excessive slowdown of requiring large caches at high loop order.

\FloatBarrier

\subsection{Sampling algorithm and numerical results}\label{sec:implementation}

As a proof-of-concept, we computed the primitive contribution $\beta^{\text{prim}}$ to the beta function  for $L\in \left \lbrace 13,14,15,16 \right \rbrace $ loops. Details and background on the beta function can be found in \cite{kompaniets_minimally_2017,balduf_statistics_2023}.  
We implemented the cut-cycle approximation (\cref{sec:cuts}), using best fit coefficients (\cref{tab:cuts_cycles_parameters}) for each loop order, in a C++ program to draw weighted samples of graphs  for $L\in \left \lbrace 14,15,16,17 \right \rbrace $ loops.

The graphs were generated with the \texttt{nauty} \cite{mckay_practical_2014} tool \texttt{geng}, version 2.8080. These graphs are distributed proportional to their symmetry factor. The weighting $\left \langle~\right \rangle _{\Aut, \bar \period}$ for \cref{PAut_average_product} can be done if we draw a sample proportional to $\bar \period$ out of these graphs. A standard way to do this is the Metropolis-Hastings algorithm\cite{metropolis_equation_1953,hastings_monte_1970}: Starting from some graph $G_1$, generate a new candidate graph $G_2$ and draw $x\in [0,1]$ uniform at random. If $\frac{\bar \period(G_2)}{\bar \period(G_1)}>x$, take $G_2$ as the next graph, otherwise keep $G_1$ and generate a new candidate $G_3$, etc.. This algorithm produces a sequence of graphs with repetitions, where the expected number of occurrences of a graph is proportional to $\bar \period(G)$. 

The repetitions in the generating sequence are no problem in our application because only one in $F_s$ graphs will actually be numerically integrated, and following \cref{sec:accuracy_performance}, we require a sampling factor $F_s \gg 1$. On the other hand, the Metropolis Hastings algorithm has the immense advantage that we do not need to determine any normalization factors before the actual sampling can start.

\begin{figure}[htb]
	
	\begin{subfigure}{ .49 \linewidth}
		\centering
		\includegraphics[width=\linewidth]{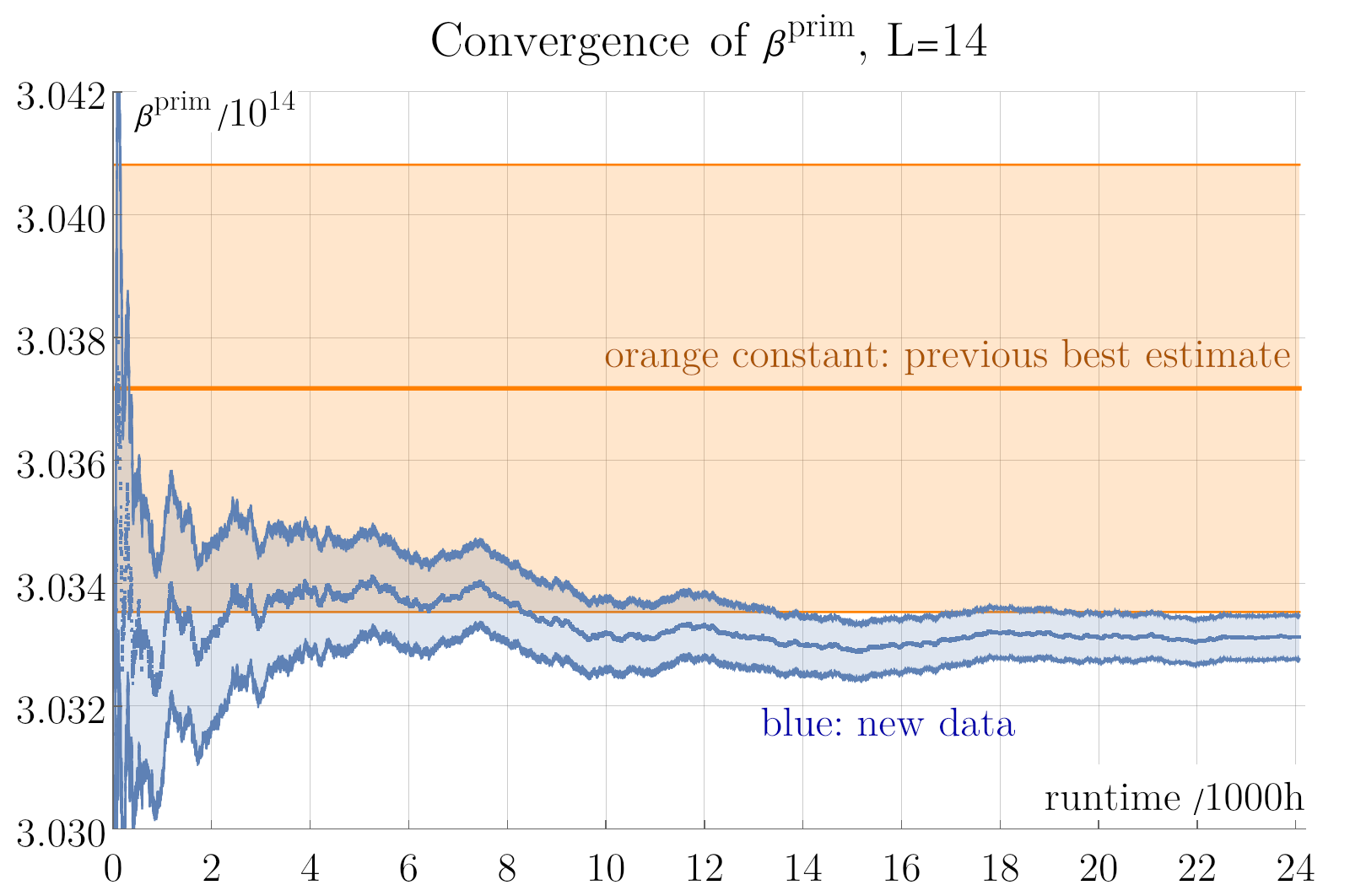}
		\subcaption{}
		\label{fig:beta14_convergence}
	\end{subfigure}
	\begin{subfigure}{ .49 \linewidth}
		\centering
		\includegraphics[width=\linewidth]{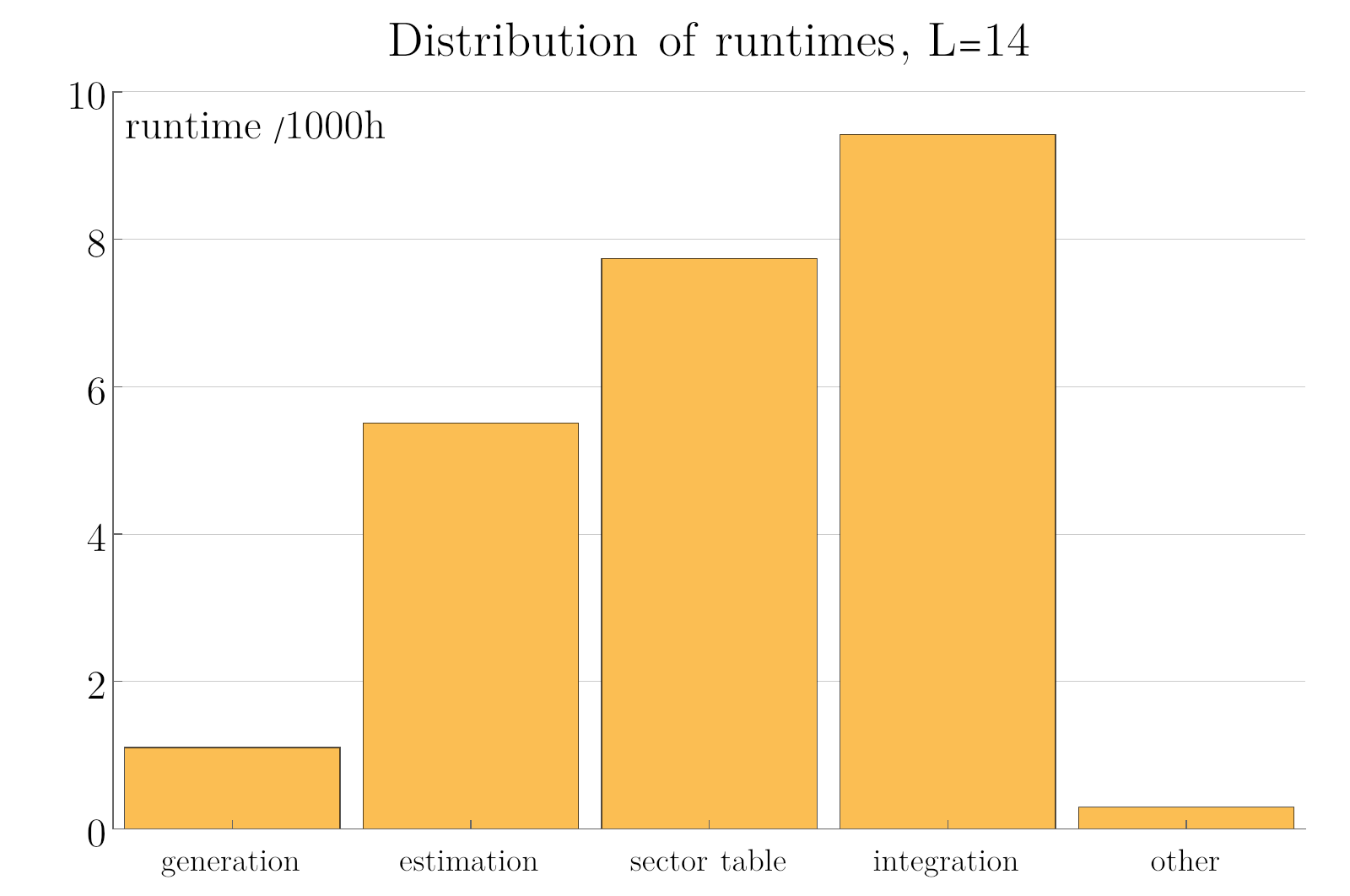}
		\subcaption{}
		\label{fig:beta14_runtimes}
	\end{subfigure}
	\caption{ 
		\textbf{(a)}  Convergence fo the beta primitive beta function (blue) over the course of the program run, for 14 loops. The orange area is the previous best estimate from \cite{balduf_statistics_2023}. Accuracies refer to 1 standard deviation.  The accuracy of the new algorithm is better than the previous one already after less than 1000 CPU core hours runtime. 
		\textbf{(b)} Times spent by the algorithm on different tasks, for $L=14$.  }
	
\end{figure}

For each loop order, we used up to 150 cores at 2.1 GHz for around one week (recall that 1 week at 150 cores are roughly 25,000 CPU core hours). While the program was running, we continuously monitored the current best estimate and uncertainty of the primitive beta function. We observe, as expected, that the fluctuations get smaller like $\frac{1}{\sqrt {N_s}}$ as the sample size $N_s$ increases.  \Cref{fig:beta14_convergence} shows the convergence for $L=14$ as a blue trajectory, where the previous best estimate and 1-sigma uncertainty (from \cite{balduf_statistics_2023}) is shown as an orange band. We see that our new algorithm reaches an accuracy comparable to the previous estimate already after a few hundred CPU hours, that is, within one afternoon running on a server. In the particular case shown in \cref{fig:beta14_convergence}, the true value of the beta function seems to be at the lower boundary of the 1-sigma confidence interval of the previous estimate, which is acceptable. 

\Cref{fig:beta14_runtimes} shows that at 14 loops, the algorithm spends most of its CPU-core time, roughly in similar proportion,    on estimation of periods, generating the subgraph table for numerical integration, and on the  numerical integration itself. At higher loop order, generating the subgraph table becomes increasingly dominant. In practice, this table is the limiting factor for large loop orders, both in terms of runtime and in terms of working memory, because the memory demand grows by a factor of 4 for each loop order.

\begin{table}[htbp]
	\centering
	\begin{tblr}{
		vlines,
		hline{1}={solid},
		hline{3,Z}={solid},
		rowsep=1pt,
		columns={halign=r},
		column{1}={halign=c,mode=math },
		cells={font=\fontsize{10pt}{11pt}\selectfont},
		row{1}={halign=c,rowsep=2pt, font=\fontsize{12pt}{14pt}\selectfont},
		row{2}={halign=c,rowsep=2pt},	
		cell{1}{2}={c=3}{c},
		cell{1}{5}={c=6}{c}, 
	}
	 & previous data \cite{balduf_statistics_2023} & & & new data & &  \\
		L & $\beta^{\text{prim}}_L$  & $\Delta \beta^{\text{prim}}_L$ & time  & $\beta^{\text{prim}}_L$ & $\Delta \beta^{\text{prim}}_L$ &  time & $\delta$ & int.   & app..  \\
		13 & $ 1.431996 \cdot 10^{13}$ & 0.25 &  228k & $1.43208 \cdot 10^{13}$ & 71 & 23k & 1.48\%  & 170k  & 382M \\
		14 & $ 3.0367  \cdot 10^{14}$ & 1063 &  400k &$3.03313 \cdot 10^{14}$ & 117 &  24k & 1.64\%  & 105k & 136M  \\
		15 & $ 6.636 \cdot 10^{15}$ &2090 & 488k & $6.6552\cdot 10^{15}$ & 146 & 43k & 1.78\% & 42k & 224M \\
		16 & $ 1.496  \cdot 10^{17}$ & 8999 &  24k & $1.5114  \cdot 10^{17}$ &366 &  26k & 1.95\% & 5814& 73M \\
	\end{tblr}
	\caption{Numerical estimates for the primitive beta function in $\phi^4$-theory. Given are best estimate $\beta^{\text{prim}}_L$,   uncertainty (standard deviation, measured in ppm) $\Delta \beta^{\text{prim}}_L$, and runtime (measured in CPU-core hours).  The first three columns show the combined results from \cite{balduf_statistics_2023}. The last six columns refer to the new data.  $\delta$ is the observed approximation accuracy (\cref{def:relative_standard_deviation}), \enquote{int.} and \enquote{est.} are the numbers of numerically integrated, and approximated, graphs. k=$10^3$, M=$10^6$. }
	\label{tab:beta_results}
\end{table}

The numerical outcomes for the primitive beta function for $L\in \left \lbrace 13,14,15,16 \right \rbrace $ loops are shown in \cref{tab:beta_results}, subject to the following comments:
\begin{itemize}
	\item The new numerical values of the beta function are compatible, within error margins (all errors are one standard deviation) to the previous data from \cite{balduf_statistics_2023}.
	\item For $L=13$ loops, the previous data from \cite{balduf_statistics_2023} is free of sampling uncertainty because \emph{all} periods had been computed. Consequently, the accuracy 0.25ppm is much better than in our new data.
	\item The various computations ran on similar, but not quite identical  servers. The \enquote{new} runtimes include generation of the graphs and we did not optimize the program to always use all 150 cores (e.g. while generating subgraph tables), whereas the \enquote{previous} runtimes  refer only to generating the subgraph table and doing the numerical integration itself. Hence, all runtimes are rough estimates.
	\item The obtained prediction accuracy $\delta$ is slightly worse than, but comparable to, the values from \cref{fig:cut_cycle_relative_standard_deviation_reducible}. This shows that the cut+cycle approximation works reliably even for new graphs that had not been in the data set used to determine the fit parameters.
	\item Since we sample with repetition, the number of approximated periods often exceeds the total number of primitive graphs.   Conversely, the integrated periods are not necessarily distinct graphs. A more refined version of the new algorithm could make use of this fact to obtain better sampling accuracy  as soon as a large proportion of all graphs have been integrated.
	\item In the new algorithm, the accuracy of the numerical integration of individual graphs is chosen dynamically and sometimes lower than in \cite{balduf_statistics_2023}. This   is not the decisive factor to explain the greater speed of the new algorithm. For example, for $L=16$ the old data set had 10,196 integrated graphs, i.e. almost twice as many as the new one, but the same runtime. 
\end{itemize}

The accuracy of a Monte Carlo integration is expected to scale like $\Delta \beta \sim \frac{1}{\sqrt t}$ (\cref{sampling_uncertainty}), where $t$ is the runtime or number of periods sampled. Consequently, $q:=\Delta \beta  \cdot \sqrt t$ is a measure for the quality of the algorithm,   independent of the runtime. Measuring $\Delta \period$ in parts per million, and runtime in 1000 core hours like in \cref{tab:beta_results}, the quality of the \enquote{previous data} naive sampling for $L\in \left \lbrace 13,14,15,16 \right \rbrace $ was $q=\left \lbrace 21270,46140,43700  \right \rbrace $. The new algorithm achieves $q=\left \lbrace 561,957,1866 \right \rbrace $, which is an improvement by  factors of $\left \lbrace 37,48, 23 \right \rbrace    $. That means that in the same runtime, the new weighted sampling algorithm produces a result whose uncertainty is smaller by a factor of $\sim 35$, or equivalently, the new algorithm reaches the same  accuracy in a runtime that is smaller by a factor $\sim 35^2\approx 1000$. This is visible from  \cref{fig:beta14_convergence}: We indeed reach an accuracy comparable to the orange band already after roughly $ \frac{400 \text k}{1000}=400$ CPU core hours.

\newpage 
\section{Period prediction by machine learning}\label{sec:machine_learning}

\subsection{Machine learning for graph-valued input data} \label{sec:ML_introduction}

All predictions discussed in \cref{sec:estimation} were based on manual fits of peculiar graph-theoretic quantities or invariants of the period. In the present section, we use more broad machine learning approaches to construct such predictions. The goal is to understand to which extent the results of \cref{sec:estimation} can be replicated without putting in physical background knowledge and tedious manual selection of fit functions.

Although there is an enormous body of literature on algorithms and applications for machine learning on graphs, our concrete setting is different from most problems considered in the literature. For our present project, we aim to use a graph as an input to a model, and obtain a real number -- the period (\cref{def:period}) -- as an output. Conversely, problems discussed in the literature include, for example, \emph{classification} of graphs, or using data \emph{on} a graph (such as values assigned to vertices) as an input, or using a graph to describe the \emph{structure} of a neural network itself, or using models that have graphs as \emph{output}. The algorithms developed for these problems are not directly applicable to our case. 

It is non-trivial to use an unlabeled graph as an input to a machine learning model. For example, one might want to use the graph matrices such as the vertex-vertex adjacency matrix, but they depend on the labeling of vertices. Hence, many different matrices correspond to the same unlabeled graph, and a model has no way of knowing that these should result in the same output. In our case,  a single $L=13$ loop graph has up to $15!>10^{12}$ different orderings of vertices, which makes it practically impossible to use all of them for training the model. In all our algorithms, we use  \texttt{nauty} \cite{mckay_practical_2014} to fix a canonical labeling of vertices. This solves the problem of arbitrary relabelings, but it imposes an artificial structure on the graph, which is actually unrelated to the desired prediction of the period. 

In the machine learning literature, the term \emph{embedding} is used for a mapping of a graph to some vector space of features, a survey of common embedding techniques can be found in \cite{yan_graph_2007,goyal_graph_2018,khosla_comparative_2020,cai_comprehensive_2018}. Typically, one demands that the vectors produced by an embedding function for two distinct input graphs are close if the graphs are \enquote{similar}, in a sense that depends on the context. 
Another approach to represent a graph by a vector, or by a sequence of numbers, goes by the name of \emph{graph signature}, see e.g. \cite{toth_capturing_2022,caudillo_signatures_2023}. Here, the focus is usually on algebraic properties of the output, rather than on preserving \enquote{similarity}. For our own models, the input features are described in  \cref{sec:features}.  Although several of them have properties usually associated with embeddings or signatures, we will not use these terms to avoid misconceptions.

A second approach to doing machine learning on graph-valued data is to use the input graph itself as a structural part of the algorithm, see e.g.  \cite{wu_comprehensive_2021,ying_gnnexplainer_2019,bevilacqua_equivariant_2022}. Among others, a typical architecture of this type is a  \emph{graph convolutional network} (GCN). 
GCNs are used in many fields, e.g.  in medicine \cite{ahmedt-aristizabal_survey_2022} or chemistry \cite{kroll_general_2023}, however, often the application is to make a binary classification, and not to predict a numerical quantity with high accuracy. We briefly consider GCNs and other architectures in \cref{sec:graph_structure}.

\subsection{Input features and data sets}\label{sec:features}
For the various machine learning models, we used all the features that have been found helpful in \cref{sec:estimation}, and a few additional ones. The following is an overview of all 194 features, where references indicate where this feature has been used earlier to predict periods.
\begin{enumerate}
	\item Loop number.
    \item Hepp bound (\cref{sec:hepp}, \cite[Sec. 6.5]{balduf_statistics_2023}, \cite{panzer_hepp_2022,kompaniets_minimally_2017}).
    \item Whether the graph is planar.
    \item Size of the automorphism group (\cite[Sec. 6.1]{balduf_statistics_2023}).
    \item Number of non-isomorphic decompletions (\cite[Sec. 6.1]{balduf_statistics_2023}).
    \item Number of planar decompletions (\cite[Sec. 6.1] {balduf_statistics_2023}).
    \item Symmetry factor of only the planar decompletions. 
    \item Number of $n$-vertex cuts for $n = 3, 4, 5$.
    \item Number of $n$-edge cuts for $n = 6, 8, 10, 12, 14$ (\cref{sec:cuts}, \cite[Sec. 6.4]{balduf_statistics_2023}).
    \item Number of connected $n$-edge cuts for $n = 6, 8, 10, 12, 14$ (\cref{sec:cuts}).
    \item Radius and diameter of the graph (i.e. minimum and maximum of the distance between vertices, \cite[Sec. 6.2]{balduf_statistics_2023}).
    \item Mean, and first moments, of the distance between vertices (\cite[Sec. 6.2]{balduf_statistics_2023}).
    \item  Eigenvalues of the distance matrix.
    \item 22 binary parameters corresponding to whether or not an matrix eigenvalue is present for the loop order under consideration (see below).
    \item Eigenvalues of the Laplacian matrix.
    \item Scaled traces $ \frac{\operatorname{Tr}(A^k)}{(L+2) 2^k} $, for $k\leq 20$, where $A$ is the adjacency matrix.
    \item Number of $n$-cycles for $n = 3, \dots, 10$ (\cref{sec:cycles}).
    \item Mean electrical resistance (=Kirchhoff index \cref{def:kirchhoff_index}). 
    \item Minimum, maximum, and standard deviation, of electrical resistances between pairs of vertices.  Absolute mean, minimum, and maximum, of the first 4 Ursell functions. Mean, and mean of absolute value, of the entries of the transfer current matrix (\cref{sec:resistance}).
    \item First 9 entries of the autocorrelation function of random walks of length $10^4$, averaged over 5000 random relabelings of the vertices.  
    \item Scaled logarithms of the Martin invariants $\frac{\ln (M^{[k]})}{2^{k+1}}$   (\cref{sec:martin}, \cite[Sec. 6.6]{balduf_statistics_2023}, \cite{panzer_feynman_2023}), with upper limit $k \leq 5$ depending on loop order.  
    \item Coefficients of the Martin polynomial (\cref{sec:martin}, \cite{panzer_feynman_2023}). 
\end{enumerate}

The size of the matrices, and therefore the number of eigenvalues, depends on the loop order of the graph. In order to have a consistent number of 22 features for each matrix, across all loop orders, we have padded the missing entries with zeros. Even if the loop order already implies how many of the features are non-trivial, we supplied the models with an explicit vector of ones and zeros to indicate whether the corresponding entries contain information or are merely padded zeros (feature 14 in the list).

In the present \cref{sec:machine_learning}, our focus is to evaluate the performance of prediction algorithms for periods that have not been used for training. Consequently,  we split the data so that only 80\% of the data was used for training, and the reported performance of the models refers to testing the model on the remaining 20\% (\enquote{validation}). 

The number of Feynman graphs grows factorially with loop order \cite{cvitanovic_number_1978,borinsky_renormalized_2017}, compare \cref{sec:data_set} and the counts given in \cite{balduf_statistics_2023}.   The number of available graphs in our data set varies strongly with the loop order, as illustrated by the red dots in \cref{fig:number_of_graphs}. The neural networks considered in \cref{sec:NN} are highly non-linear and therefore prone to overfitting. 
If we train them on \emph{all} available data, these models are effectively mostly being trained on $L\in \left \lbrace 12, 13,14,15 \right \rbrace $ loops, since most of the data has this loop order. They then show very poor performance for smaller or larger loop orders. To remedy this imbalance, we only used a random subset of the graphs for training. Concretely, at 12 loops 50\% of the graphs were used for training, at 13 loops 33\%, and at 14 and 15 loops 10\%, and the remaining graphs were used for verification. We found that this restriction has very little influence on the performance at the central loop orders, but it improves accuracy for the loop orders where only few graphs are available.  Conversely, the regression models   (\cref{sec:standard_machine_learning}) show relatively consistent performance across all loop orders, hence we used all available graphs for them.

\subsection{Standard Regression Techniques}\label{sec:standard_machine_learning}

For the standard machine learning techniques in this paper, we used the python library  \texttt{scikit-learn}  \cite{pedregosa_scikitlearn_2012} and its auxiliary functions such as \texttt{PolynomialFeatures} and \texttt{MinMaxScaler}.  

\subsubsection{Linear Regression}\label{sec:linear_regression}
We observed in \cref{sec:estimation} that in various cases, a simple linear fit of the features already gives a good prediction of the period. In the present section, we construct a  linear regression model (see e.g. \cite{maulud_review_2020}) that takes \emph{all} the features described in \cref{sec:features}, not just manually selected subsets, as an input.
Linear regression aims to approximate a quantity $y$  by a multi-linear function $\bar y$, given by
\begin{align}\label{def:linear_regression}
y  \approx \bar y  &:= \beta_0 + \vec \beta_1 \vec x, 
\end{align}
where   $\vec x$ is a vector of input features, and $\left \lbrace \beta_0, \vec \beta_1 \right \rbrace $ are parameters to be fitted .

We constructed three distinct linear regression models. For all three of them, we  normalized all features $x \in \vec x$ (\cref{sec:features}) to lie within the interval $(0, 1)$. The first model \textbf{LR1} aims to predict the period $\period$ directly. The second model \textbf{LR2} instead scales the period and Hepp bound to their leading asymptotic growth with loop order according to \cref{period_scaling} such that the Hepp bounds and periods across all loop orders are of similar magnitude. Finally, the third model \textbf{LR3} aims to predict the natural logarithm $y= \ln   \period $ of the period, and uses the logarithm of the Hepp bound as an input, both scaled to their asymptotic growth (\cref{period_scaling}).

The coefficient of determination for a $N$-element data set is defined as
\begin{align}\label{def:R}
	R^2 := 1 - \frac{\sum_{i = 1}^N (y_i - \bar y(\vec x_i))^2}{\sum_{i = 1}^N (y_i - \left \langle  y \right \rangle  )^2},
\end{align}
where $\left \langle y \right \rangle $ denotes arithmetic mean. $R^2$ is
a measure for how much better the multi-linear function \cref{def:linear_regression} describes the data, compared to a function that just returns the average $\left \langle y \right \rangle $ as an approximation.

\begin{table}[htbp]
	\centering 
	\begin{tblr}{
			vlines,
			hline{1}={solid},
			hline{2,Z}={solid},
			rowsep=1pt,
			columns={halign=r},
			column{1}={halign=c },
			column{2, 3, 4} = {halign=c},
		}
		\text{Model} & $R^2$ Training & $R^2$ Testing &  $\Delta$ (\%) \\ 
		LR1 &  0.9981745 &	0.9982034 & 3.37  \\
		LR2 &   0.9996490 & 0.9996573 & 0.04  \\ 
		LR3 &  0.9999994 &	0.9999994 & 0.04 \\ 
\end{tblr}
\caption{Coefficient of determination (\cref{def:R}) for the three models using linear regression; average relative difference $\Delta$  (\cref{def:relative_difference}). The models LR2 and LR3, which include scaling to the asymptotics (\cref{period_scaling}), are by far superior to the unscaled model LR1.}
\label{tab:linear_regression_results}
\end{table}

The values shown in \cref{tab:linear_regression_results} suggest that scaling to the leading asymptotics by \cref{period_scaling} is very helpful in order to construct a model that works across all loop orders. 
Consequently, used this scaling and normalization for all remaining machine learning models in the present article.

\begin{figure}[htb]
	
	\begin{subfigure}{ .49 \linewidth}
		\centering
		\includegraphics[width=\linewidth]{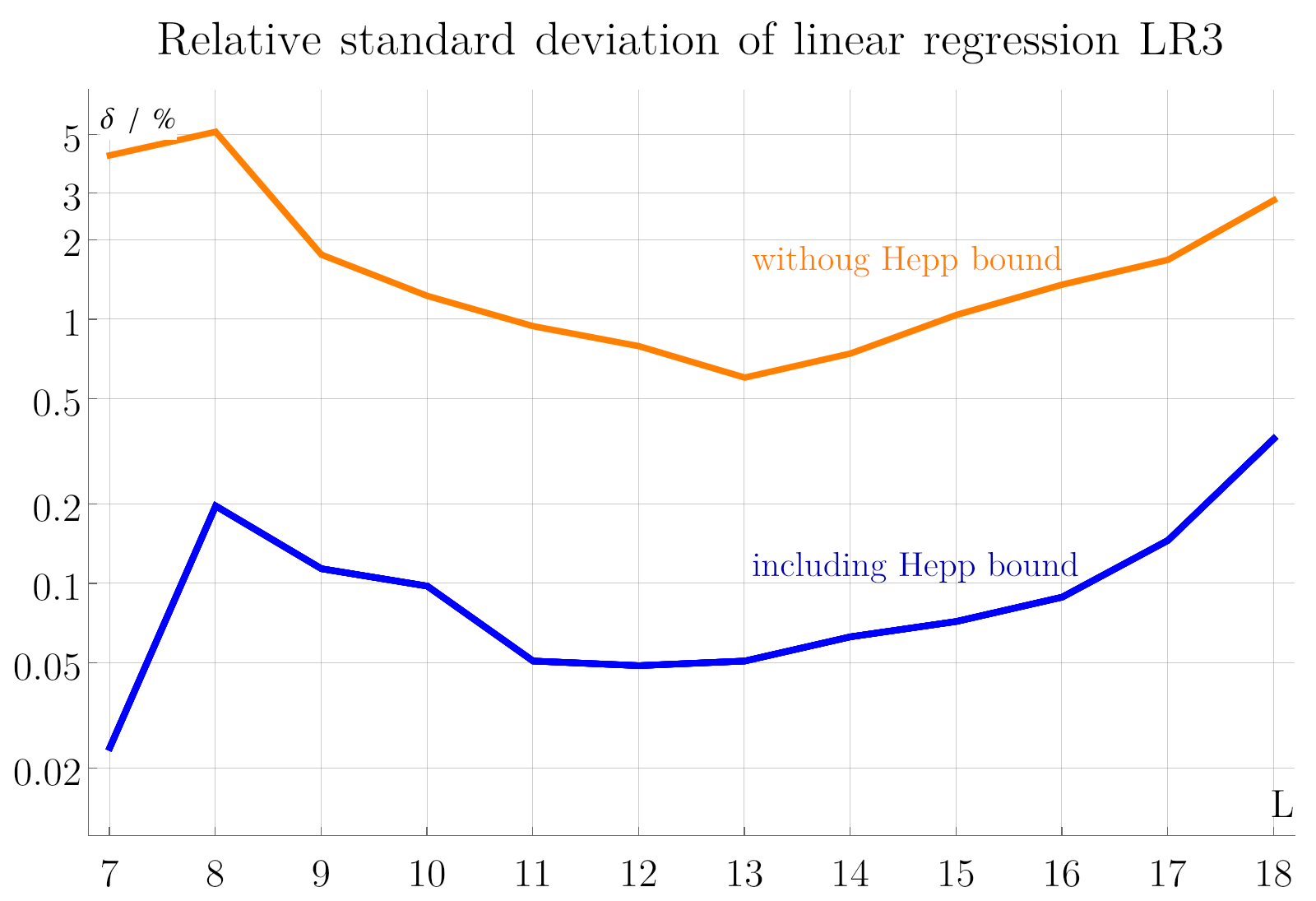}
		\subcaption{}
		\label{fig:LR3_delta}
	\end{subfigure}
	\begin{subfigure}{ .49 \linewidth}
		\centering
		\includegraphics[width=\linewidth]{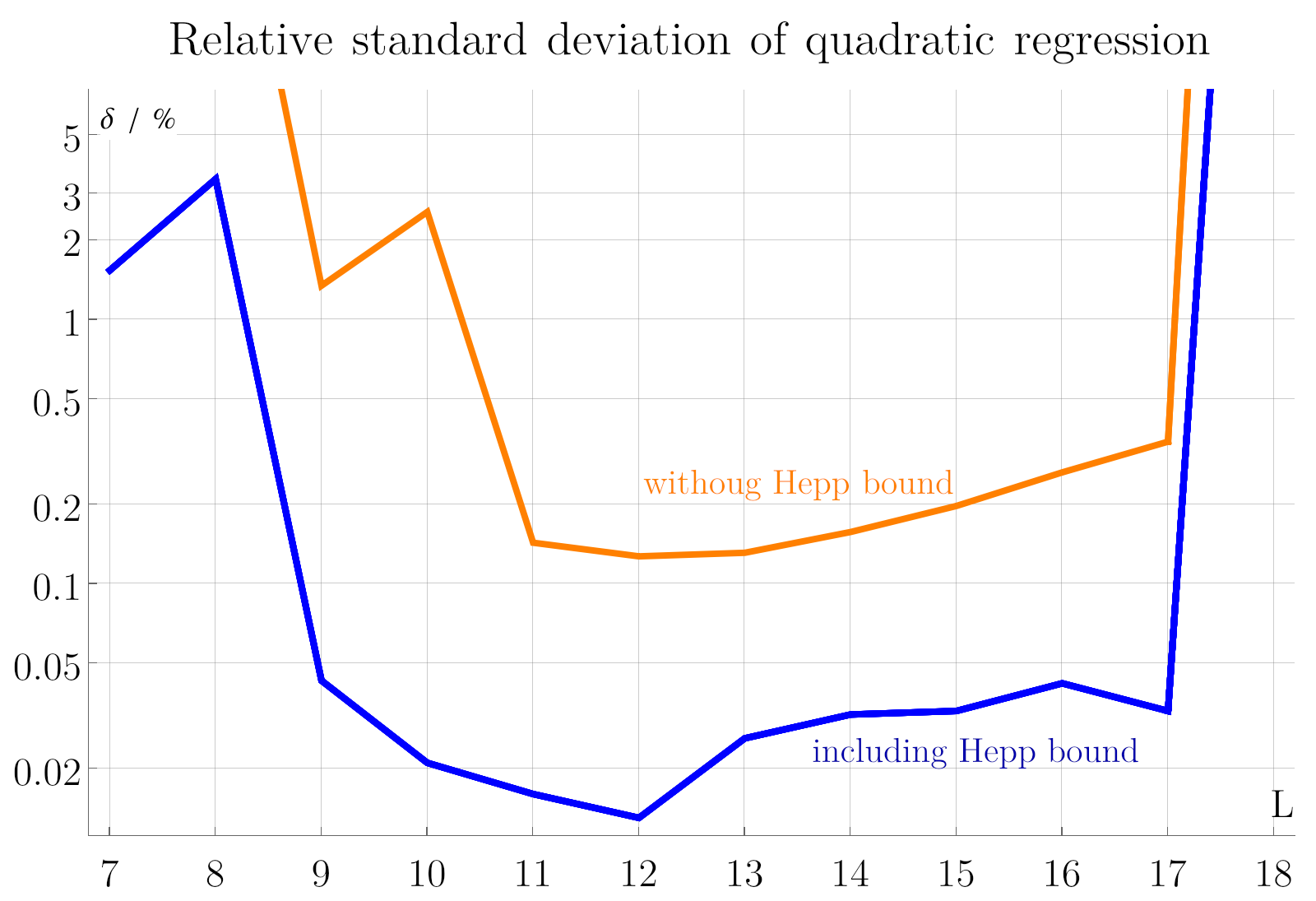}
		\subcaption{}
		\label{fig:QR_delta}
	\end{subfigure}
	\caption{ 
		\textbf{(a)}  Relative standard deviation $\delta$ (\cref{def:relative_standard_deviation}) for the linear regression model LR3. Including the Hepp bound improves accuracy by roughly one order of magnitude. Overall, $\delta$ is lower for larger the data sets than for smaller ones. 
		\textbf{(b)} Relative standard deviation $\delta$ for quadratic regression. If the data set is large enough, then the accuracy is   better than the linear model (a), but the quadratic model suffers more strongly from having only small, or non-complete, data sets. }
	\label{fig:regression}
\end{figure}

The plot in \cref{fig:LR3_delta} shows how the linear regression model LR3 performs when it is trained with data of only one particular loop order. Performance tends to be overall better for the large data sets around $L=13$ loops than for the smaller ones, but the differences are moderate. The Hepp bound  as an input feature has a large influence, improving the accuracy by roughly a factor of 10. This is expected from the findings of \cref{sec:hepp}, that the Hepp bound alone is sufficient for a relatively accurate prediction. Nevertheless, even without the Hepp bound, the linear regression model reaches an accuracy of   $\delta\leq 1\%$ if the data set is large enough.

\subsubsection{Quadratic Regression}\label{sec:quadratic_regression}
We extended standard regression techniques, to train models using quadratic regression  \cite{maulud_review_2020}. Instead of \cref{def:linear_regression}, this model fits the function
\begin{align}\label{def:quadratic_regression}
\ln \period  \approx \ln \bar \period:= \beta_0 + \sum_i \beta_{1,i} x_i + \sum_{i,j} \beta_{2,i,j}x_i x_j.
\end{align}
For $N$ input parameters, this model involves not just $N$ quadratic terms of the individual parameters, but $\frac{N(N-1)}{2}$  potentially mixed quadratic terms and $N$ linear ones, resulting in $\frac{N^2+N+2}{2}$ fit parameters. If we were to use all 194 input features of \cref{sec:features}, then \cref{def:quadratic_regression} would have 18914 free fit parameters, making the model numerically unstable and prone to overfitting. We excluded list items 13, 14, 15, 16, 21 and 22 of \cref{sec:features}. The resulting model uses   only 83 input features and has 3487 fit parameters, it obtains  $R^2$ values (\cref{def:R}) of better than $0.999$. Including the data from all loop orders, quadratic regression reached $\Delta = 0.021\%$, and $\delta = 0.079 \%$. For models trained on one particular loop order, the performance is shown in \cref{fig:QR_delta}. Similar to the linear regression model (\cref{sec:linear_regression}), we observe that the accuracy of the quadratic model decreases by approximately one order of magnitude if the Hepp bound is not used as an input feature.

\subsection{Neural Networks} \label{sec:NN}

In the present section, we discuss several neural network models  to predict the  period. Some of them are based on the features listed in \cref{sec:features}, while others additionally incorporate  the structure of the Feynman graphs themselves, as explained below. 

All neural networks  were implemented using Python 3.10.4, and \texttt{PyTorch} \cite{paszke_pytorch_2019} version 2.0.0, along with mean squared error (MSE) loss function (which essentially is $\Delta$ from \cref{def:relative_difference}). We used the \texttt{RobustScaler} from \texttt{scikit learn}\cite{pedregosa_scikitlearn_2012}, which scales the data using the quantile range. 

\subsubsection{Single-layer Perceptron}

A  perceptron \cite{rosenblatt_perceptron_1957} is the basic historical type of  neural network. It only consist of an input and an output layer, without any hidden layers in between.  This type of neural network algebraically represents a multi-linear function, equivalent to linear regression (\cref{def:linear_regression}). The difference lies in the training algorithm: While linear regression algebraically finds the exact least-squares solution, the perceptron is trained iteratively. We trained a single layer perceptron and confirmed that the resulting standard deviation $\delta$  that is similar to \cref{fig:LR3_delta} for large data sets, but inferior for small data sets. Presumably, the difference is due to the iterative training being more susceptible to  outliers in small data sets.

\subsubsection{Basic Model}\label{sec:basicmodel}

\begin{figure}[htb]
	\begin{center} 
		
		\begin{tikzpicture}[scale=1]
		 
		 	\coordinate (xstep) at (3 ,0);
		 	\coordinate (ystep) at (0,-1 );
		 	\def\drawarrow(#1,#2){
		 		\draw[white,-{Latex[length=4mm, width=3mm]},line width=.5mm] (#1) -- (#2);
		 		\draw[black,-{Latex[length=3mm, width=2mm]},line width=.2mm] (#1) -- (#2);
		 	}
			
			\node[anchor=north,align=left,color=inputlayercolor!80!black, font={ }] at (.1,3.2) {input\\[-0.2em]features};
			
			\node[input feature] (in1) at ($(-.0,0) -1.5*(ystep)$) {\#1};
			\node[input feature] (in2) at ($(-.0,0) -.5*(ystep)$)  {\#2};
			\node[input feature] (in3) at ($ (-.0,0)+.5*(ystep)$) {\#3};
			\node[input feature] (in4) at ($ (-.0,0)+1.5*(ystep)$) {\#4};
			\node (inX) at ($(-.0,0)+2.5*(ystep)$) {$\vdots$};
 			\node[input feature] (inN) at  ($(-.0,0)+3.5*(ystep)$) { \#{194}};
			
			\node[anchor=north,align=center,color=inputlayercolor!80!black, font={ }] at ($ 1*(xstep)+(0,3)$) {input layer};
			
			\node[anchor=north,align=center,color=hiddenlayercolor!60!black,font={ }] at ($ 2.5*(xstep)+(0,3)$) {2 hidden layers};
			
			\node[node in] (h11) at ($(xstep) -1*(ystep)$){1};
			\node[node in] (h12) at ($(xstep)  $){2};
			\node[node in] (h13) at ($(xstep)+1*(ystep)$){3};
			\node (h1X)  at ($(xstep) +2*(ystep)$){$\vdots$};
			\node[node in] (h1N) at ($(xstep) +3*(ystep)$){128};
			
			\node[node hidden] (h21) at ($2*(xstep) -.5*(ystep)$){1};
			\node[node hidden] (h22) at ($2*(xstep) +.5*(ystep) $){2};
			\node (h2X)  at ($2*(xstep) +1.5*(ystep)$){$\vdots$};
			\node[node hidden] (h2N) at ($2*(xstep) +2.5*(ystep)$){64};
			
			\node[node hidden] (h31) at ($3*(xstep) $){1};
			\node (h3X)  at ($3*(xstep) +1*(ystep)$) {$\vdots$};
			\node[node hidden] (h3N) at ($3*(xstep) +2*(ystep)$){32};
			
			\node[anchor=north,align=right,color=outputlayercolor!80!black,font={ }] at ($ 3.8*(xstep)+(0,3)$) {period\\[-.2em]approximation\\[-.2em]output};
			\node[node out] (out) at ($4*(xstep) +1*(ystep)$) {$\ln \bar \period$};

			\drawarrow(in1.east,h11);
			\drawarrow(in1.east,h12);
			\drawarrow(in1.east,h13);
			\drawarrow(in1.east,h1N);
			\drawarrow(in2.east,h11);
			\drawarrow(in2.east,h12);
			\drawarrow(in2.east,h13);
			\drawarrow(in2.east,h1N);
			\drawarrow(in3.east,h11);
			\drawarrow(in3.east,h12);
			\drawarrow(in3.east,h13);
			\drawarrow(in3.east,h1N);
			\drawarrow(in4.east,h11);
			\drawarrow(in4.east,h12);
			\drawarrow(in4.east,h13);
			\drawarrow(in4.east,h1N);
			\drawarrow(inX.east,h11);
			\drawarrow(inX.east,h12);
			\drawarrow(inX.east,h13);
			\drawarrow(inX.east,h1N);
			\drawarrow(inN.east,h11);
			\drawarrow(inN.east,h12);
			\drawarrow(inN.east,h13);
			\drawarrow(inN.east,h1N);
			
			\drawarrow(h11,h21);
			\drawarrow(h11,h22);
			\drawarrow(h11,h2N);
			\drawarrow(h12,h21);
			\drawarrow(h12,h22);
			\drawarrow(h12,h2N);
			\drawarrow(h13,h21);
			\drawarrow(h13,h22);
			\drawarrow(h13,h2N);
			\drawarrow(h1X,h21);
			\drawarrow(h1X,h22);
			\drawarrow(h1X,h2N);
			\drawarrow(h1N,h21);
			\drawarrow(h1N,h22);
			\drawarrow(h1N,h2N);
			
			\drawarrow(h21,h31);
			\drawarrow(h21,h3N);
			\drawarrow(h22,h31);
			\drawarrow(h22,h3N);
			\drawarrow(h2X,h31);
			\drawarrow(h2X,h3N);
			\drawarrow(h2N,h31);
			\drawarrow(h2N,h3N);
			
			\drawarrow(h31,out);
			\drawarrow(h3X,out);
			\drawarrow(h3N,out);
			
			\node[draw,thick,minimum height=.2cm,minimum width=6.66cm,scale=.9,fill=gray!10,rotate=90,font={\small}] at ($(xstep)+(.8,0)+1*(ystep)$){dropout 0.3};
			
			\node[draw,thick,minimum height=.2cm,minimum width=6cm,fill=yellow!10,rotate=90,font={\small},rounded corners=0cm ] at ($(xstep)+(1.8,0)+1*(ystep)$){normalization};
		
			\node[draw,thick,minimum height=.2cm,minimum width=5.55cm,scale=.9,fill=gray!10,rotate=90,font={\small}] at ($2*(xstep)+(.8,0)+1*(ystep)$){dropout 0.2};
			
			\node[draw,thick,minimum height=.2cm,minimum width=5cm,fill=yellow!10,rotate=90,font={\small},rounded corners=0cm ] at ($2*(xstep)+(1.8,0)+1*(ystep)$){normalization};
				
			\node[draw,thick,minimum height=.2cm,minimum width=4cm,fill=yellow!10,rotate=90,font={\small},rounded corners=0cm ] at ($3*(xstep)+(1.5,0)+1*(ystep)$){normalization};
			
		\end{tikzpicture}
	\end{center}
	\caption{Sketch of the architecture of the basic model used in \cref{sec:basicmodel}.  }
	\label{fig:basicmodel_architecture}
\end{figure}
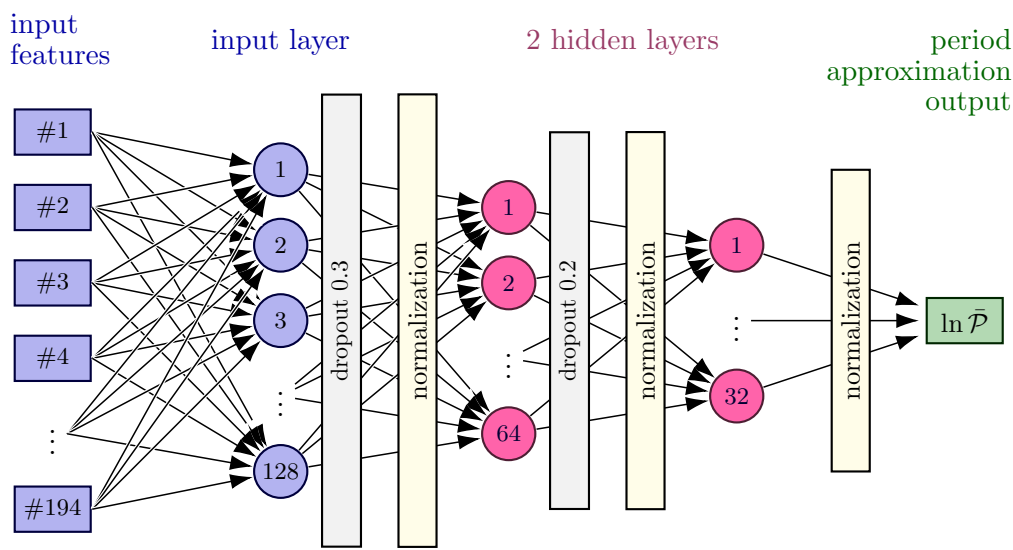

Under the name \enquote{basic model}, we use a standard architecture for a neural network, shown in \cref{fig:basicmodel_architecture}. The basic model is a feed-forward neural network with four fully connected linear layers, using \emph{Leaky Rectified Linear Unit (ReLU)} activation functions and batch normalization layers between the neuron layers. Two dropout layers were included in order to reduce susceptability to overfitting \cite{goodfellow_deep_2016}. We use the \emph{Adam} \cite{kingma_adam_2014} optimizer provided with \texttt{PyTorch}, with a mean squared error loss function. The input of the basic model are  all 194 features listed in Section~\ref{sec:features}.

The performance of the basic model is shown in \cref{fig:basic_stack_delta}. We reach accuracies of $\delta \approx 10\%$, which is similar to the resistance approximations in \cref{sec:resistance}. 
 Unlike with the regression models in \cref{sec:standard_machine_learning}, the performance of the basic model is only reduced marginally if the Hepp bound is not included. 
We observed that on the one hand, the regression models of \cref{sec:standard_machine_learning} are by far more accurate than the basic model, and on the other hand, they are much less susceptible to outliers and random fluctuations. Different runs of the neural network models (on different randomly selected subsets of graphs used for training) resulted in considerably different performance, including occasional a breakdowns in accuracy as for $L=10$ in the basic model in \cref{fig:basic_stack_delta}.

\subsubsection{Stack Model}\label{sec:stackmodel}
The \enquote{stack model} has the same architecture as the basic model (\cref{sec:basicmodel}), shown in \cref{fig:basicmodel_architecture}, but additionally to the 194 features of \cref{sec:features}, it also  uses  the adjacency  matrix of the graph  as input feature. To that end, the adjacency matrix is padded with zeros to a $20\times 20$ matrix,  flattened into a 1-dimensional vector \cite{goyal_graph_2018}, and concatenated with the initial 194 graph features to produce a vector of 594 input features per graph. The model was trained using the same optimizer, loss function and activation function as the basic model (\cref{sec:basicmodel}).

\begin{figure}[htb]
	
	\begin{subfigure}{ .49 \linewidth}
		\centering
		\includegraphics[width=\linewidth]{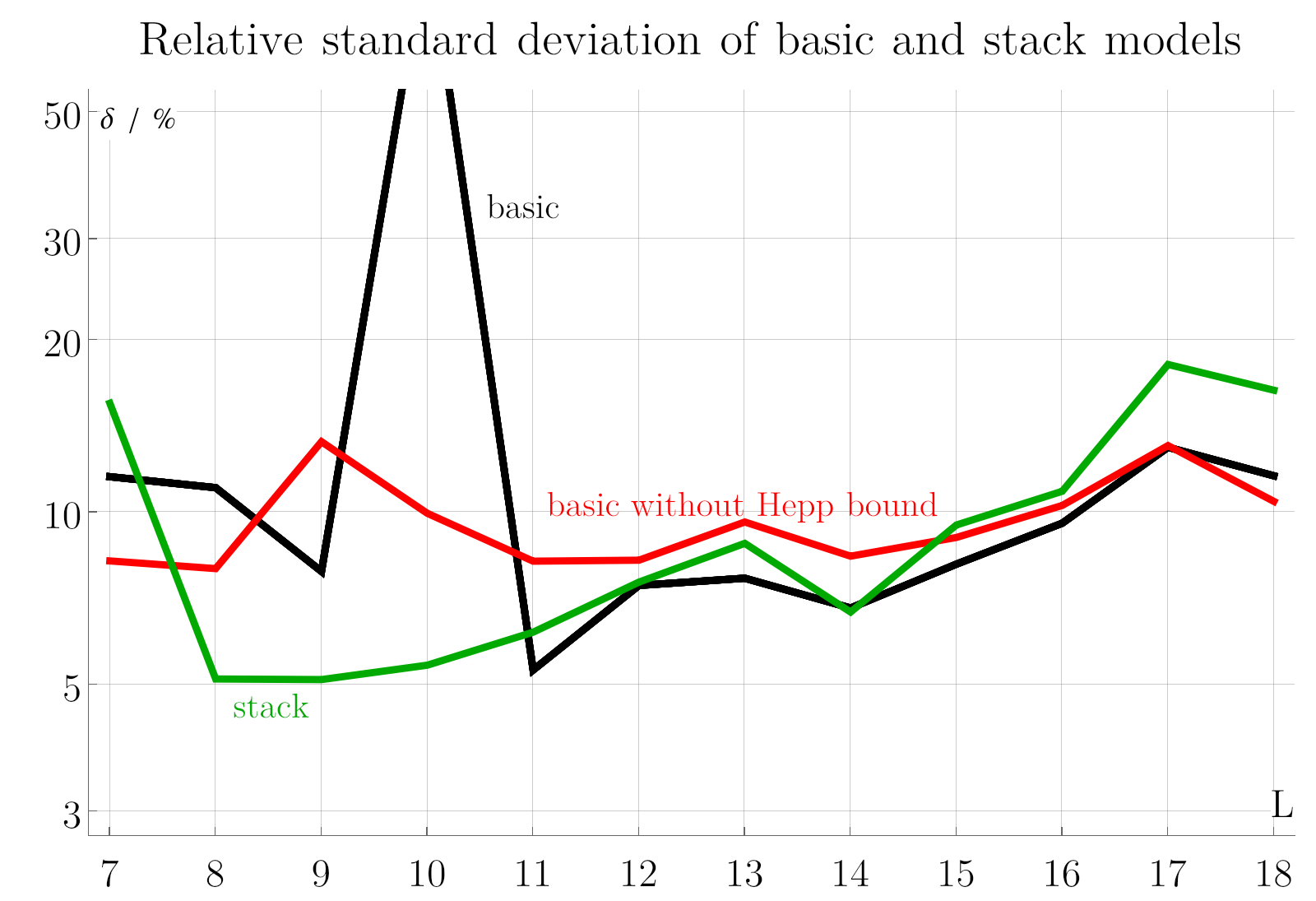}
		\subcaption{}
		\label{fig:basic_stack_delta}
	\end{subfigure}
	\begin{subfigure}{ .49 \linewidth}
		\centering
		\includegraphics[width=\linewidth]{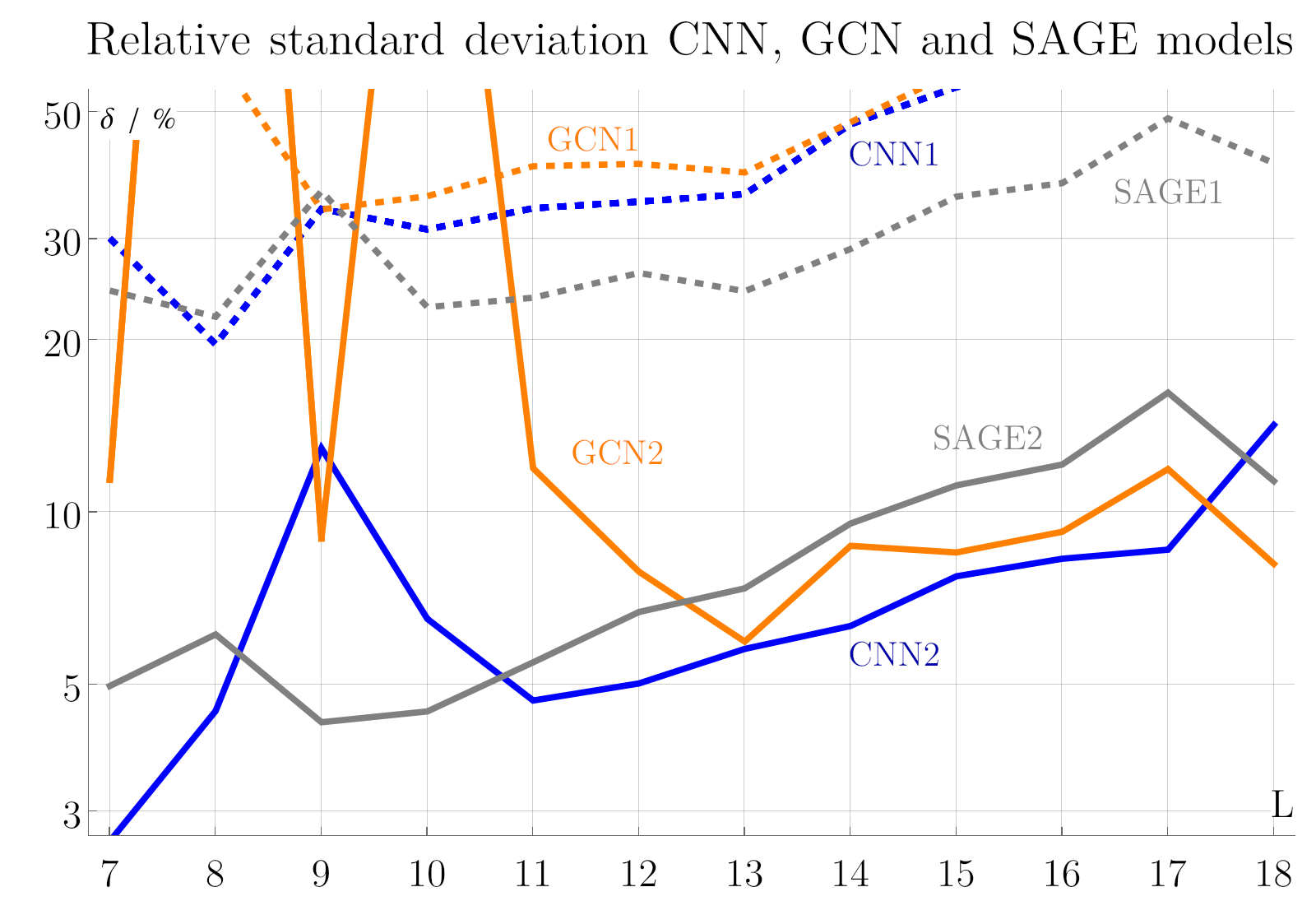}
		\subcaption{}
		\label{fig:GCN_CNN_delta}
	\end{subfigure}
	\caption{ 
		\textbf{(a)}  Relative standard deviation $\delta$ (\cref{def:relative_standard_deviation}) for the basic model (\cref{sec:basicmodel}) and the stack model (\cref{sec:stackmodel}). These models overall perform similarly, except from random fluctuation such as $L=10$ for the basic model.  
		\textbf{(b)} Relative standard deviation $\delta$ for CNN, GCN, and SAGE (\cref{sec:graph_structure}) models. For each of the three, the \enquote{1}-version (dashed) is the \enquote{pure} model, whereas the \enquote{2}-version additionally use the input features of the basic model. We see that adding the more advanced models gives only small increases in accuracy compared to the basic model, and all these models are by far inferior to regression models (compare  \cref{fig:regression}) 
		}
	
\end{figure}

As shown in \cref{fig:basic_stack_delta}, the stack model overall shows similar performance as the basic model, and is superior only for $L \leq 10$ loops.

\subsubsection{Models based on the graph structure}\label{sec:graph_structure}

The fact that the stack model (\cref{sec:stackmodel}) did not perform better than the basic model (\cref{sec:basicmodel}) suggests that the adjacency matrix, when flattened to a 1-dimensional vector, is not useful as an input feature. In this final section, we briefly comment on  three different types of artificial neural networks that operate more directly on the graph structure itself. 

A   \textbf{convolutional neural network} (CNN) processes a 2-dimensional input array  by convoluting it with a 2-dimensional kernel function \cite{goodfellow_deep_2016}, where the kernel parameters   are subject to training. 
CNNs are commonly used for image classification and computer vision, see e.g.  \cite{alzubaidi_review_2021}. In our case, we used the adjacency matrix, padded with zeros to a size of $20\times 20$, as an input \enquote{image}. Our model CNN1 has  4 convolutional layers,   pooled twice using a $2 \times 2$ pooling layer. Pooling can reduce computational effort and help eliminate noise from the data. For the model CNN2, we combined CNN1 with the basic model (\cref{sec:basicmodel}) such that the final output is a weighted average of the two.

The performance of the two CNN models is shown in \cref{fig:GCN_CNN_delta}. CNN2 reaches $\delta$ between 5\% and 10\%, similar to the basic model in \cref{fig:basic_stack_delta}, whereas CNN1 alone performs notably worse. From this we conclude that the CNN itself is essentially unable to predict the period. A probable reason for this is that the adjacency matrix is not invariant under vertex relabelings, and the CNN, which essentially treats the adjacency matrix as an \enquote{image}, was unable to incorporate this fact and therefore failed to infer the properties of the underlying graph from the given matrix.

In a \textbf{graph convolutional network} (GCN), see e.g. \cite{kipf_semisupervised_2017}, the convolution is computed for adjacent vertices, following the edges of the graph, instead of for adjacent \enquote{pixels} of an input array.  
For our model GCN1, we used 4 \texttt{GCNConv} layers with leaky ReLU activation function from the \texttt{PyTorch Geometric} library  as convolutional layers, and additionally 1 pooling layer and 3 dropout layers. The vertices where labeled with integers. The model GCN2 is a combination of GCN1 and the basic model (\cref{sec:basicmodel}). As shown in \cref{fig:GCN_CNN_delta}, the GCN models perform similar to the CNN models. The probable cause for the low accuracy of GCN1 is that the graphs in our application don't have any meaningful node features apart from the (random) node label. A GCN excels at processing data \emph{on} a graph, but in our case, the graph itself is the input data.

A \textbf{Graph Sample and Aggeregrate} (GraphSAGE) model is an architecture that samples neighbours from the nodes of the graph and aggregates features from the node's neighbor to create an embedding.   \cite{hamilton_inductive_2017a}.  We used a GraphSAGE implementation from \texttt{Stanford Network Analysis Project (SNAP)}  \cite{leskovec_snap_2016}. The SAGE1 model used 4 GraphSAGE convolutional layers, 1 pooling layer and 3 dropout layers. The SAGE2 model combines SAGE1 with a basic model.   As for the GCN, the node features were taken to be integer vertex labels. The performance of the SAGE models is similar to the previous two models, as shown in \cref{fig:GCN_CNN_delta}.

\bigskip
\noindent
This concludes our investigation of period prediction models. A summary of the results has been given above in \cref{sec:results}, discussion and outlook  in \cref{sec:discussion}.

\newpage 

\appendix

\section{Best fit parameters}

\begin{table}[h!]
		\centering
		\begin{tblr}{hlines,vlines,	rowsep=1pt,leftsep=2pt, rightsep=2pt,
				cells={r,font=\small},
				row{1-2}={c,rowsep=2pt},
				column{1}={r},
			}			
			L& $g_0$     & $g_6$ &$g_8$ & $g_{10} $ & $f_3$ &$f_4$ & $f_5 $ & $f_6$ &$f_7$  & $f_8 $ & $f_9$ &$f_{10}$   \\
			~8 &  6.1651 & 1.7734 & -0.0819 & -0.1463 & -0.7397 &-0.6443 &-0.6400 & -0.7738 & -1.0458 & -1.4507 & -1.9083 & -1.1101 \\
			~9 &  2.4183 & 1.1918 &  0.0336 & -0.0151 &  0.7795 & 0.5200 & 0.2788 & -0.0291 & -0.4214 & -0.9447 & -1.5452 & -1.9044  \\
			10 & -4.0960 & 0.7576 &  0.1042 &  0.0905 &  2.4795 & 1.9688 & 1.5752 &  1.1815 &  0.7015 &  0.1927 & -0.5441 & -1.0731 \\
			11 & -8.3113 & 0.5519 &  0.0949 &  0.1838 &  3.4224 & 2.8048 & 2.3731 &  1.9684 &  1.5019 &  0.9696 &  0.4036 & -0.4078 \\
			12 & -9.9380 & 0.4932 &  0.0536 &  0.1917 &  3.7959 & 3.1491 & 2.7340 &  2.3771 &  1.9635 &  1.4584 &  0.8666 &  0.2169 \\
			13 & -9.7290 & 0.4327 & -0.0309 &  0.1573 &  3.9281 & 3.2675 & 2.8647 &  2.5482 &  2.1997 &  1.7446 &  1.1759 &  0.5196 \\
			14 & -8.8307 & 0.3881 & -0.1209 &  0.1017 &  3.9640 & 3.2988 & 2.9029 &  2.6139 &  2.3199 &  1.9332 &  1.4043 &  0.7366 \\
			15 & -7.3938 & 0.3322 & -0.2268 &  0.0268 &  3.9669 & 3.2990 & 2.9040 &  2.6268 &  2.3655 &  2.0373 &  1.5740 &  0.9464 \\
			16 & -5.3670 & 0.2652 & -0.3588 & -0.0767 &  3.9526 & 3.2849 & 2.8867 &  2.6081 &  2.3599 &  2.0630 &  1.6603 &  1.1355\\
			17 & -3.6042 & 0.2735 & -0.4454 & -0.1798 &  3.8841 & 3.2312 & 2.8387 &  2.5651 &  2.3222 &  2.0469 &  1.7302 &  1.2539 \\
			18 & -2.5465 & 0.4568 & -0.4545 & -0.2252 &  3.6645 & 3.0543 & 2.6844 &  2.4343 &  2.2097 &  1.9558 &  1.6239 &  1.3253 \\
		\end{tblr}
		\caption{Best fit parameters of the combined cuts+cycles model (\cref{def:cut_cycle_approximation}) for  irreducible periods. The parameters are such that they predict $\ln \period$, where $\period$ is not scaled to its leading growth \cref{period_scaling}. This scaling would be a constant shift in $g_0$. }
		\label{tab:cuts_cycles_parameters}
\end{table}
\begin{table}[h!]
	\centering
	\begin{tblr}{hlines,vlines,	rowsep=1pt,
			cells={r,font=\small},
			row{1-2}={c,	rowsep=1pt},
			column{1}={r},
		}
		L  & $h_0$   & $h_1$  & $h_2$    & $h_3 $    & $g_6$    & $g_8$       & $g_{10} $    \\
		8  & -48.878 & 8.9692 & -0.51616 & 0.011953  & -0.32812 &  0.015376   & 0.007724    \\
		9  & -39.434 & 5.9285 & -0.26167 & 0.0051195 & -0.28710 &  0.012356   & 0.005760    \\
		10 & -38.298 & 4.9412 & -0.17819 & 0.0030215 & -0.28472 &  0.017506   & 0.012503     \\
		11 & -39.865 & 4.5608 & -0.14125 & 0.0021208 & -0.31109 &  0.015840   & 0.016871     \\
		12 & -42.145 & 4.3456 & -0.11844 & 0.0015961 & -0.34334 &  0.012780   & 0.022433     \\
		13 & -44.928 & 4.2327 & -0.10377 & 0.0012729 & -0.38356 &  0.007306   & 0.028879    \\
		14 & -49.246 & 4.3328 & -0.10024 & 0.0011512 & -0.43013 & -0.000599   & 0.036169    \\
		15 & -54.287 & 4.5082 & -0.10020 & 0.0010906 & -0.48282 & -0.013399   & 0.045720     \\
		16 & -64.665 & 5.2563 & -0.12138 & 0.0012971 & -0.53634 & -0.028454   & 0.054880    \\
		17 & -79.925 & 6.4321 & -0.15498 & 0.0016232 & -0.60232 & -0.050717   & 0.065161    \\
		18 & -94.114 & 7.3643 & -0.17654 & 0.0017757 & -0.65995 & -0.072315   & 0.069051   \\
	\end{tblr}
	\caption{Best fit parameters of the combined Hepp+cuts model (\cref{def:hepp_cut_approximation}) for irreducible periods. The parameters are such that they predict $\ln \period$, not scaled to \cref{period_scaling}. }
	\label{tab:hepp_parameters}
\end{table}

\FloatBarrier

\printbibliography

\end{document}